\documentclass[11pt, oneside]{article}   	% use "amsart" instead of "article" for AMSLaTeX format
\usepackage{geometry}                		% See geometry.pdf to learn the layout options. There are lots.
\usepackage{graphicx}				% Use pdf, png, jpg, or eps§ with pdflatex; use eps in DVI mode
\usepackage{amssymb,amsmath,bbm,bm}
\usepackage{setspace,booktabs}
\allowdisplaybreaks

\usepackage[longnamesfirst]{natbib}
\usepackage[noblocks]{authblk}
\newtheorem{theorem}{Theorem}[section]

\newtheorem{assumption}{Assumption}
\newtheorem{definition}{Definition}

\newtheorem{lemma}[theorem]{Lemma}
\newtheorem{remark}{Remark}

\newcommand{\norm}[1]{\left\lVert#1\right\rVert}

\title{A Design-Based Approach to Spatial Correlation}
\author{Ruonan Xu}
\affil{Rutgers University}
\author{Jeffrey M. Wooldridge}
\affil{Michigan State University}	
\date{}						

\begin{document}
\maketitle
\begin{abstract}
When observing spatial data, what standard errors should we report? With the finite population framework, we identify three channels of spatial correlation: sampling scheme, assignment design, and model specification. The Eicker-Huber-White standard error, the cluster-robust standard error, and the spatial heteroskedasticity and autocorrelation consistent standard error are compared under different combinations of the three channels. Then, we provide guidelines for whether standard errors should be adjusted for spatial correlation for both linear and nonlinear estimators. As it turns out, the answer to this question also depends on the magnitude of the sampling probability.
\end{abstract}

\section{Introduction}

A common concern among empirical researchers is spatial correlation, as policy or event treatments are often spatially correlated. A merger between two gasoline companies, for example, could affect the local gasoline retail market spatially and disproportionately in the neighborhoods close to the rebranded stations \citep{houde2012spatial}. Closure and demolition of public housing affects the areas closest to the projects more than those farther away \citep{aliprantis2015blowing}. Along with spatially correlated treatments, there can also be spillover effects to adjacent entities that diminish with distance. One common point of confusion is whether standard errors should be adjusted for spatial correlation. In this paper, we answer this question by disentangling sampling scheme, assignment design, and model specification under a finite population framework.

Adopting the finite population paradigm when studying spatial data has clear advantages. For one, any cases where spatial effects are of interest, the unit of observations is determined by geography or generalized distance measure, and we often sample all units in the population. For example, typically we can collect information on all counties in the United States.  As pointed out by \cite{pinkse2007central}, ``with spatial data, it is common for the sample and the population to be the same (e.g., the set of all firms in a market)" (p.~216). The sampling uncertainty underlying the superpopulation approach is unnatural for thinking about uncertainty in such settings; see also \cite{abadie2020sampling}. 

Second, with a well-defined finite population, we can explicitly introduce different sampling schemes after imposing spatial correlations on the population units. Rather than fixing the sites and treating different realizations of the data at all sites as random draws from a superpopulation, it is more practical to treat the lattice as a finite population and draw different collections of sites from it. The latter is something we can actually do in real life, but the former is merely a thought experiment. Along with sampling, we observe only one potential outcome at each location. Therefore, our framework combines sampling- and design-based uncertainty. We can also differentiate between spatial correlation among all neighbors in the population and correlation within a subset of neighbors observed in the sample, which has a major impact on statistical inference. With that said, because we allow the sampling probability to shrink to zero, our asymptotic theory can also accommodate sampling from superpopulations as a special case.

By means of a set of newly developed limit theorems, we derive the finite population spatial heteroskedasticity and autocorrelation consistent (SHAC) variance-covariance matrix of M-estimators. We show that, in the case where a nontrivial fraction of the population is sampled, the new asymptotic variance matrix is smaller than the superpopulation SHAC variance matrix. This finding generalizes \cite{neyman1923application}'s result on conservativeness of the finite population variance estimation and the extension to regression settings in \cite{abadie2020sampling}. Based on the alternate finite population variance-covariance matrix, we provide guidelines on whether standard errors should be adjusted for spatial correlation instead of relying on heuristic arguments of spatial correlation of unobserved characteristics. 

To summarize, whenever there is spatial assignment, meaning that the ``treatment” variables are spatially correlated, or a spillover effect is estimated, even in the absence of spillover, one should adjust standard errors for spatial correlation, for instance, by reporting the SHAC standard errors. However, there are a few exceptions. When we independently sample a negligible portion of the population, the Eicker-Huber-White (EHW) standard errors would suffice irrespective of the existence of spatial correlation. Similarly, if we sample a small fraction of clusters from all the clusters in the population, the cluster-robust standard errors would suffice.

Our paper contributes to three strands of literature. Our first contribution is to the literature on limit theorems for random fields. There is a very general asymptotic theory for spatial processes under either the mixing condition or near-epoch dependence in \cite{jenish2009central} and \cite{jenish2012spatial}. Unfortunately, due to the introduction of sampling indicators necessary for our paper, their theory cannot incorporate sampling from a superpopulation. Recently, \cite{bradley2017central} develop a central limit theorem for nonstationary random fields using strong mixing conditions with certain restrictions. By borrowing the techniques in the aforementioned articles, we derive new laws of large numbers and a central limit theorem for near-epoch dependent (NED) processes. We apply these limit theorems to finite population asymptotic theory but they also accommodate superpopulations by including zero sampling probability in the limit. Meanwhile, we allow for nonstationary processes with unbounded moments on irregularly spaced lattices. We also introduce cluster correlation on top of spatial correlation by explicitly including the sampling indicators and allowing for cluster sampling.

Second, we contribute to the literature on finite population inference. Several articles under the finite population framework study the EHW standard error and the cluster-robust standard error for both linear and nonlinear estimators. See, for instance, \cite{abadie2020sampling}, \cite{abadie2017should}, \cite{xu2021potential}, and \cite{xu2021asymptotic}. \cite{bojinov2021panel} extend the finite population framework to panel experiments. They allow for spillovers across time periods but maintian the assumption of no spillover across units. \cite{savje2021average}, \cite{savje2021causal}, and \cite{leung2022causal} consider network/spatial interference in randomized experiments with access to the entire population. Their work focuses on estimating the treatment effect or exposure effect consistently. Our paper is the first to examine the necessity of spatial-correlation robust inference under different scenarios for a general class of estimators. Lastly, we contribute to the literature on spatial econometrics. A summary of recent developments can be found in \cite{xu2019theoretical}.

The remainder of the paper is organized as follows. We derive the asymptotic distribution for M-estimators under finite populations with spatial correlation in Section 2. In Section 3, the asymptotic theory is extended to functions of M-estimators. Among the leading examples are average partial effect (APE) estimators resulting from nonlinear models. Using simulation studies in Section 4, we compare the small sample performance of the EHW, cluster-robust, and SHAC standard errors. The research is concluded in Section 5. Proofs are collected in the appendix.

%%%%%%%%%%%%%%%%%%%%%%%%%%%%%%%%%%%%%%%%%%%%%%%%%%%%%%%%%%%%%%%%%%%%%%%%%%%%%%%%%%%%%
\section{Asymptotic Properties of M-estimators}

\subsection{Setup}

Let $D\subseteq \mathbb{R}^d$, $d\geq 1$, be a lattice of (possibly) unevenly placed locations in $\mathbb{R}^d$. Consider a sequence of finite subsets of $D$, $\{D_M\}$, where $M$ indexes the sequence of finite populations. $|D_M|$ diverges to infinity in deriving the asymptotic properties, where $|V|$ denotes the cardinality of a finite subset $V\subseteq D$. We adopt the metric $\nu(i,j)=\max_{1\leq l\leq d}|j_l-i_l|$ in space $\mathbb{R}^d$, where $i_l$ is the $l$-th component of $i$. The distance between any subsets $K,V\subseteq D$ is defined as $\nu(K,V)=\inf\{\nu(i,j): i\in K\mbox{ and } j\in V\}$. For any random vector W, $\norm{W}_p=(E\norm{W}^p)^{1/p}$, $p\geq 1$, denotes its $L_p$-norm. Let $\mathcal{F}_{iM}(s)=\sigma(U_{jM}; j\in T_M: \nu(i,j)\leq s)$ be the $\sigma$-field generated by the random vectors $U_{jM}$ located in the $s$-neighborhood of location $i$. Lastly, $C$ denotes a generic positive constant that may be different in different circumstances.

Let $(X, z, Y)=\{(X_{iM}, z_{iM}, Y_{iM}), i\in D_M, M\geq 1\}$ and $U=\{U_{iM}, i\in T_M, M\geq 1\}$ be triangular arrays of random fields defined on a probability space $(\Omega, \mathcal{F}, P)$. For each unit $i$, we observe $(X_{iM}, z_{iM}, Y_{iM})$, where $X_{iM}$ is the vector of assignment variables, $z_{iM}$ is a set of attributes, and $Y_{iM}$ is the realized outcome. The categorization of assignments and attributes is based on the posed empirical question. Typically, the key variables of interest in an empirical study could be viewed as assignment variables, and the remaining covariates as attribute variables. There is no restriction in terms of the nature of the triple above: they can be discrete, continuous, or mixed. 

We relax the stable unit treatment value assumption in the standard potential outcome framework by allowing for interference among individuals. There exists a mapping, denoted by the potential outcome function $y_{iM}(\bm{x}_M)$, from a vector of assignment variables of all population units to the potential outcomes, where $\bm{x}_M=\{x_{iM}, i\in D_M, M\geq 1\}$.\footnote{I emphasize $\bm{x}_M$ as the argument of the potential outcome function because its realization, $\bm{X}_M$, is the only stochastic vector in the function.} The potential outcome function, $y_{iM}(\cdot)$, along with the observed attributes $z_{iM}$, are non-stochastic. By contrast, the assignment vector $\bm{X}_M$ is random, where $\bm{X}_M=\{X_{iM}, i\in D_M\}$ is the realization of $\bm{x}_M$. As a result, the realized potential outcome, $Y_{iM}=y_{iM}(\bm{X}_M)$, is random. Alternatively, the finite population setting can be understood as conditioned on the potential outcomes and attributes of the $|D_M|$ units in the population. For the most part, we denote $W_{iM}=(\bm{X}_M, Y_{iM})$ for brevity.

We do not take a stance on whether an underlying model is correctly specified. There currently exist discussions on incorporating interference into causal inference and how to consistently estimate the exposure effect with misspecified exposure mappings; see, for instance, \cite{hudgens2008toward} and \cite{savje2021causal}. With misspecified interference, we estimate some approximation of the spillover effects in the sample. \cite{manski2013identification} discusses the identification of potential outcome distributions with social interactions, which is out of the scope of the current paper. Throughout, we assume that the finite population parameters are identified whether or not any feature of the model is correctly specified. 

According to the sampling scheme, the population $M$ can be partitioned into $G_M$ mutually exclusive clusters, $\{D_{gM}: g=1,2,\dots,G_M\}$, based on the primary sampling units. $C_{iM}\in\{1,2,\dots,G_M\}$ denotes the cluster that unit $i$ belongs to. %Let $C_{igM}=\mathbbm{1}\{C_{iM}=g\}$ be the cluster indicator, equal to one if unit $i$ in population $M$ belongs to cluster $g$ and zero otherwise. 
Each cluster size is denoted by $|D_{gM}|$ %$M_g=\sum_{i\in D_M}C_{igM}$, 
with $|D_M|=\sum_{g=1}^{G_M}|D_{gM}|$, $g\in\{1,2,\dots,G_M\}$. When the primary sampling units are individual entities, random sampling is included as a special case. 

As is the starting point in the superpopulation paradigm, we study solutions to a population minimization problem, where the estimand of interest is a $k\times 1$ vector denoted by $\theta^*_M$:
\begin{equation}\label{eq13}
\begin{aligned}
\theta^*_M&=\arg\min_\theta \frac{1}{|D_M|}\sum_{g=1}^{G_M}\sum_{j\in {D_{gM}}} \mathbb{E}\big[q_{jM}(W_{jM}, \theta)\big]\\
&=\arg\min_\theta \frac{1}{|D_M|}\sum_{i\in {D_M}} \mathbb{E}\big[q_{iM}(W_{iM}, \theta)\big]
\end{aligned}
\end{equation}
The interpretation of the finite population parameter is up to the researcher; see for instance, \cite{rambachan2020design}, for discussion on when the finite population estimand carries a causal interpretation.  
Notice that the expectation $\mathbb{E}$ in equation (\ref{eq13}) is taken over the distribution of $X_M$ since $X_M$ is the source of randomness here. 
The function $q_{iM}(\cdot, \cdot)$ is the objective function for a single unit. The subscripts of the objective function indicate its dependence on the non-stochastic attribute variables $\{z_{iM},i\in D_M\}$. 

Let $R_{iM}$ denote the binary sampling indicator, which is equal to one if unit $i$ is sampled. Hence, the sample size is $|D_N|=\sum^{G_M}_{g=1}\sum_{i\in D_{gM}}R_{iM}=\sum_{i\in D_M}R_{iM}$. Below we will be precise about the nature of the sampling scheme but, unless the sample equals the population, the sample size is random. The spatial M-estimator of $\theta^*_M$ is denoted by $\hat{\theta}_N$, which solves the minimization problem in the sample.
\begin{align*}
\hat{\theta}_N&=\arg\min_\theta\frac{1}{|D_N|}\sum_{g=1}^{G_M}\sum_{j\in {D_{gM}}} R_{jM}q_{jM}(W_{jM}, \theta)\\
&=\arg\displaystyle \min_\theta\frac{1}{|D_N|}\sum_{i\in {D_M}} R_{iM}q_{iM}(W_{iM}, \theta)
\tag{\stepcounter{equation}\theequation}
\end{align*}

In order to establish desirable asymptotic properties of the spatial M-estimator, we need to impose restrictions on the spatial dependence of the stochastic components in the estimation problem. We adopt the definition of mixing coefficients in \cite{bradley2017central}, near-epoch dependent (NED) random fields in \cite{jenish2012spatial}, and $m$-dependent random fields in \cite{moricz2008strong}.
\begin{definition}
Let $\mathcal{A}$ and $\mathcal{B}$ be two sub-$\sigma$-algebras of $\mathcal{F}$, and let 
\begin{equation*}
\alpha(\mathcal{A}, \mathcal{B})=\sup(|P(AB)-P(A)P(B)|, A\in\mathcal{A}, B\in\mathcal{B})
\end{equation*}
and 
\begin{equation*}
\rho(\mathcal{A}, \mathcal{B})=\sup|corr(f,g)|, f\in L^2_{real}(\mathcal{A}), g\in L^2_{real}(\mathcal{B}).
\end{equation*}
For $K\subseteq D_M$ and $V\subseteq D_M$, let $\sigma_M(K)=\sigma(U_{iM}, i\in K)$ and $\alpha_M(K,V)=\alpha(\sigma_M(K), \sigma_M(V))$. Then, the $\alpha$-mixing coefficient for the random field $U$ is defined as:
\begin{equation*}
\bar{\alpha}(r)=\sup_{M}\sup_{K,V}(\alpha_M(K,V), \nu(K,V)\geq r).
\end{equation*}
The maximal correlation coefficient is defined as:
\begin{equation*}
\bar{\rho}(r)=\sup_{M}\sup_{K,V}(\rho_M(K,V), \nu(K,V)\geq r).
\end{equation*}

\end{definition}

\begin{definition}
Let $W=\{W_{iM}, i\in D_M, M\geq 1\}$ be a random field%with $\norm{W_{iM}}_p<\infty$, $p\geq 1$, 
, let $U=\{U_{iM}, i\in T_M, M\geq 1\}$ be another random field, where $|T_M|\to \infty$ as $M\to\infty$, and let $d=\{d_{iM}, i\in D_M, M\geq 1\}$ be an array of finite positive constants. Then the random field $W$ is said to be $L_p(d)$-near-epoch dependent on the random field $U$ if 
\begin{equation*}
\norm{W_{iM}-E(W_{iM}|\mathcal{F}_{iM}(s))}_p\leq d_{iM}\psi(s)
\end{equation*}
for some sequence $\psi(s)\geq 0$ with $\lim_{s\to\infty}\psi(s)=0$. The $\psi(s)$ are called the NED coefficients, and the $d_{iM}$ are called the NED scaling factors. $W$ is said to be $L_p$-NED on $U$ of size $-\lambda$ if $\psi(s)=O(s^{-\mu})$ for some $\mu>\lambda>0$.
\end{definition}

\begin{definition}
A random field $U=\{U_{iM},i\in D_M, M\geq 1\}$ is called $m$-dependent if for all finite subsets $K, V \subset D$ with $\nu(K,V)>m$ the $\sigma$-algebras $\sigma(U_{iM}, i\in K)$ and $\sigma(U_{iM}, i\in V)$ are independent. 
\end{definition}

We make the following assumptions. Detailed regularity assumptions are listed in Appendix A. 
\begin{assumption}\label{assump1}
Suppose $\{D_M\}$ is a sequence of finite subsets of $D$ such that $|D_M|\to \infty$ as $M\to \infty$, where the lattice $D\subseteq \mathbb{R}^d$, $d\geq 1$, is infinitely countable. All elements in $D$ are located at distances of at least $\nu_0>0$ from each other, i.e., for all $i,j\in D$: $\nu(i,j)\geq \nu_0$; w.l.o.g. we assume that $\nu_0>1$. 
\end{assumption}

\begin{assumption}\label{assump2}
(\romannumeral 1) The sampling scheme consists of two steps. In the first step, a random group of clusters is drawn according to Bernoulli sampling with probability $\rho_{cM}>0$; in the second step, units are independently sampled, according to a Bernoulli trial with probability $\rho_{uM}>0$, from the subpopulation consisting of all the sampled clusters. (\romannumeral 2) The sequence of sampling probabilities $\rho_{cM}$ and $\rho_{uM}$ satisfies $\rho_{cM }\to \rho_c \in [0,1]$, $\rho_{uM}\to\rho_u \in [0,1]$, and $|D_M|\rho_{uM}\rho_{cM}\to\infty$ as $M\to\infty$.
\end{assumption}

\begin{assumption}\label{assump3}
$\max_{1\leq g\leq G_M} |D_{gM}|\leq C<\infty$ as $M\to\infty$.
\end{assumption}

\begin{assumption}\label{assump4}
The sampling indicators, $R=\{R_{iM}, i\in D_M, M\geq 1\}$, are independent of the assignment variables, $X=\{X_{iM}, i\in D_M, M\geq 1\}$, and the underlying mixing random fields, $U=\{U_{iM}, i\in T_M, M\geq 1\}$, where $D_M\subseteq T_M\subseteq D$.
\end{assumption}

\begin{assumption}\label{assump5} (Mixing condition)
For the input random field $U$: (\romannumeral 1) $\bar{\alpha}(r)\to 0$ as $r\to \infty$; (\romannumeral 2) $\displaystyle\lim_{r\to\infty}\bar{\rho}(r)<1$.
\end{assumption}

\begin{assumption}\label{assump6} (NED condition)
The random field $g=\{g_{iM}(W_{iM},\theta), i\in D_M, M\geq 1\}$ is $L_2$-NED on $U=\{U_{iM},i\in T_M, M\geq 1\}$ with the scaling factors $d_{iM}$ and the NED coefficients $\psi(s)$ of size $-2d(r-1)/(r-2)$ for some $r>2$. The $g(\cdot)$ function includes $q_{iM}(W_{iM},\theta)$, $m_{iM}(W_{iM},\theta)$, $\nabla_{\theta} m_{iM}(W_{iM},\theta)$, $f_{iM}(W_{iM},\theta)$, and $\nabla_{\theta} f_{iM}(W_{iM},\theta)$ defined in Appendix A.
\end{assumption}

\begin{assumption}\label{assump7}
The weights satisfy: $\omega (0)=1$; $\omega\Big(\frac{\nu(i,j)}{b_M}\Big)=0$ for any $\nu(i,j)>b_M$; $\Big|\omega\Big(\frac{\nu(i,j)}{b_M}\Big)\Big|<\infty$, $\forall$ $\nu(i,j)$ and $\forall$ $M$; $\lim_{M\to\infty}\frac{1}{|D_M|}\bigg|\sum_{i\in D_M}\sum_{j\in D_M: \nu(i,j)\leq b_M}\Big[\omega\Big(\frac{\nu(i,j)}{b_M}\Big)-1\Big]\cdot Cov\Big(\frac{R_{iM}}{\sqrt{\rho_{uM}\rho_{cM}}}h_{iM}, \frac{R_{jM}}{\sqrt{\rho_{uM}\rho_{cM}}}h_{jM}\Big)\bigg|=0$, where $b_M=o\big((|D_M|\rho_{uM}\rho_{cM})^{1/2d}\big)$ and the $h(\cdot)$ function includes $m_{iM}(W_{iM},\theta^*_M)/J_M$ and $\big(f_{iM}(W_{iM},\theta^*_M)-F_M(\theta^*_M)H_M(\theta^*_M)^{-1}m_{iM}(W_{iM},\theta^*_M)\big)/J_M$ as defined in Appendix A.
\end{assumption}

Assumption 1 is taken from \cite{jenish2012spatial}. Consistent with the increasing domain asymptotics, the assumption of the minimum distance ensures the expansion of the sample region. In Assumption 2, we introduce the possibility of cluster correlation via the two-stage sampling scheme. It is helpful to go through various sampling schemes resulted from different values of the sampling probabilities. With $\rho_{cM}=\rho_{uM}=1$, we observe the entire population; $\rho_{cM}=1$ and $\rho_{uM}<1$ means random sampling; $\rho_{cM}<1$ and $\rho_{uM}\leq 1$ implies cluster sampling. In particular, sampling from a superpopulation is nested in our unified theory since the sampling probabilities in both steps are allowed to be zero in the limit. When $\rho_c=0$, only a negligible fraction of clusters are sampled from a population of a large number of clusters; while when $\rho_c=1$ and $\rho_u=0$, a negligible portion of units are randomly drawn from a large population. Regardless, the expected sample size, $|D_M|\rho_{uM}\rho_{cM}$, diverges to infinity. 

Assumption 3 imposes boundedness of the cluster sizes whenever the population units are partitioned into clusters because of either cluster assignment or cluster sampling. As a result, the number of clusters in the population diverges to infinity along with the population size. Assumption 4 implies that the sampling process and the assignment process are independent of each other, which rules out sample selection due to assignment status.

Assumption 5 is the key weak dependence assumption in the proof of the asymptotic properties. Assignment variables are allowed to be spatially correlated as long as the spatial correlation dies out along with distance. In addition to the mixing condition, we also require the maximal correlation coefficient to be less than one in the limit. %Explicit formulas for maximal correlation are rarely known. As a special case, if the assignment variables are multivariate normal, then Assumption \ref{assump5}(\romannumeral 2) requires that the assignment variables for units far away are not perfectly correlated (either positively or negatively). 
With the extension to NED processes in Assumption 6, we can study a more generalized class of random fields.\footnote{Assumption 6 is more than sufficient because in some cases, we only need $L_1$-NED, which can be implied by $L_2$-NED. For certain functions, the NED coefficients are only required to be of size $-d$.} Here, we impose NED conditions on the objective function, score function, and the Hessian matrix directly so that both continuous and discrete variables are allowed. See, for example, Chapter 4 in \cite{gallant1988unified} for primitive conditions that ensure preservation of the NED property under transformations. Assumption 7 is required to establish consistency of the SHAC variance estimator. The last part of Assumption 7 is a high-level condition, which requires that the kernel weights $\omega\Big(\frac{\nu(i,j)}{b_M}\Big)$ converge to one sufficiently fast as $M\to\infty$.

%%%%%%%%%%%%%%%%%%%%%%%%%%%%%%%%%%%%%%%%%%%%%%%%%%%%%%%%%%%%%%%%%%%%

\subsection{Asymptotic Distribution}\label{sec2.2}
We introduce the following notation for the variance-covariance matrix of M-estimators. 
Define 
\begin{align*}
S_M=&\Delta_{ehw,M}(\theta^*_M)+\rho_{uM}\Delta_{cluster,M}(\theta^*_M)+\rho_{uM}\rho_{cM}\Delta_{spatial,M}(\theta^*_M)\\
&-\rho_{uM}\rho_{cM}\Delta_{E,M}-\rho_{uM}\rho_{cM}\Delta_{EC,M}-\rho_{uM}\rho_{cM}\Delta_{ES,M}
\tag{\stepcounter{equation}\theequation}
\end{align*}
and
\begin{equation}
V_M=H_M(\theta^*_M)^{-1}S_MH_M(\theta^*_M)^{-1},
\end{equation}
where 
\begin{equation}
\Delta_{ehw,M}(\theta)=\frac{1}{|D_M|}\sum_{i\in D_M}\mathbb{E}\big[m_{iM}(W_{iM},\theta)m_{iM}(W_{iM},\theta)'\big],
\end{equation}
\begin{equation}
\Delta_{E,M}=\frac{1}{|D_M|}\sum_{i\in D_M}\mathbb{E}\big[m_{iM}(W_{iM},\theta^*_M)\big]\mathbb{E}\big[m_{iM}(W_{iM},\theta^*_M)\big]',
\end{equation}
\begin{equation}
\Delta_{cluster,M}(\theta)=\frac{1}{|D_M|}\sum_{i\in D_M}\sum_{j\in D_M, j\neq i}\mathbbm{1}(C_{iM}=C_{jM})\mathbb{E}\big[m_{iM}(W_{iM},\theta)m_{jM}(W_{jM},\theta)'\big],
\end{equation}
\begin{equation}
\Delta_{EC,M}=\frac{1}{|D_M|}\sum_{i\in D_M}\sum_{j\in D_M, j\neq i}\mathbbm{1}(C_{iM}=C_{jM})\mathbb{E}\big[m_{iM}(W_{iM},\theta^*_M)\big]\mathbb{E}\big[m_{jM}(W_{jM},\theta^*_M)\big]',
\end{equation}
\begin{equation}
\Delta_{spatial,M}(\theta)=\frac{1}{|D_M|}\sum_{i\in D_M}\sum_{j\in D_M, j\neq i}\mathbbm{1}(C_{iM}\neq C_{jM})\mathbb{E}\big[m_{iM}(W_{iM},\theta)m_{jM}(W_{jM},\theta)'\big],
\end{equation}
\begin{equation}
\Delta_{ES,M}=\frac{1}{|D_M|}\sum_{i\in D_M}\sum_{j\in D_M, j\neq i}\mathbbm{1}(C_{iM}\neq C_{jM})\mathbb{E}\big[m_{iM}(W_{iM},\theta^*_M)\big]\mathbbm{E}\big[m_{jM}(W_{jM},\theta^*_M)\big]',
\end{equation}
and 
\begin{equation}
H_M(\theta)=\frac{1}{|D_M|}\sum_{i\in D_M}\mathbb{E}\big[\nabla_\theta m_{iM}(W_{iM},\theta)\big].
\end{equation}

\begin{theorem}\label{thm1}
Under Assumptions \ref{assump1}-\ref{assump6}, and Assumption \ref{assumpA1} in Appendix A, $V_M^{-1/2}|D_N|^{1/2}(\hat{\theta}_N-\theta^*_M)\overset{d}\to \mathcal{N}(\textbf{0}, I_k)$.
\end{theorem}

%\begin{remark}
%The EHW standard errors are sufficient unless there is cluster sampling, cluster assignment, or spatial assignment.
%\end{remark}

Theorem \ref{thm1} is derived using a set of new limit theorems given in the Appendix. It shows that the M-estimators are asymptotically normal with an alternative finite population SHAC variance-covariance matrix. The result holds for spatial assignments both at the individual level or the cluster level. For the latter assignment design, there is cluster assignment in the first place, where assignments within clusters can be strongly correlated. On top of that, the assignments across clusters are allowed to be spatially correlated as long as the key mixing condition in Assumption \ref{assump5} is satisfied. With bounded cluster sizes, units far away from each other fall into different clusters and the spatial correlation eventually can die out. For instance, while studying individual outcomes, policies are imposed at the school district level and the school districts nearby may coordinate in certain ways.

%\begin{remark} (Convergence rate)\label{remark2}
%$\hat{\theta}_N$ is $|D_N|^{1/2}$-consistent for $\theta^*_M$.
%\end{remark}

\begin{remark}\label{remark1}
We should only adjust standard errors for spatial correlation if: (\romannumeral 1) assignment variables are spatially correlated; or (\romannumeral 2) spillover effects are specified in the model.
\end{remark}

Let us first focus on positive sampling probabilities. Based on the variance-covariance matrix in (3) and (4), we can see that sampling scheme plays a role in cluster correlation but not in spatial correlation. The two terms involving spatial correlation, $\Delta_{spatial,M}(\theta^*_M)$ and $\Delta_{ES,M}$, cancel out whenever the assignment variables are independent. Therefore, it is only necessary to adjust the standard errors for spatial correlation if the assignment variables are spatially correlated either because of the assignment design or the inclusion of the spillover effects. Also notice that, even in the absence of spillover effects in the potential outcome function, we manually introduce spatial correlation among assignment variables when we explicitly specify spillover effects in the model. Here, assignment variables include your own ``treatment" and your neighbors' ``treatments." Another implication of Remark \ref{remark1} is that we can test the necessity of spatial-correlation robust inference since assignment variables are observable to us. 

We report the SHAC standard errors in the simulation below as one way to adjust standard errors for spatial correlation, as this is the dominant approach in the literature so far. Nevertheless, there are alternative ways to construct standard errors and confidence intervals robust to spatial correlation; see for instance, \cite{muller2022spatiala} and \cite{muller2022spatialb}. Regardless, Remark \ref{remark1} goes through in the population variance-covariance matrix.

\begin{remark}\label{remark2}
(\romannumeral 1) When $\rho_u=0$, reporting the EHW standard error would suffice; (\romannumeral 2) When $\rho_c=0$, reporting the cluster-robust standard error would suffice.
\end{remark}

When we switch to zero sampling probabilities, there are exceptions. In theory, when we observe the entire population or sample a large portion from a finite population, the spatial-correlation robust standard errors should be reported to account for spatial assignments. However, if we sample a minimal amount from the population, either the cluster-robust or the EHW standard errors would suffice under spatial assignments. Intuitively, when we independently sample a small proportion of units from the population, the majority of your neighbors would not be observed. Hence, there would be little difference between the EHW, cluster-robust, and the SHAC standard errors. In another case, when we randomly draw a small fraction of clusters from the population, most of your neighbors would be contained in the clusters drawn. Therefore, the cluster-robust standard errors take the majority of the spatial correlation into account and would deviate little from the SHAC standard errors with a sufficient bandwidth. In addition, when either of the sampling probability is zero, we are essentially sampling from a superpopulation, so the usual EHW standard error or the usual cluster-robust standard error would no longer be conservative. 

\subsection{Estimation of the Variance-Covariance Matrix}
Define 
\begin{equation}
\hat{V}_{SN}=\hat{H}_N(\hat{\theta}_N)^{-1}\hat{S}_N(\hat{\theta}_N)\hat{H}_N(\hat{\theta}_N)^{-1},
\end{equation}
where 
\begin{equation}
\hat{H}_N(\theta)=\frac{1}{|D_N|}\sum_{i\in D_M}R_{iM}\nabla_\theta m_{iM}(W_{iM},\theta)
\end{equation}
and 
\begin{equation}
\hat{S}_N(\theta)=\frac{1}{|D_N|}\sum_{i\in D_M}\sum_{j\in D_M}R_{iM}R_{jM}\cdot \omega\bigg(\frac{\nu(i,j)}{b_M}\bigg)m_{iM}(W_{iM},\theta)m_{jM}(W_{jM},\theta)'.
\end{equation}
\begin{theorem}\label{thm3}
Under Assumptions \ref{assump1}-\ref{assump7}, and Assumptions \ref{assumpA1}-\ref{assumpA2} in Appendix \ref{appendix1}, $\hat{V}_{SN}-(V_M+\rho_{uM}\rho_{cM}V_E)\overset{p}\to \textbf{0}$, where $V_E=H_M(\theta^*_M)^{-1}S_E H_M(\theta^*_M)^{-1}$ and \\$S_E=\frac{1}{|D_M|}\sum_{i\in D_M}\sum_{j\in D_M}\omega\bigg(\frac{\nu(i,j)}{b_M}\bigg)\mathbb{E}\big[m_{iM}(W_{iM},\theta^*_M)\big]\mathbb{E}\big[m_{jM}(W_{jM},\theta^*_M)\big]'$.
\end{theorem}

\begin{remark}
The usual SHAC variance estimator is conservative for the finite population SHAC variance-covariance matrix unless the sampling probabilities are zero. 
\end{remark}

In Section \ref{sec2.2}, we have summarized when you should and should not adjust the standard errors for spatial correlation. In terms of the estimation of the variance matrix, Theorem \ref{thm3} shows that when spatial correlation needs to be accounted for, there is an upward bias of the usual SHAC variance estimator. \cite{leung2022causal} reaches a similar conclusion for a weighted difference-in-means estimator for network data with interference, assuming the entire population is observed. We additionally introduce sampling probabilities and study both linear and nonlinear estimators. 

We do not discuss the superpopulation limit of the extra term, $S_E$, as this requires weak dependence assumptions on the superpopulation, which we have not touched upon yet. However, since the usual SHAC variance estimator is consistent for the superpopulation variance-covariance matrix, the finite population SHAC variance matrix should be smaller than the superpopulation version of the variance matrix, in the matrix sense.     

\section{Asymptotic Distribution of Functions of M-estimators}\label{sec3}

Given that we study M-estimators, the APE estimator from nonlinear models would be of great interest. As a result, we study the asymptotic distribution of a generic function of M-estimators to complete the discussion.

Let $f_{iM}(W_{iM},\theta^*_M)$ be a $q \times 1$ function of $W_{iM}$ and $\theta^*_M$. We wish to estimate $\gamma^*_M=\frac{1}{|D_M|}\sum_{i\in D_M}\mathbb{E}\big[f_{iM}(W_{iM},\theta^*_M)\big]$. Let $\hat{\gamma}_N=\frac{1}{|D_N|}\sum_{i\in D_M}R_{iM}f_{iM}(W_{iM},\hat{\theta}_N)$ be the estimator of $\gamma^*_M$. 
Denote the finite population variance matrix by
\begin{align*}
V_{f,M}=&\Delta^f_{ehw,M}+\rho_{uM}\Delta^f_{cluster,M}+\rho_{uM}\rho_{cM}\Delta_{spatial,M}^f\\
&-\rho_{uM}\rho_{cM}\Delta^f_{E,M}-\rho_{uM}\rho_{cM}\Delta^f_{EC,M}-\rho_{uM}\rho_{cM}\Delta^f_{ES,M}.
\tag{\stepcounter{equation}\theequation}
\end{align*}
And the usual SHAC variance estimator is denoted by $\hat{V}_{f,SN}$.
The detailed definition of each term can be found in Appendix \ref{appendix1}. 

\begin{theorem}\label{thm4}
Under Assumptions \ref{assump1}-\ref{assump7}, and Assumptions \ref{assumpA1}-\ref{assumpA3} in Appendix \ref{appendix1}, (1) $V_{f,M}^{-1/2}|D_N|^{1/2}(\hat{\gamma}_N-\gamma^*_M)\overset{d}\to \mathcal{N}(\textbf{0}, I_q)$; (2) $\hat{V}_{f,SN}-(V_{f,M}+\rho_{uM}\rho_{cM}V_{f,E})\overset{p}\to \textbf{0}$.
\end{theorem} 

We see that the same results for M-estimators carry over to functions of M-estimators. For the APE estimator, we should only report the spatial-correlation robust standard errors if assignments are spatially correlated or spillover effects are estimated. Additionally, the usual SHAC standard errors are supposed to be conservative. However, as a well-known fact, the SHAC standard errors suffer from downward bias when the spatial correlation is high. Hence, the actual finite sample performance of the usual SHAC standard errors remains unclear. 

\section{Simulation Designs}
We consider an uneven lattice. The population units lie within a square of dimension $\sqrt{M}\times \sqrt{M}$, where M is the population size. The locations $(s_{1,iM}, s_{2,iM})$ are drawn once and kept fixed across designs. $s_{1,iM}\sim \mathcal{U}(0,\sqrt{M})$, $s_{2,iM}\sim \mathcal{U}(0,\sqrt{M})$, and they are independent of each other. The distance between units $i$ and $j$ is measured by $\nu(i,j)=\max(|s_{1,iM}-s_{1,jM}|,|s_{2,iM}-s_{2,jM}|)$. We have four major designs involving spatial assignments at the individual level, spatial assignments at the cluster level, spatial assignments allowing for spillover effects, and spatial assignments in nonlinear models. 

\subsection{Spatial Correlation at the Individual Level}
The potential outcome function is given below:
\begin{equation}\label{s1}
y_{ig}(x_{ig})=a\cdot\beta_{ig} x_{ig}+c_g+u_{ig},
\end{equation}
where half of $\beta_{ig}$ are equal to 1 and the other half -1. $a$ is a constant scalar. The group unobserved heterogeneity $c_g$ is independently drawn from the standard normal distribution and remains fixed across replications. Without cluster sampling, we can ignore the $g$ subscript. The individual unobservables are generated from the linear spatial autoregressive model below,
\begin{equation}\label{s2}
u_M=p_u W_u u_M+\epsilon_M,
\end{equation} 
where $u_M$ and $\epsilon_M$ are $M\times 1$ vectors. $\epsilon_M$ are i.i.d. draws from a standard normal distribution and kept fixed. $W_u$ is a contiguity matrix and units $i$ and $j$ are neighbors if $\nu(i,j)\leq \sqrt{2}$. It is row-standardized with the diagonal elements being zero.

There are two sub-designs. In the first design, the assignment variables are i.i.d. draws from a Bernoulli distribution with a probability of 0.5 for each replication. The spatial correlation of the individual unobservables (in the sense of superpopulation) depends on $p_u$, which varies from 0 to 0.9 with an increment of 0.1. In the second sub-design, 
the assignment variables are binary variables equal to one when the input value $\xi_{iM}$ is greater than or equal to its population average $\sum^M_{i=1}\xi_{iM}/M$, where $\xi_M$ is an $M\times 1$ vector drawn from a multivariate normal distribution with mean zero and a variance-covariance matrix equal to $p_x$ raised to the power of the distance. $p_u$ is fixed at 0.3, while $p_x$ takes value from 0 to 0.9. Therefore, individual assignments can be spatially correlated. 

The expected size of each dimension of the lattice is 18, unless otherwise noted, which leads to an expected sample size of 324. The clusters in the sampling scheme are formed by grouping the consecutive three units by order, resulting in an expected number of 108 clusters in the sample. There are five sampling schemes: (\romannumeral 1) we observe the entire population; (\romannumeral 2) we independently sample clusters from all the clusters in the population with a probability of 0.25; (\romannumeral 3) we independently draw units from the entire population with a probability of 0.25; (\romannumeral 4) we independently sample clusters from all the clusters in the population with a probability of 0.01; (\romannumeral 5) we independently draw units from the entire population with a probability of 0.01. The last two sampling schemes mimic cluster sampling and independent sampling from the infinite population, respectively. For the last two sampling schemes, the expected size of each dimension of the lattice is decreased to 12 to reduce the computational burden. The constant $a=2$ for the first three sampling schemes and $a=1$ for the last two.

We first regress $Y_i$ on 1 and $X_i$ and report different standard errors of the slope coefficient estimator. For cluster sampling with probability 0.25, we also report results for fixed effect by demeaning variables within clusters. The standard errors among comparison are the EHW standard errors, the cluster-robust standard errors, and the SHAC standard errors. We use the Parzen kernel to estimate the SHAC standard errors. The SHAC standard errors are highly sensitive to the choice of bandwidth. In the first sub-design, where assignments are independent, we report the SHAC standard errors with bandwidth $d^*\in \{1, 2, 3\}$. In the second sub-design, we report the SHAC standard errors with two bandwidths chosen in the following way. The SHAC standard errors are estimated using bandwidth up to 20 with a distance increment of one. Among the 20 bandwidths, We choose either the one that minimizes the mean square error or the one that minimizes the bias of the SHAC standard error with respect to the Monte Carlo standard deviation of the slope coefficient estimator. 
The number of iterations is 1,000.

\subsubsection{Independent Assignments with Spatially Correlated Unobservables}

\begin{figure}[htbp] %  figure placement: here, top, bottom, or page
   \centering
   \includegraphics[width=5.5in]{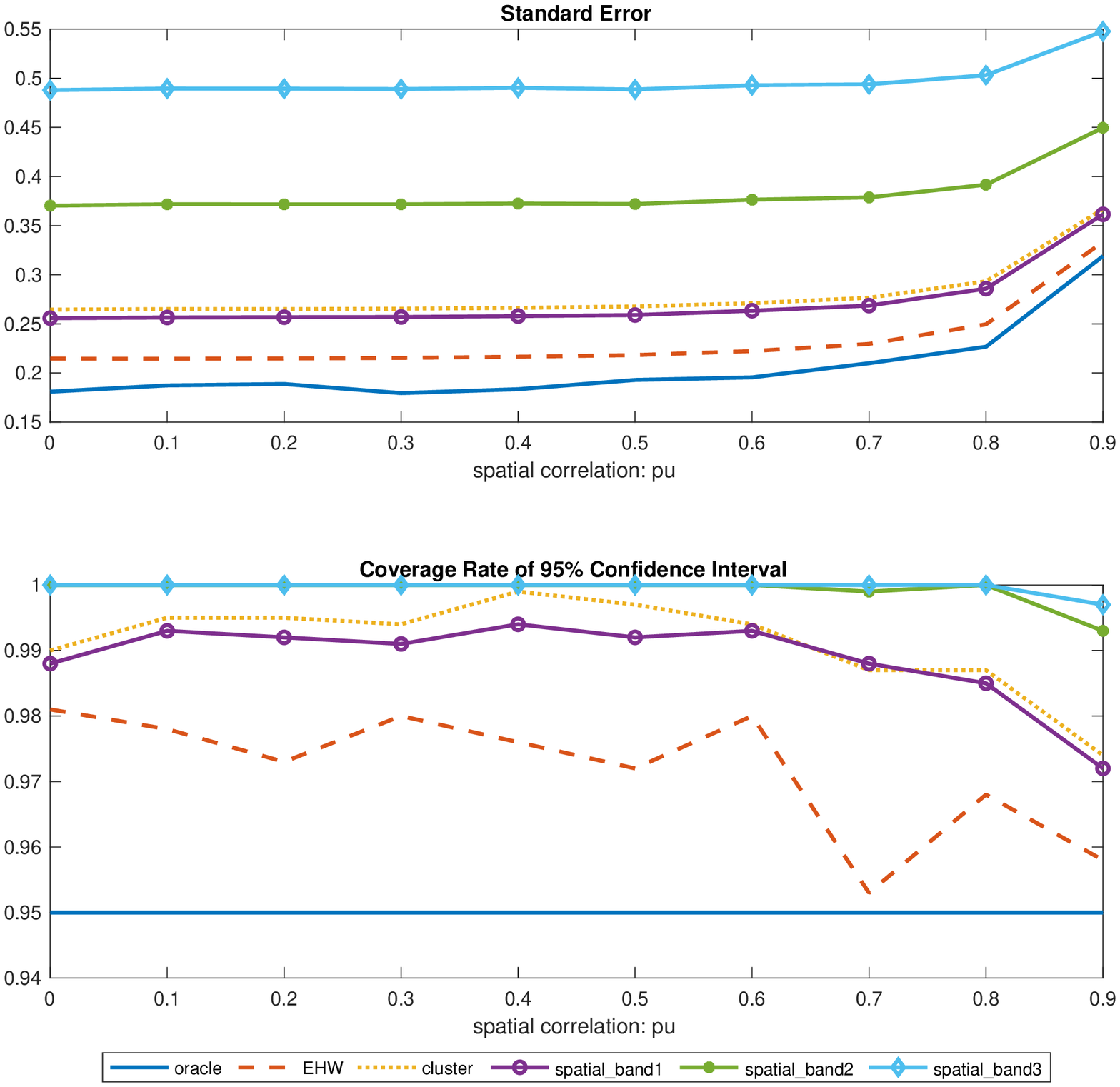} 
   \caption{Independent Assignments Observing Entire Population}
   \label{fig:example}
\end{figure}

When the assignments are independent with the entire population observed, we see from Figure 1 that all standard errors are larger than the Monte Carlo standard deviation of the slope coefficient estimator and the corresponding coverage rates of the 95\% confidence interval are almost all above the benchmark line of 0.95.\footnote{In the top panel of all figures below, ``oracle" means the Monte Carlo standard deviation of the coefficient estimator or the APE estimator. While in the bottom panels, ``oracle" stands for the benchmark coverage rate of the 95\% confidence interval, 0.95.} This is expected because the superpopulation standard errors are supposed to be conservative. Among all the standard errors reported, the EHW standard errors are the closest to the Monte Carlo standard deviation, regardless of the spatial correlation of the unobservables in the superpopulation. As well, the coverage rate of the 95\% confidence interval based on the EHW standard errors is closest to the theoretical level. The cluster-robust standard errors are generally too large and the SHAC standard errors increase along with the bandwidth. With independent assignments, the latter two standard errors are unnecessarily conservative. 

\begin{figure}[htbp] %  figure placement: here, top, bottom, or page
   \centering
   \includegraphics[width=5.5in]{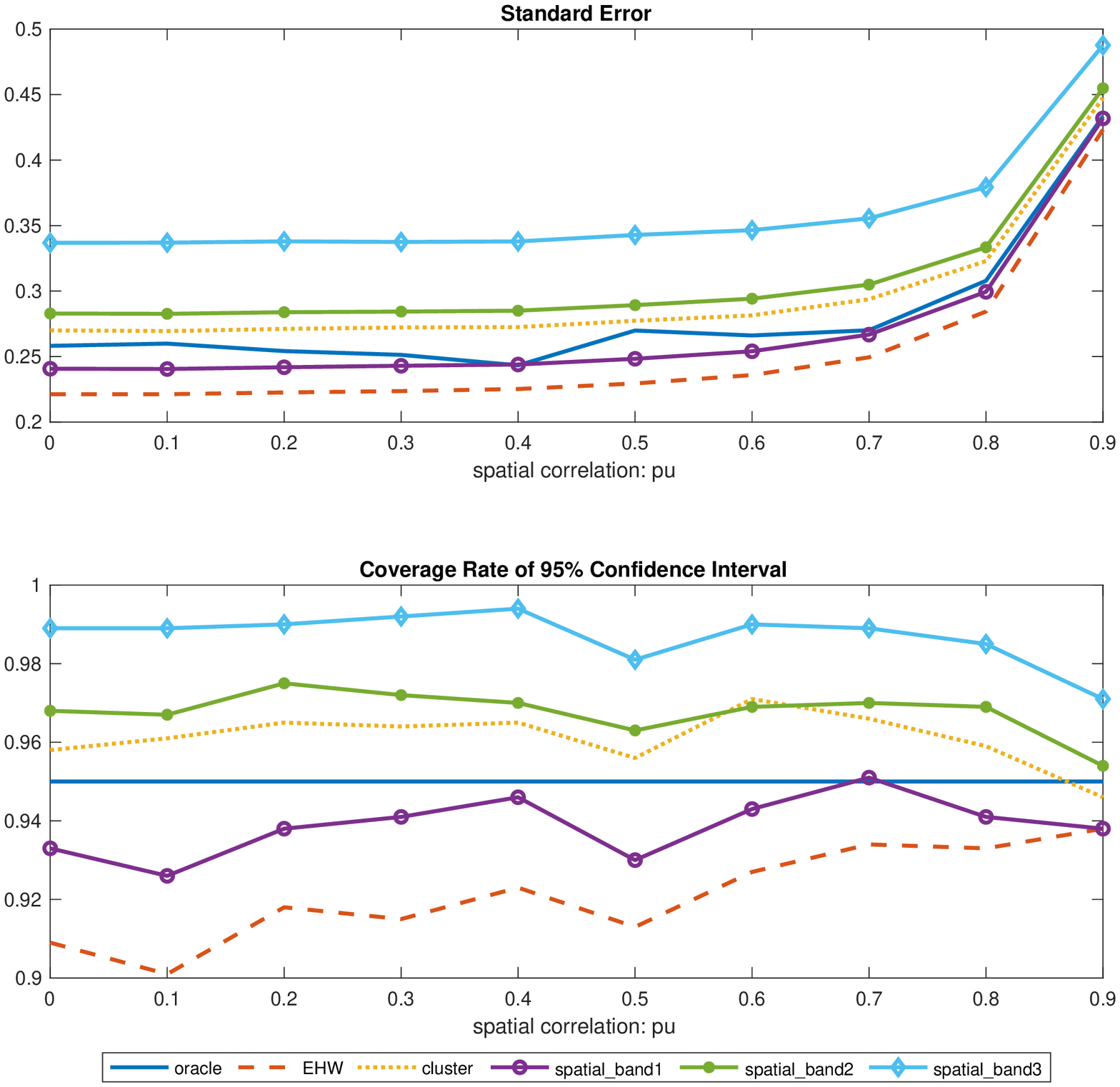} 
   \caption{Independent Assignments with Cluster Sampling 0.25}
   \label{fig:example}
\end{figure}

\begin{figure}[htbp] %  figure placement: here, top, bottom, or page
   \centering
   \includegraphics[width=5.5in]{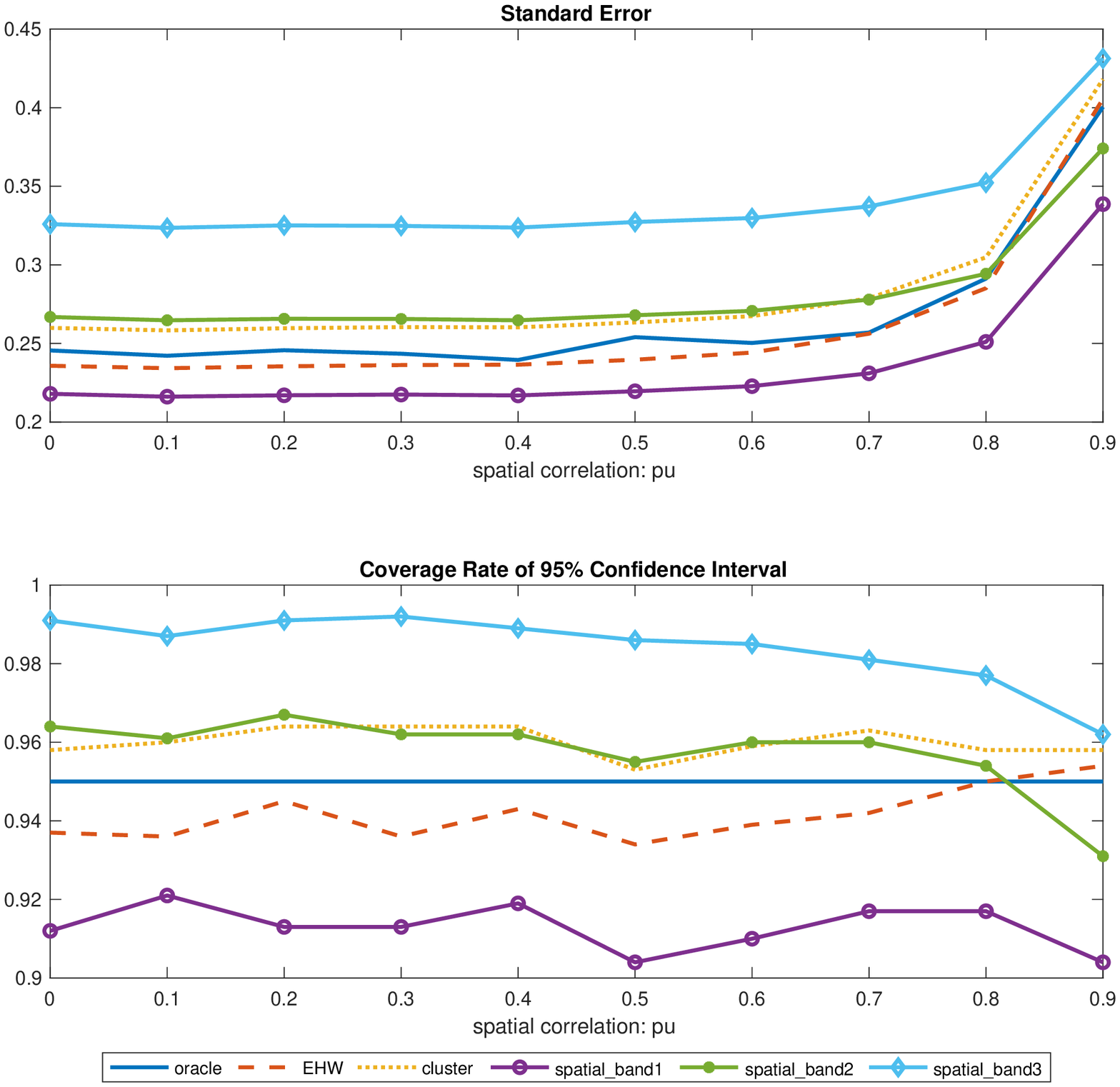} 
   \caption{Independent Assignments with Cluster Sampling 0.25: fixed effects}
   \label{fig:example}
\end{figure}

\begin{figure}[htbp] %  figure placement: here, top, bottom, or page
   \centering
   \includegraphics[width=5.5in]{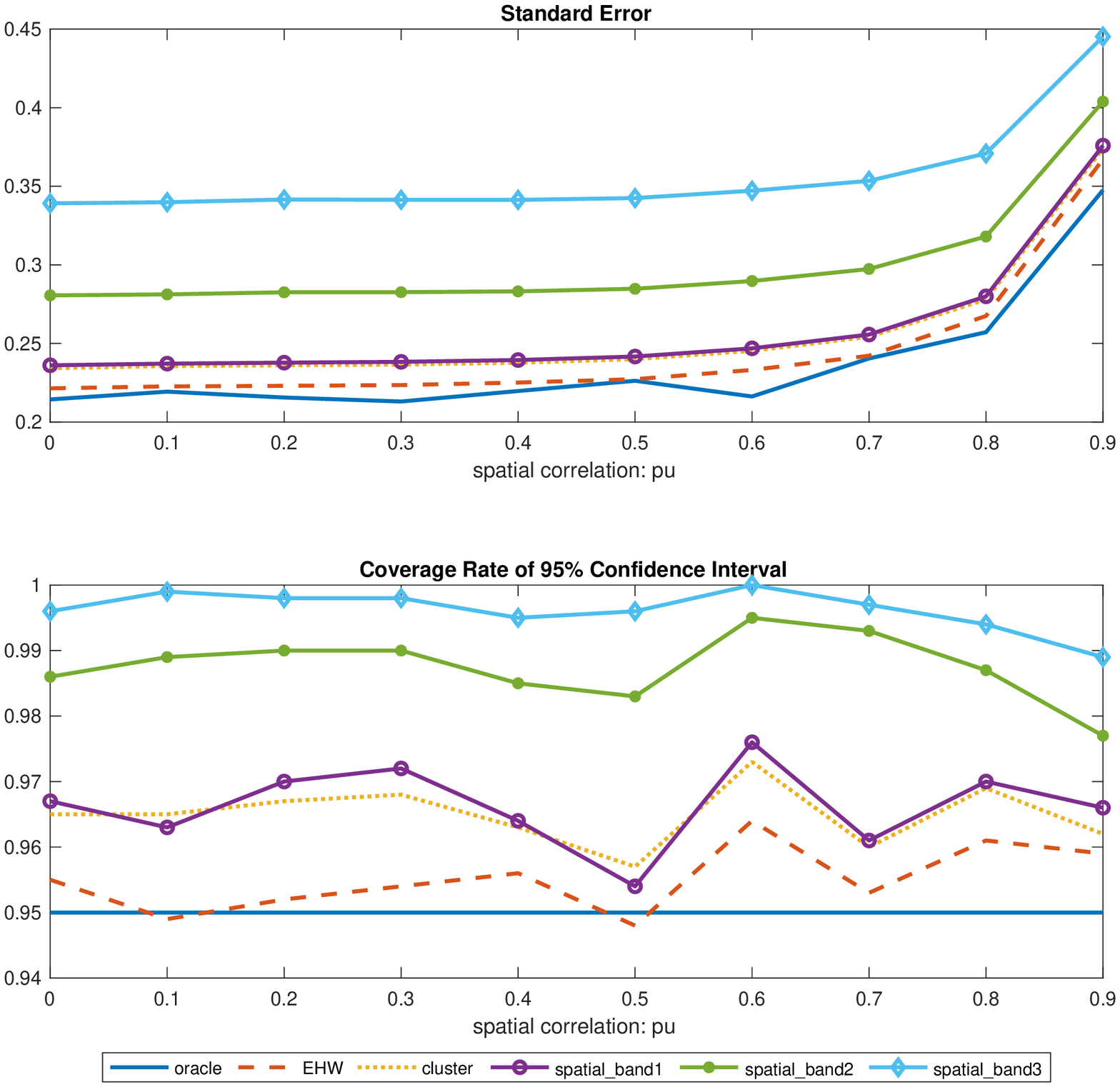} 
   \caption{Independent Assignments with Independent Sampling 0.25}
   \label{fig:example}
\end{figure}

\begin{figure}[htbp] %  figure placement: here, top, bottom, or page
   \centering
   \includegraphics[width=5.5in]{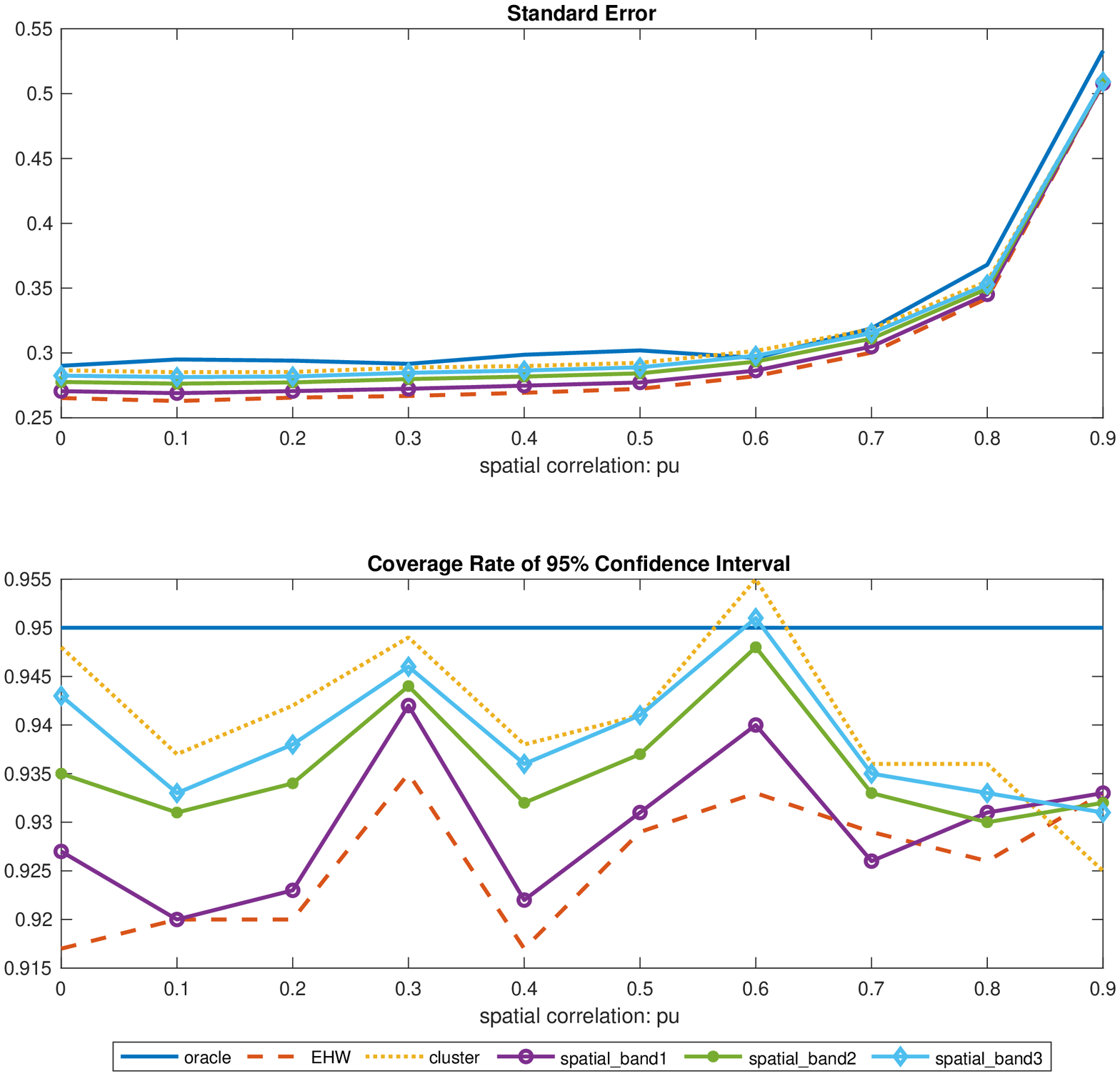} 
   \caption{Independent Assignments with Cluster Sampling 0.01}
   \label{fig:example}
\end{figure}

\begin{figure}[htbp] %  figure placement: here, top, bottom, or page
   \centering
   \includegraphics[width=5.5in]{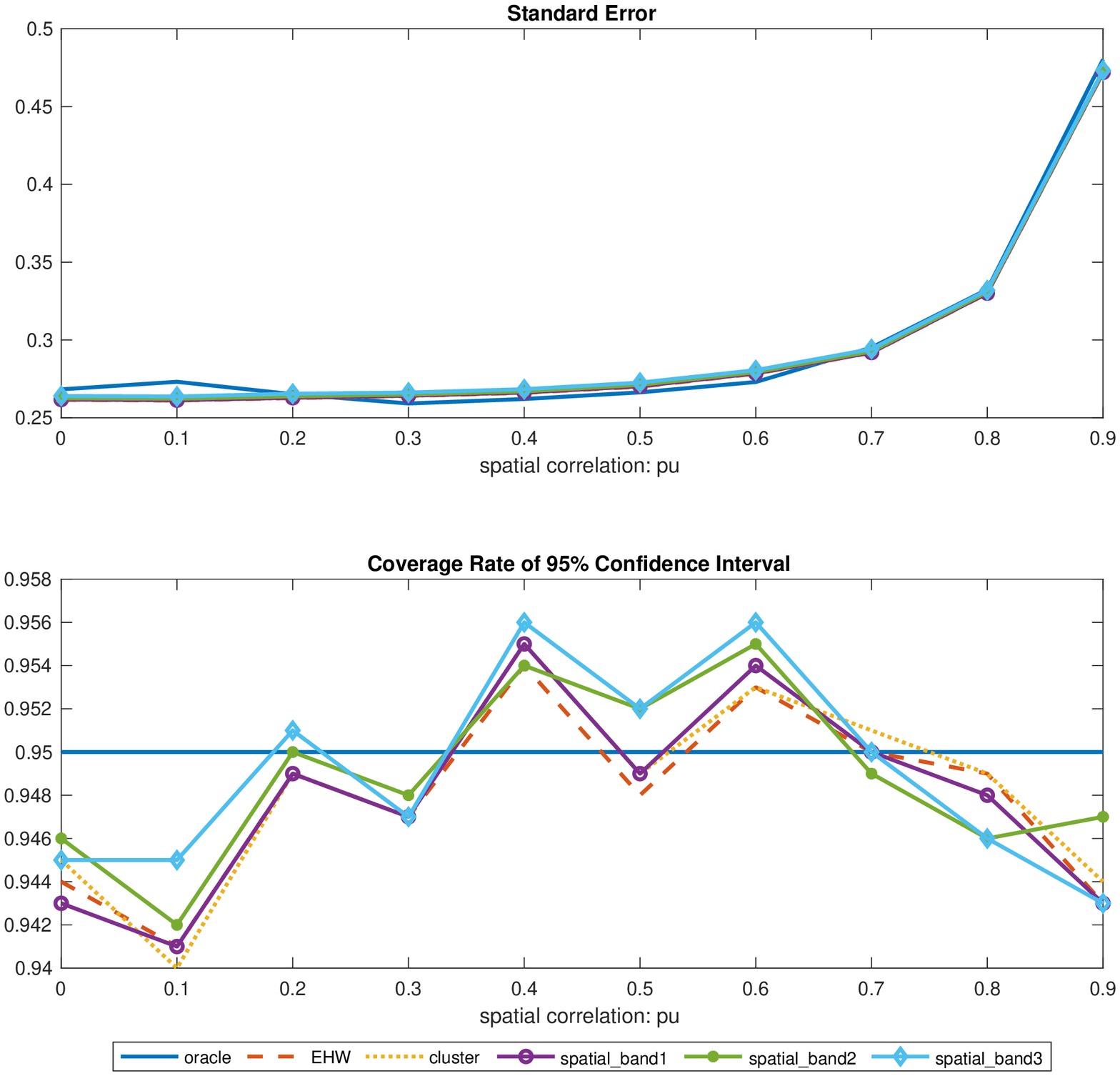} 
   \caption{Independent Assignments with Independent Sampling 0.01}
   \label{fig:example}
\end{figure}

When we sample clusters from the population, things become slightly different. Although assignments are still independent, cluster sampling introduces cluster correlation. As a result, the cluster robust standard error is the one closest to the Monte Carlo standard deviation and the coverage rate of the corresponding confidence interval hovers around the benchmark line. The EHW standard errors are too small, as shown in Figures 2 and 5. We observe the same pattern for the standard errors of the fixed effect estimators. However, the comparison between the cluster robust standard errors and the SHAC standard errors depends on the sampling probability. When the sampling probability of each cluster is nonnegligible, the SHAC standard errors grow along with the bandwidth and eventually become too conservative. On the other hand, when the sampling probability is as small as 0.01, resembling cluster sampling from an infinite population, the difference between the cluster-robust standard errors and the SHAC standard errors narrows down. This is especially true when the chosen bandwidth contains enough nearby units. 

When we independently draw units from the population, the EHW standard errors again turn out to be the appropriate one to report. Similarly, when the sampling probability is 0.25, you see a clear discrepancy among the standard errors. When the sampling probability is 0.01, which resembles the case of independent sampling from an infinite population, the EHW, cluster, and SHAC standard errors are almost identical to each other.%\footnote{The intuition of these observations will be explained once we introduce spatial assignments into the simulation design.}   

\subsubsection{Spatial Assignments}

\begin{figure}[htbp] %  figure placement: here, top, bottom, or page
   \centering
   \includegraphics[width=5.5in]{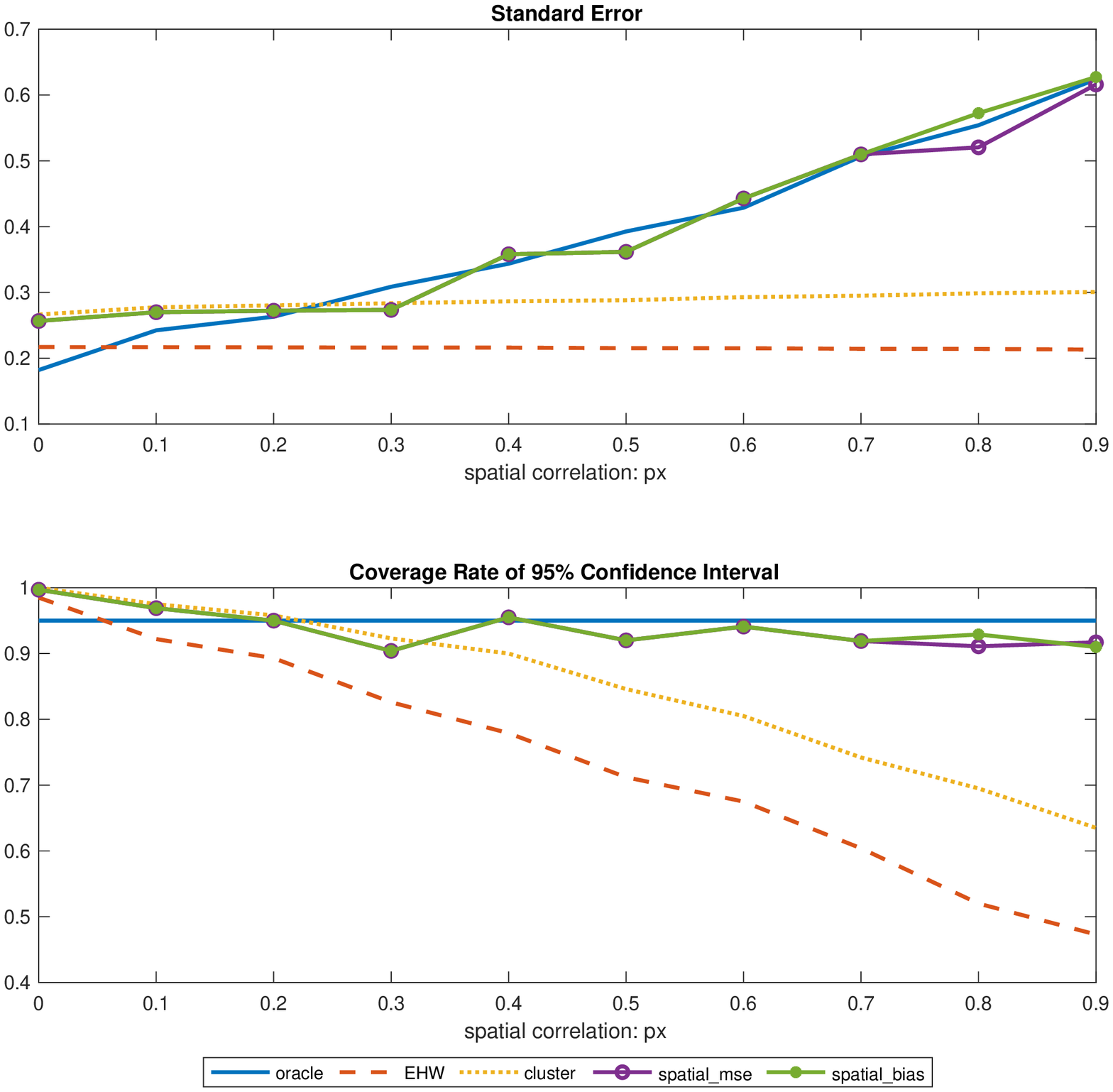} 
   \caption{Spatial Assignments Observing Entire Population}
   \label{fig:example}
\end{figure}

\begin{figure}[htbp] %  figure placement: here, top, bottom, or page
   \centering
   \includegraphics[width=5.5in]{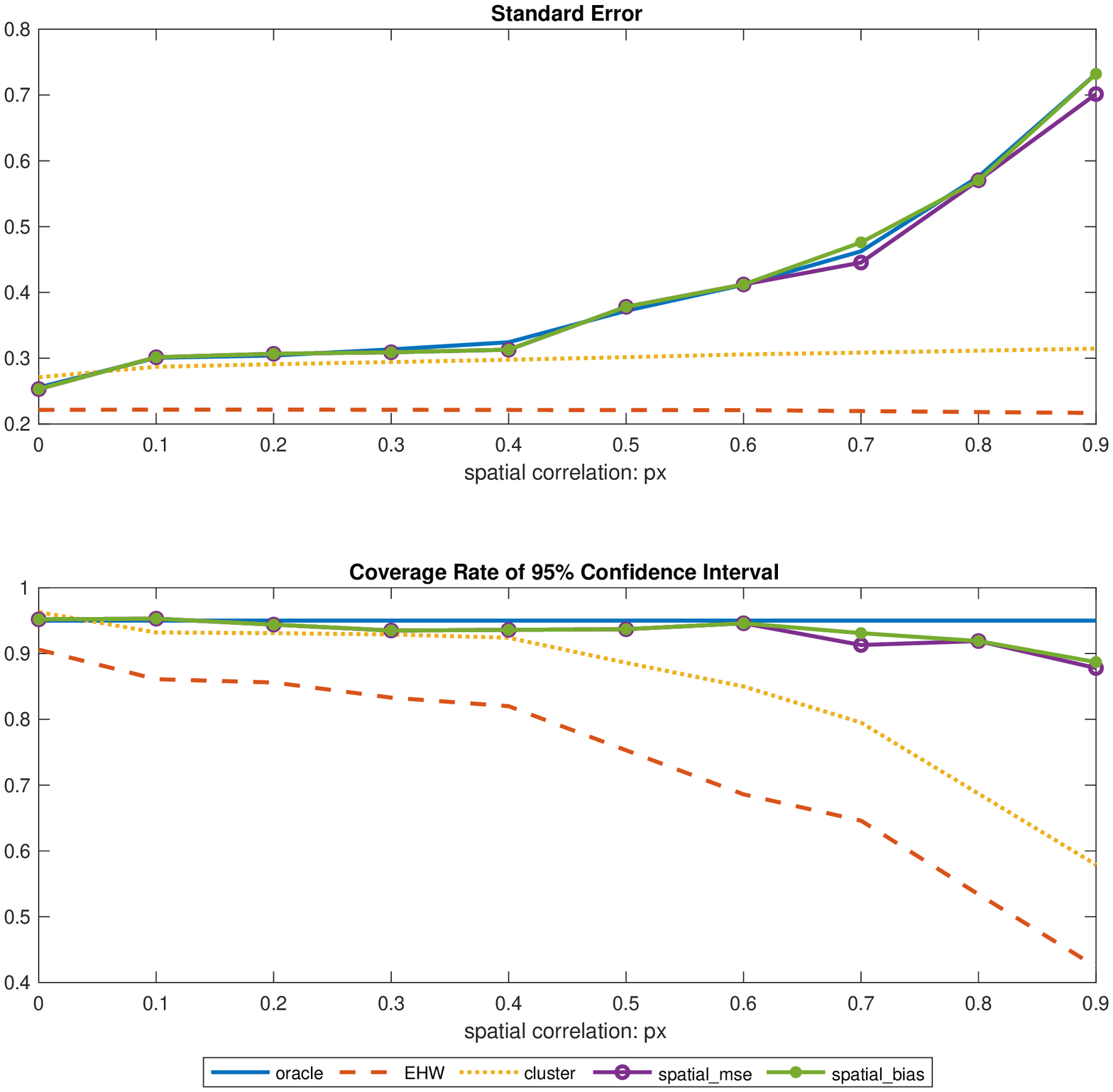} 
   \caption{Spatial Assignments with Cluster Sampling 0.25}
   \label{fig:example}
\end{figure}

\begin{figure}[htbp] %  figure placement: here, top, bottom, or page
   \centering
   \includegraphics[width=5.5in]{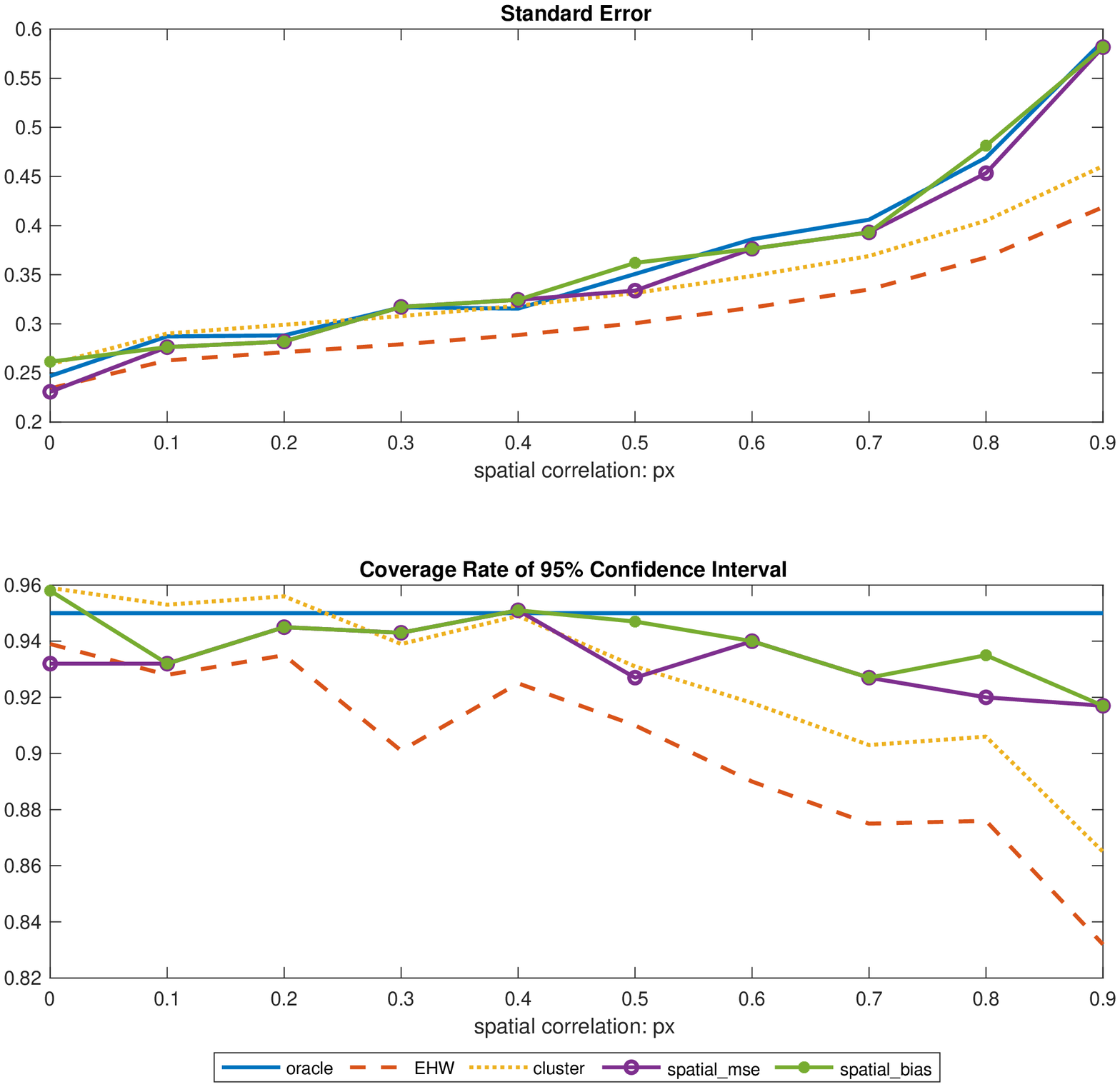} 
   \caption{Spatial Assignments with Cluster Sampling 0.25: fixed effects}
   \label{fig:example}
\end{figure}

\begin{figure}[htbp] %  figure placement: here, top, bottom, or page
   \centering
   \includegraphics[width=5.5in]{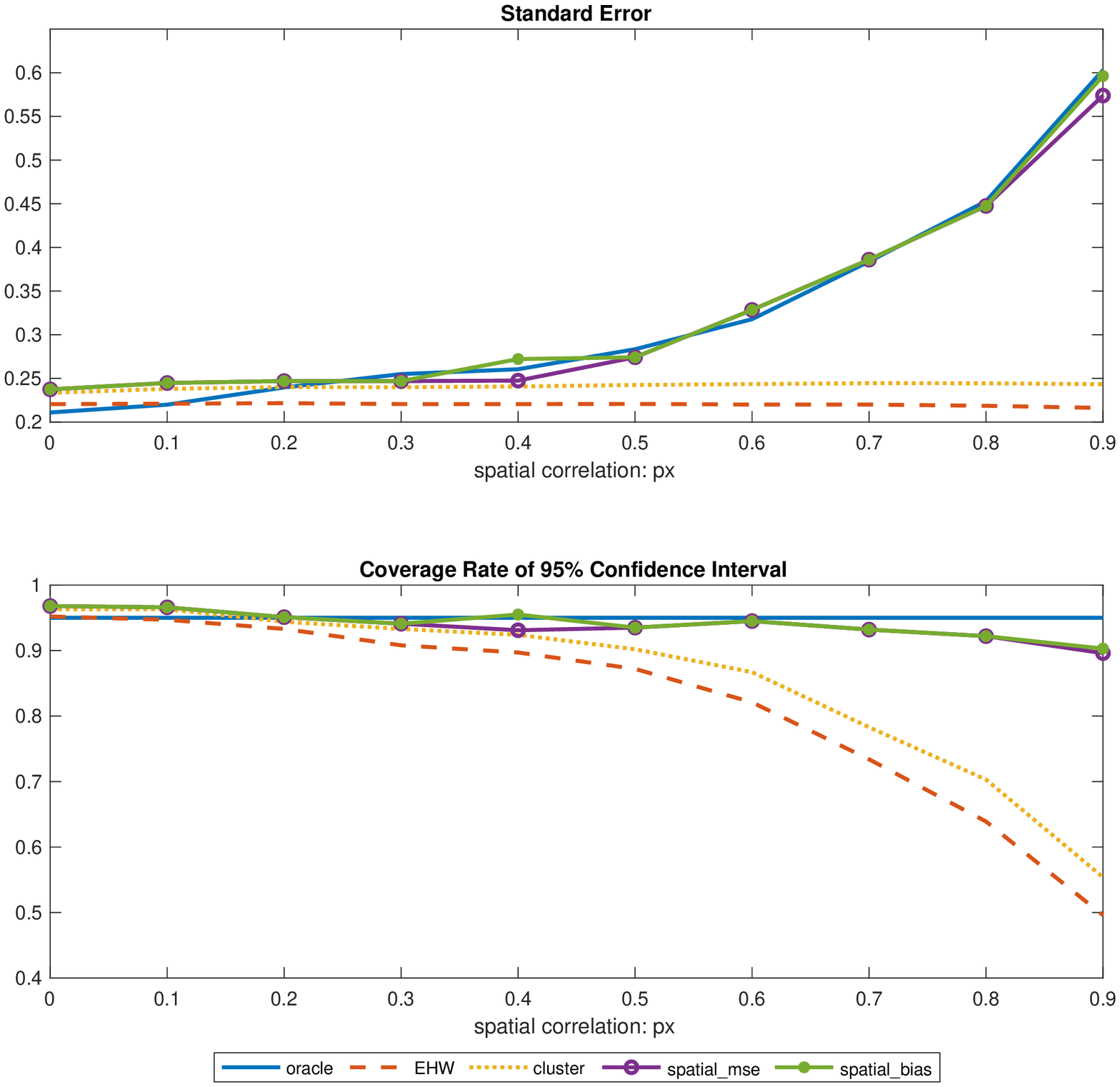} 
   \caption{Spatial Assignments with Independent Sampling 0.25}
   \label{fig:example}
\end{figure}

\begin{figure}[htbp] %  figure placement: here, top, bottom, or page
   \centering
   \includegraphics[width=5.5in]{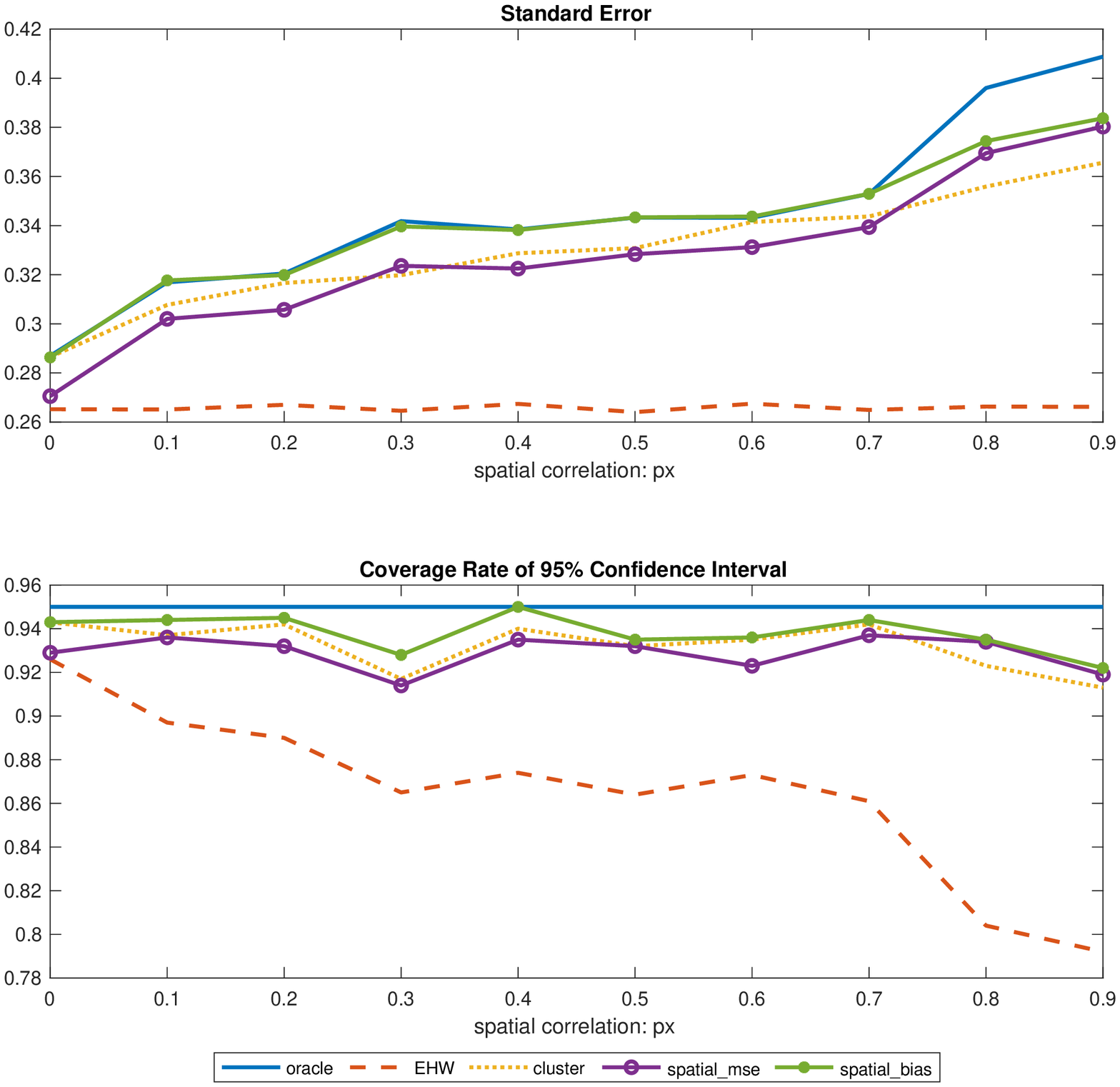} 
   \caption{Spatial Assignments with Cluster Sampling 0.01}
   \label{fig:example}
\end{figure}

\begin{figure}[htbp] %  figure placement: here, top, bottom, or page
   \centering
   \includegraphics[width=5.5in]{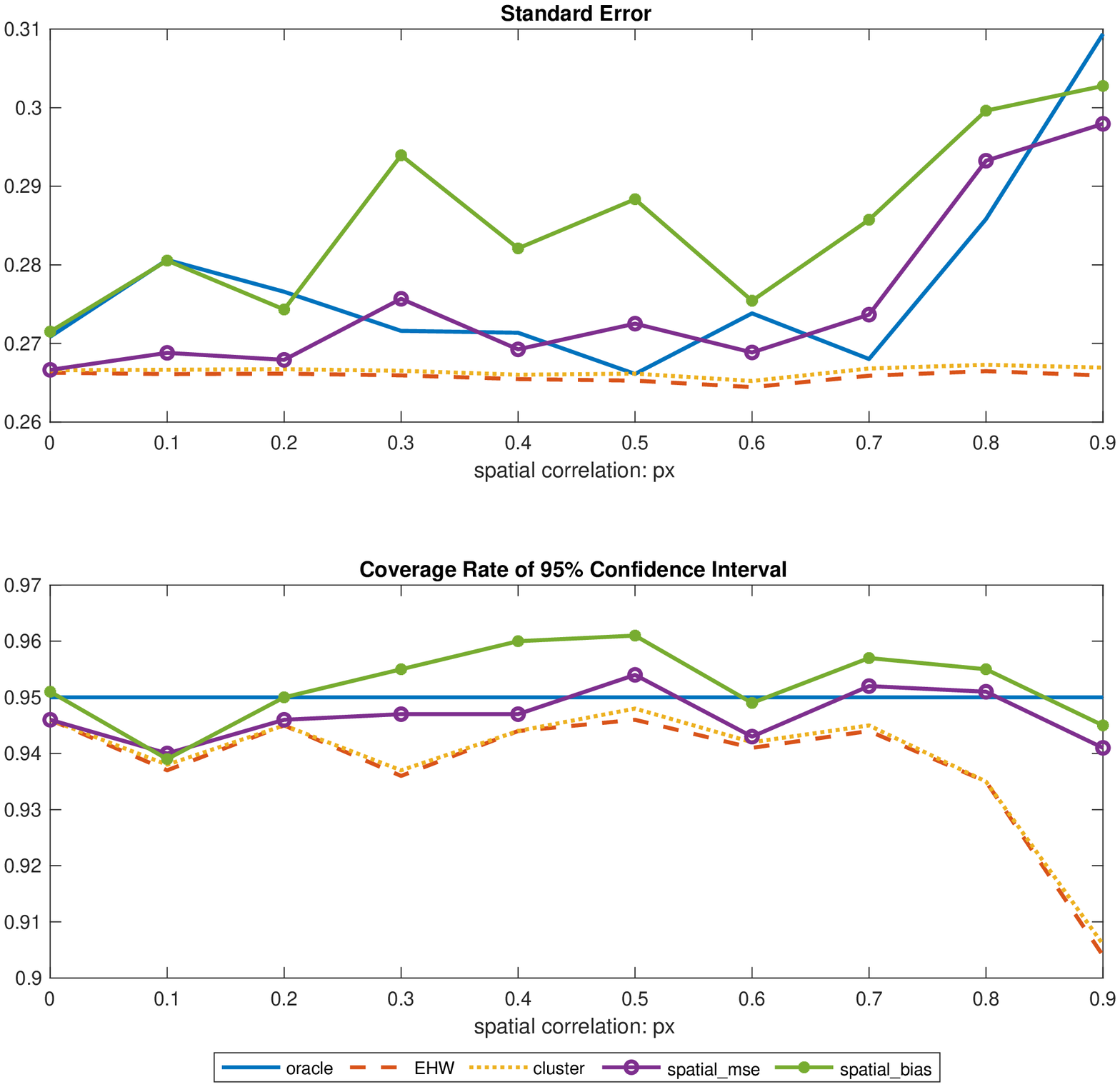} 
   \caption{Spatial Assignments with Independent Sampling 0.01}
   \label{fig:example}
\end{figure}

In the second sub-design, spatial assignments are introduced through the spatial correlation parameter, $p_x$. When $p_x=0$, we are back to the case of independent assignments with $p_u=0.3$. As we can see from Figures 7 and 10, the EHW standard errors are the ones with the best performance when we either observe the entire population or independently sample units from the population. With cluster sampling, the EHW standard errors underestimate the standard deviation and the SHAC standard errors are quite similar to the cluster-robust standard errors. Nevertheless, with spatial assignments we see that both the EHW and the cluster-robust standard errors gradually become too small when the spatial correlation among the assignment variables increases unless the sampling probability is as small as 0.01. The same observation holds for both pooled OLS and fixed effect estimators when there is cluster partition.

Though not perfect, we see patterns along the lines of the theoretic prediction in Remark \ref{remark2} when the sampling probability is 0.01. With cluster sampling, the gap between the SHAC standard errors and the cluster-robust standard errors is much smaller even with high spatial correlation compared with the case of a larger sampling probability. Similarly, with independent sampling, the difference among the EHW, cluster, and the SHAC standard errors is relatively small compared with Figures 7-10. All standard errors suffer from some downward bias. However, keep in mind that the sample size is only over 100 and we allow the correlation parameter, $p_x$, to be as large as 0.9, which results in the correlation between the assignment variables within a distance of one averaging at 0.85. I expect that the discrepancy among the different standard errors would gradually disappear when the sampling probability becomes even lower, especially when the spatial correlation is not too high. However, this would drastically increase the computation burden by enlarging the size of the lattice to a great extent. Hence, results are not reported here due to computation restrictions.     

When the ratio of the sample to the population size is small, the intuition of reporting either the EHW or the cluster-robust standard errors remains the same no matter whether we introduce spatial correlation at the cluster level, spillover effects, or nonlinearity later on. Hence, for the simulation designs below, we omit the results for sampling probabilities of 0.01. 

\subsection{Spatial Assignments at the Cluster Level}

In this design, assignments are imposed at the cluster level, and we introduce spatial correlation across clusters. The potential outcome function and the individual unobservables are the same as in equations (\ref{s1}) and (\ref{s2}) with $a=1$. $p_u$ is also fixed at 0.3.
On the contrary, the assignments are generated differently. We construct a contiguity matrix among cluster pairs, where the distance between clusters is measured as the minimum distance of units in the cluster pair. Hence, the cluster contiguity matrix $W_G$ is a $G\times G$ matrix, where $G$ is the number of clusters in the population and clusters $l$ and $m$ are neighbors if $\nu(l,m)\leq 2$. We adopt the first three sampling schemes as in the baseline design.  

The assignment variables are generated in the following way:
\begin{equation}
\tilde{X}_G=p_x W_G \tilde{X}_G+\xi_G,
\end{equation}
where $\tilde{X}_G$ and $\xi_G$ are $G\times 1$ vectors, and $\xi_g \overset{i.i.d.}\sim \mathcal{N}(0,1), g=1, \dots,G$. The cluster assignment $X_g=\mathbbm{1}\{\tilde{X}_g>\frac{1}{G}\sum^G_{l=1}\tilde{X}_l\}$, $\forall\ g=1,2,\dots, G$. Units within the same cluster receive the same assignment. 

%We apply either the distance between individual units or the distance between cluster pairs in estimating the SHAC standard errors. For the latter distance measure, units within the same cluster share the same distance. 

\begin{figure}[htbp] %  figure placement: here, top, bottom, or page
   \centering
   \includegraphics[width=5.5in]{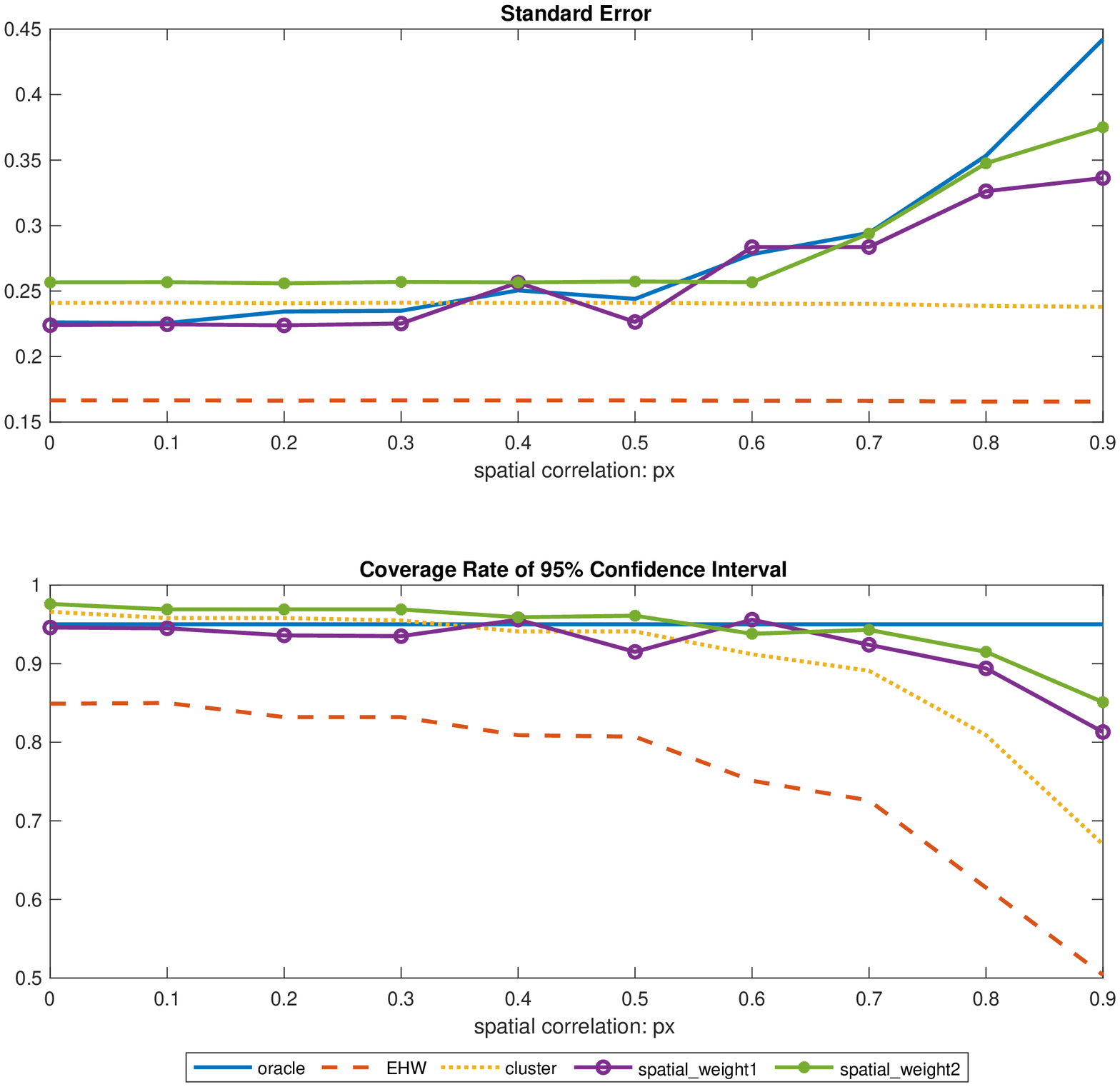} 
   \caption{Spatial Assignments at the Cluster Level Observing Entire Population}
   \label{fig:example}
\end{figure}

\begin{figure}[htbp] %  figure placement: here, top, bottom, or page
   \centering
   \includegraphics[width=5.5in]{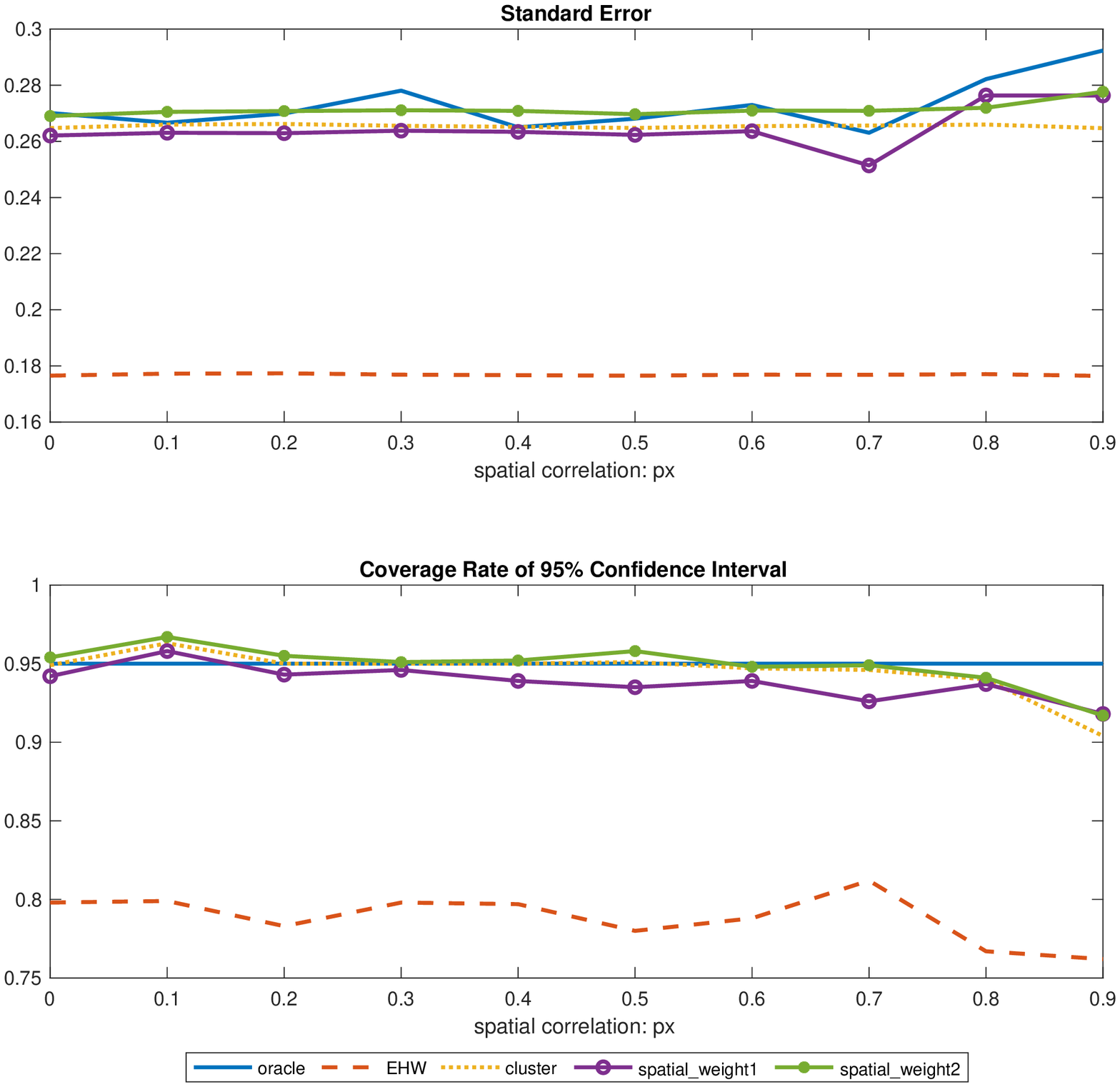} 
   \caption{Spatial Assignments at the Cluster Level with Cluster Sampling 0.25}
   \label{fig:example}
\end{figure}

\begin{figure}[htbp] %  figure placement: here, top, bottom, or page
   \centering
   \includegraphics[width=5.5in]{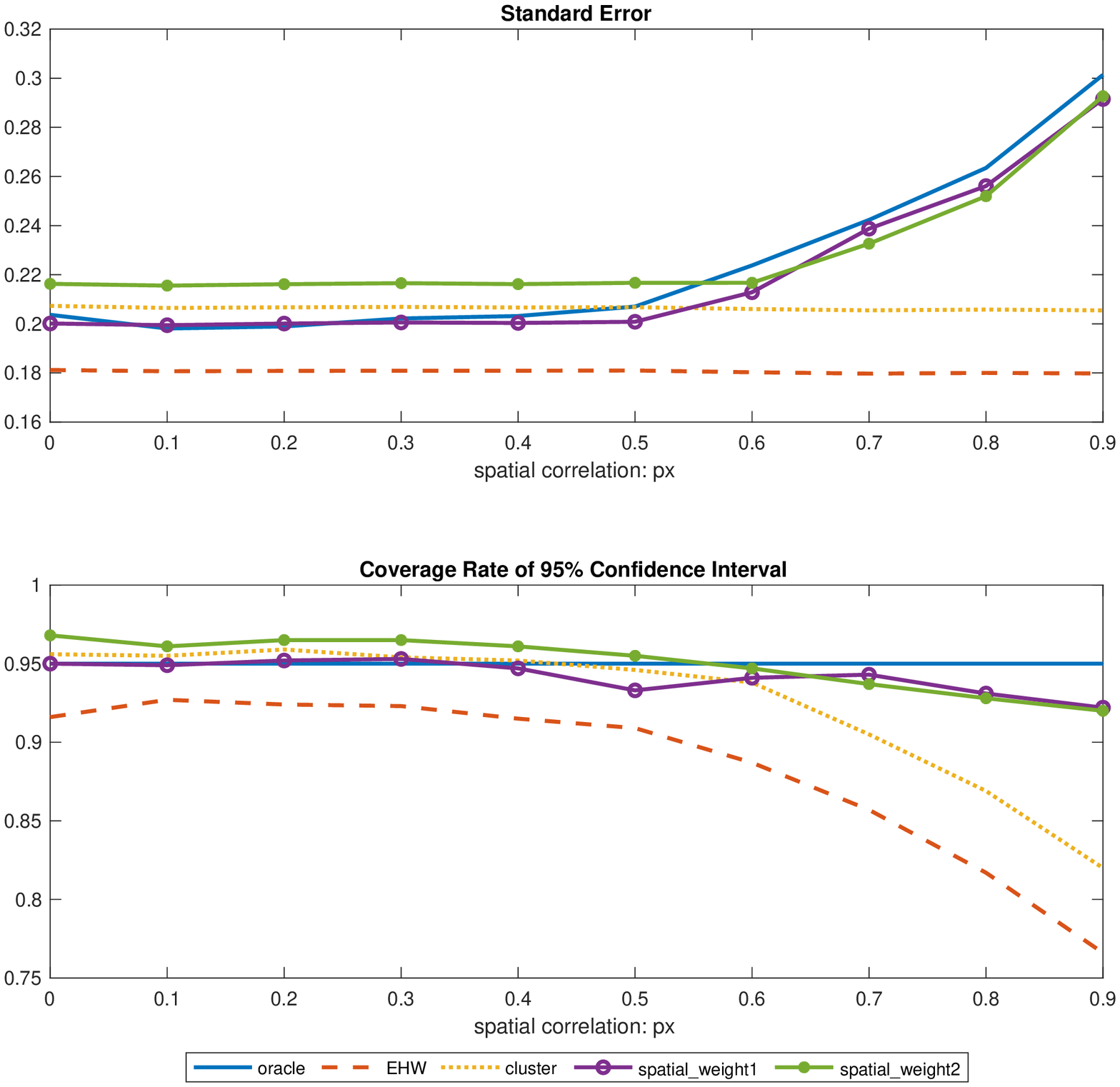} 
   \caption{Spatial Assignments at the Cluster Level with Independent Sampling 0.25}
   \label{fig:example}
\end{figure}

When the spatial correlation is imposed at the cluster assignment variables rather than the individual assignment variables, the comparison of the standard errors in the baseline design carries over. In summary, we should report the SHAC standard errors under spatial assignments. In the absence of spatial assignments, we should report the cluster-robust standard errors because cluster assignments always occur regardless of sampling scheme. However, to account for the cluster correlation in addition to the spatial correlation across clusters, I introduce a slightly different distance measure based on the distance between cluster pairs. Using this cluster distance measure, units within the same cluster receive the same weights when estimating the SHAC standard errors. The SHAC standard errors based on the cluster distance slightly outperform the ones based on the unit distance. This is because the latter typically suffer from downward bias even when the spatial correlation is low or moderate. However, the former can also become too conservative when there is a lot of heterogeneity across units.

\subsection{Spillover Effects}
In this simulation design, we allow for spillover effects in addition to spatial assignments by setting up a linear-in-means type of model. The expected sample size of each dimension of the lattice is 36.
The potential outcome function is given below:
\begin{equation}
y_{ig}(\bm{x}_M)=2\beta_{ig} x_{ig}+\gamma W_x \bm{x}_M + \epsilon_{ig},
\end{equation}
where $\bm{x}_M$ is the $M\times 1$ vector containing assignments for all units in the population. The realized assignments $\bm{X}_M$ follows a multivariate normal distribution with mean zero and a variance-covariance matrix equal to $p_x$ raised to the power of the distance. The contiguity matrix, $W_x$, is constructed in the same way as $W_u$. Elements in $W_x$ equal one if the corresponding distance is less than or equal to 0.5. $\beta_{ig}$ are defined in the same way as in Section 4.1. The unobservables $\epsilon_{ig}$ are independently drawn from a standard normal distribution and kept fixed. 

We adopt the first three sampling schemes in the baseline design. Within each sampling scheme, we have three assignment and spillover combinations: (\romannumeral 1) $p_x=0$ and $\gamma=0$; (\romannumeral 2) $p_x=0$ and $\gamma=1$; (\romannumeral 3) $p_x=0.1$ and $\gamma=1$. Each represents the case of no spatial assignments and no spillover effects, no spatial assignments with spillover effects, and spatial assignments combined with spillover effects, respectively. Regardless, we estimate spillover effects by regressing $Y_i$ on 1, $X_i$, and $W_s\bm{X}_s$, where $\bm{X}_s$ is the collection of assignments of all units in the sample and $W_s$ is a $N\times N$ contiguity matrix with a distance cutoff of 0.5.

\begin{table}[htbp]
  \centering
  \caption{Specifying Spillover Effects: $\gamma$}
    \begin{tabular}{lccccccccc}
\hline
    & \multicolumn{3}{c}{entire population} & \multicolumn{3}{c}{cluster sampling} & \multicolumn{3}{c}{independent sampling} \\
 \cmidrule(lr){2-4}  \cmidrule(lr){5-7}  \cmidrule(lr){8-10}
  & $p_x=0, $ & $p_x=0,$ & $p_x=0.1,$ & $p_x=0,$ & $p_x=0,$ & $p_x=0.1,$ & $p_x=0,$ & $p_x=0,$ & $p_x=0.1,$ \\
  & $\gamma=0$ & $\gamma=1$ & $\gamma=1$ & $ \gamma=0$ & $\gamma=1$ & $\gamma=1$ & $\gamma=0$ & $\gamma=1$ & $\gamma=1$ \\
\hline
 coeff &      0.002 & 1.002 & 1.061 & 0.004 & 0.636 & 0.546 & -0.001 & 0.500 & 0.472 \\
 std &   0.133 & 0.110 & 0.136 & 0.113 & 0.128 & 0.169 & 0.137 & 0.149 & 0.160 \\
  EHW &      0.101 & 0.083 & 0.100 & 0.087 & 0.094 & 0.123 & 0.103 & 0.112 & 0.121 \\
 EHW\_CI &       (0.863) & (0.860) & (0.855) & (0.860) & (0.843) & (0.836) & (0.854) & (0.853) & (0.848) \\
cluster&     0.107 & 0.096 & 0.109 & 0.109 & 0.115 & 0.150 & 0.111 & 0.121 & 0.129 \\
   cluster\_CI &   (0.887) & (0.910) & (0.887) & (0.938) & (0.924) & (0.911) & (0.883) & (0.883) & (0.875) \\
SHAC1  &      0.129 & 0.107 & 0.135 & 0.112 & 0.122 & 0.167 & 0.134 & 0.145 & 0.159 \\
  SHAC1\_CI &  (0.934) & (0.938) & (0.950) & (0.943) & (0.941) & (0.947) & (0.936) & (0.935) & (0.935) \\
  SHAC2 &   0.129 & 0.108 & 0.136 & 0.113 & 0.123 & 0.169 & 0.135 & 0.146 & 0.160 \\
  SHAC2\_CI&    (0.936) & (0.941) & (0.936) & (0.941) & (0.943) & (0.941) & (0.939) & (0.933) & (0.937) \\
\hline
    \end{tabular}%
  \label{tab:addlabel}%
\end{table}%

Table 1 reports the coefficient estimates and the standard errors of the spillover effects coefficient. Even though all standard errors suffer from downward bias in all designs, the SHAC standard errors perform the best among the three classes of standard errors. When we include spillover effects in our models, regardless of the sampling scheme we use or whether there are spillover effects in the potential outcome function or not, we should always report the spatial-correlation robust standard errors.

It is worth mentioning that without observing the entire population, what we are identifying is the spillover effects in the sample because we can only observe our neighbors who happen to be selected into the sample, which is an imperfect measure of the spillover in the population. As a result, the coefficient estimator on the spillover effects is biased in the latter two sampling schemes. With cluster sampling, the spillover effects estimator seems to be less biased given that more neighbors are included in the cluster sample.  

\begin{table}[htbp]
  \centering
  \caption{Coefficient on the Individual Assignment Variable with Independent Assignments}
    \begin{tabular}{lcccccc}
\hline
    & \multicolumn{2}{c}{entire population} & \multicolumn{2}{c}{cluster sampling} & \multicolumn{2}{c}{independent sampling} \\
 \cmidrule(lr){2-3}  \cmidrule(lr){4-5}  \cmidrule(lr){6-7}
  & $\gamma=0$ & $\gamma=1$  & $ \gamma=0$ & $\gamma=1$  & $\gamma=0$ & $\gamma=1$  \\
\hline
  coeff &      -0.002 & 0.000 & -0.001 & 0.001 & -0.001 & 0.009 \\
   std &  0.087 & 0.084 & 0.116 & 0.118 & 0.094 & 0.098 \\
  EHW &   0.100 & 0.100 & 0.100 & 0.101 & 0.100 & 0.101 \\
  EHW\_CI &   (0.977) & (0.979) & (0.908) & (0.900) & (0.958) & (0.957) \\
 cluster&    0.127 & 0.127 & 0.127 & 0.127 & 0.107 & 0.108 \\
   cluster\_CI &  (0.994) & (0.997) & (0.961) & (0.975) & (0.971) & (0.972) \\
  SHAC1  &  0.130 & 0.122 & 0.112 & 0.113 & 0.106 & 0.108 \\
  SHAC1\_CI &   (0.995) & (0.995) & (0.934) & (0.938) & (0.969) & (0.974) \\
SHAC2 &    0.130 & 0.122 & 0.112 & 0.113 & 0.106 & 0.108 \\
   SHAC2\_CI&  (0.995) & (0.995) & (0.934) & (0.938) & (0.969) & (0.974) \\
   \hline
    \end{tabular}%
  \label{tab:addlabel}%
\end{table}%

The coefficient estimates on the individual assignment variable are also reported in Table 2, along with the standard errors. Depending on the sampling scheme, reporting the EHW or cluster-robust standard errors is sufficient. Here, the individual assignment variables and the spillover effects are independent of each other because of independent assignments. In light of this observation, it appears unnecessary to make inference robust to spatial correlation for both coefficient estimators on two assignment variables when one is assigned independently, the other is spatially correlated, and the two assignment variables are independent of each other. We can easily verify the simulation results using the Frisch-Waugh theorem in linear models.  

\begin{table}[htbp]
  \centering
  \caption{Coefficient on the Individual Assignment Variable Omitting Spillover Effects}
    \begin{tabular}{lccccccccc}
\hline
    & \multicolumn{3}{c}{entire population} & \multicolumn{3}{c}{cluster sampling} & \multicolumn{3}{c}{independent sampling} \\
 \cmidrule(lr){2-4}  \cmidrule(lr){5-7}  \cmidrule(lr){8-10}
  & $p_x=0, $ & $p_x=0,$ & $p_x=0.3,$ & $p_x=0,$ & $p_x=0,$ & $p_x=0.3,$ & $p_x=0,$ & $p_x=0,$ & $p_x=0.3,$ \\
  & $\gamma=0$ & $\gamma=1$ & $\gamma=1$ & $ \gamma=0$ & $\gamma=1$ & $\gamma=1$ & $\gamma=0$ & $\gamma=1$ & $\gamma=1$ \\
\hline
  coeff &  -0.001 & 0.001 & 0.603 & 0.003 & 0.000 & 0.662 & -0.003 & -0.002 & 0.536 \\
  std &   0.084 & 0.088 & 0.157 & 0.119 & 0.119 & 0.148 & 0.093 & 0.100 & 0.117 \\
  EHW &   0.100 & 0.101 & 0.098 & 0.100 & 0.101 & 0.101 & 0.100 & 0.102 & 0.104 \\
   EHW\_CI &  (0.983) & (0.973) & (0.776) & (0.905) & (0.902) & (0.812) & (0.968) & (0.952) & (0.923) \\
   cluster&    0.127 & 0.128 & 0.134 & 0.127 & 0.128 & 0.138 & 0.107 & 0.109 & 0.113 \\
   cluster\_CI &   (0.998) & (0.997) & (0.905) & (0.960) & (0.960) & (0.934) & (0.977) & (0.965) & (0.942) \\
  SHAC1  &   0.130 & 0.129 & 0.145 & 0.113 & 0.114 & 0.146 & 0.107 & 0.110 & 0.116 \\
  SHAC1\_CI &   (0.998) & (0.996) & (0.933) & (0.948) & (0.935) & (0.943) & (0.977) & (0.967) & (0.945) \\
  SHAC2 &   0.130 & 0.129 & 0.145 & 0.113 & 0.114 & 0.146 & 0.107 & 0.110 & 0.116 \\
 SHAC2\_CI&   (0.998) & (0.996) & (0.933) & (0.948) & (0.935) & (0.943) & (0.977) & (0.967) & (0.945) \\
\hline
    \end{tabular}%
  \label{tab:addlabel}%
\end{table}%

In Table 3, we change $p_x$ to 0.3 for the case of spatial assignments, and omit spillover effects in the estimation by regressing $Y_i$ on 1 and $X_i$ only. Due to the omitted variable bias, the coefficient on the individual assignment variable is biased upward when the assignments are spatially correlated. Nevertheless, we must make inference robust to spatial correlation because of the assignment design. With independent assignments, however, the individual assignment variable and the constructed spillover effect are independent. Therefore, not only is the slope coefficient estimator unbiased, but the standard errors need not be adjusted for spatial correlation either. In other words, even with the spillover effect omitted from the estimation, correct inference depends solely on the nature of the individual assignment variable.  

\subsection{Nonlinear Models}
In the last design, we consider spatial assignments in a nonlinear model. 
\begin{equation}
y_{ig}(x_{ig})=\mathbbm{1}\{\beta_{ig} x_{ig}+c_g+u_{ig}\geq 0\},
\end{equation}
where $X_M$ is an $M\times 1$ continuous random vector following a multivariate normal distribution with mean zero and a variance-covariance matrix equal to $p_x$ raised to the power of the distance. Other aspects of the population generating process are the same as the baseline design. Without spatial assignments, $p_x=0$ and $p_u$ ranges from 0 to 0.9; with spatial assignments, $p_u$ is fixed at 0.3 while $p_x$ varies from 0 to 0.9. 

We run probit regressions of $Y_i$ on 1 and $X_i$ and report standard errors of the APE estimator of $X_i$. We report the results when the entire population is observed, which are identical to the linear case with or without spatial assignments. Simulation results from other sampling schemes are also similar to the linear case and hence are omitted. 

\begin{figure}[htbp] %  figure placement: here, top, bottom, or page
   \centering
   \includegraphics[width=5.5in]{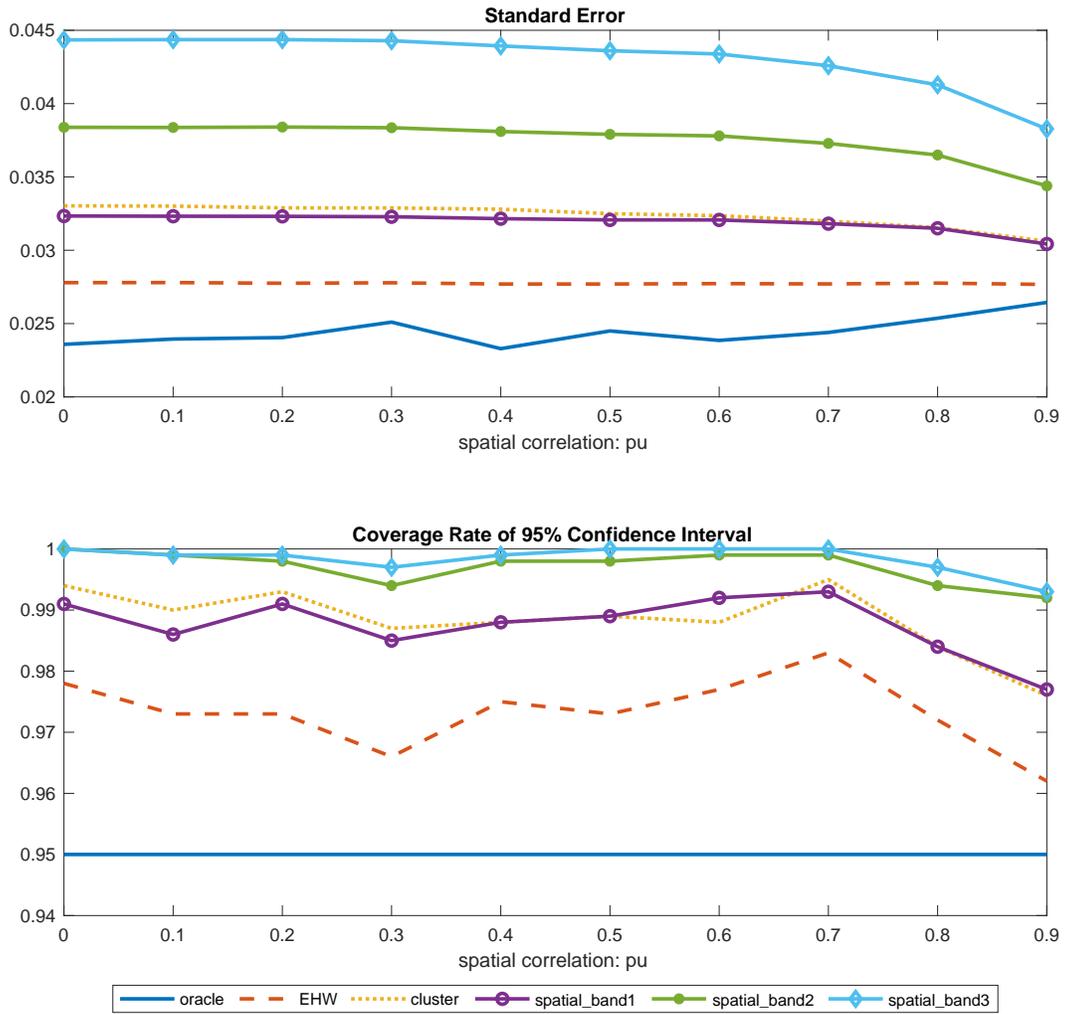} 
   \caption{Independent Assignments Observing Entire Population: APE}
   \label{fig:example}
\end{figure}

\begin{figure}[htbp] %  figure placement: here, top, bottom, or page
   \centering
   \includegraphics[width=5.5in]{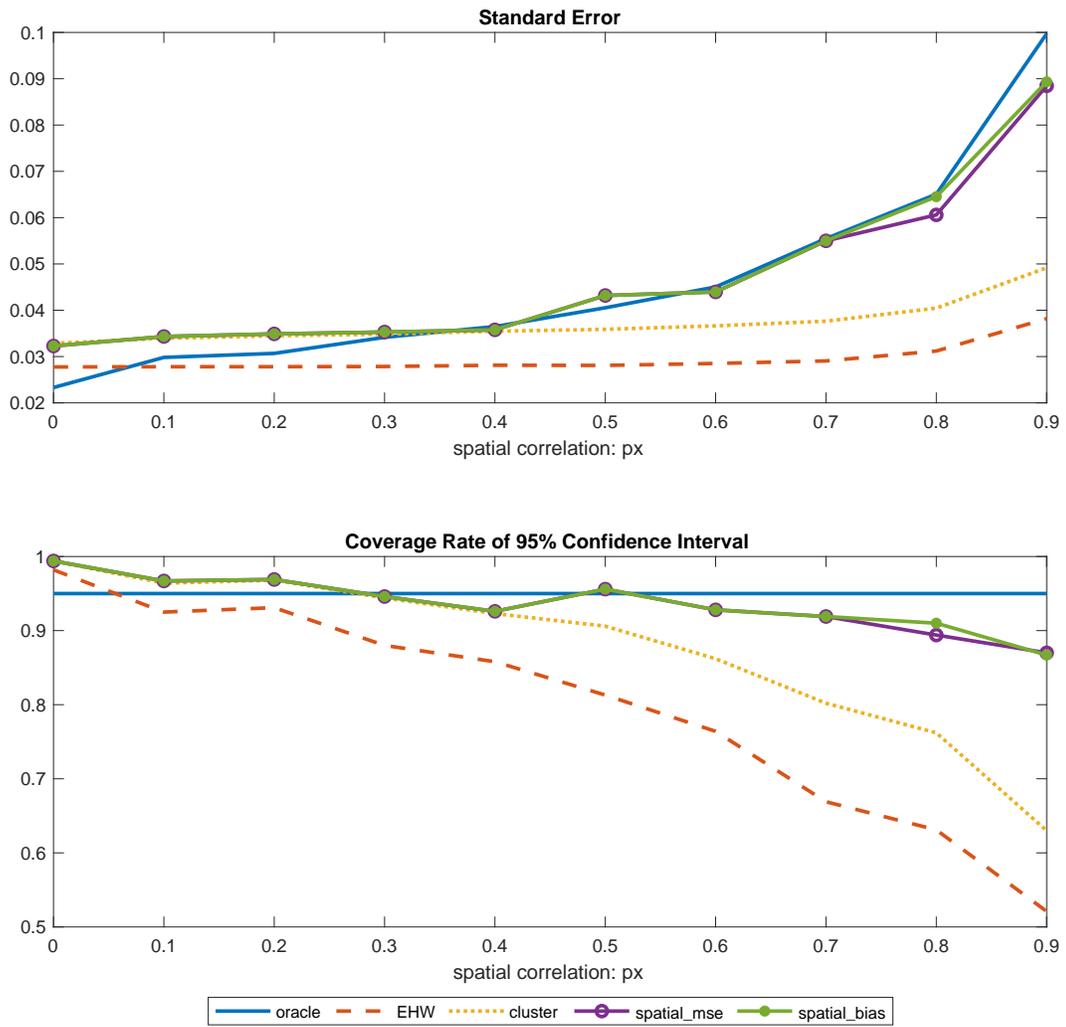} 
   \caption{Spatial Assignments Observing Entire Population: APE}
   \label{fig:example}
\end{figure}

\section{Conclusion}
Using a design-based approach, we identify the sources of uncertainty underlying spatial data. Whenever there are spatial assignments or when spillover effects are estimated, we must make inference robust to spatial correlation, unless the sampling probability is negligible. Since we allow for cluster partition, our results can be applied to short panel data as well with both spatial and temporal dependence. 

\appendix
\numberwithin{equation}{section}
\numberwithin{assumption}{section}

% Do we need to distinguish between endogenous and exogenous effects (spillover)? Or treat the potential outcome function as a reduced form solution to the system of structural equations as in Manski (2013)?
% The model may not necessarily be identified. 

\section{Notation and Regularity Conditions}\label{appendix1}

The following notation provides details of the variance-covariance matrix and the variance estimator in Section \ref{sec3}: 
\begin{align*}
\Delta^f_{ehw,M}=&\frac{1}{|D_M|}\sum_{i\in D_M}\mathbb{E}\Big\{\big[f_{iM}(W_{iM},\theta^*_M)-\gamma^*_M-F_M(\theta^*_M)H_M(\theta^*_M)^{-1}m_{iM}(W_{iM},\theta^*_M)\big]\cdot\\
&\big[f_{iM}(W_{iM},\theta^*_M)-\gamma^*_M-F_M(\theta^*_M)H_M(\theta^*_M)^{-1}m_{iM}(W_{iM},\theta^*_M)\big]'\Big\},
\tag{\stepcounter{equation}\theequation}
\end{align*}
\begin{equation}
\begin{aligned}
\Delta^f_{E,M}=&\frac{1}{|D_M|}\sum_{i\in D_M}\Big\{\mathbb{E}\big[f_{iM}(W_{iM},\theta^*_M)-\gamma^*_M-F_M(\theta^*_M)H_M(\theta^*_M)^{-1}m_{iM}(W_{iM},\theta^*_M)\big]\cdot\\
&\mathbb{E}\big[f_{iM}(W_{iM},\theta^*_M)-\gamma^*_M-F_M(\theta^*_M)H_M(\theta^*_M)^{-1}m_{iM}(W_{iM},\theta^*_M)\big]'\Big\},
\end{aligned}
\end{equation}
\begin{equation}
\begin{aligned}
\Delta^f_{cluster,M}=&\frac{1}{|D_M|}\sum_{i\in D_M}\sum_{j\in D_M,j\neq i}\mathbbm{1}(C_{iM}=C_{jM})\mathbb{E}\Big\{\big[f_{iM}(W_{iM},\theta^*_M)-\gamma^*_M\\
&-F_M(\theta^*_M)H_M(\theta^*_M)^{-1}m_{iM}(W_{iM},\theta^*_M)\big]\cdot\\
&\big[f_{jM}(W_{jM},\theta^*_M)-\gamma^*_M-F_M(\theta^*_M)H_M(\theta^*_M)^{-1}m_{jM}(W_{jM},\theta^*_M)\big]'\Big\},
\end{aligned}
\end{equation}
\begin{equation}
\begin{aligned}
\Delta^f_{EC,M}=&\frac{1}{|D_M|}\sum_{i\in D_M}\sum_{j\in D_M, j\neq i}\mathbbm{1}(C_{iM}=C_{jM})\Big\{\mathbb{E}\big[f_{iM}(W_{iM},\theta^*_M)-\gamma^*_M\\
&-F_M(\theta^*_M)H_M(\theta^*_M)^{-1}m_{iM}(W_{iM},\theta^*_M)\big]\cdot\\
&\mathbb{E}\Big[f_{jM}(W_{jM},\theta^*_M)-\gamma^*_M-F_M(\theta^*_M)H_M(\theta^*_M)^{-1}m_{jM}(W_{jM},\theta^*_M)\big]'\Big\},
\end{aligned}
\end{equation}
\begin{equation}
\begin{aligned}
\Delta^f_{spatial,M}=&\frac{1}{|D_M|}\sum_{i\in D_M}\sum_{j\in D_M,j\neq i}\mathbbm{1}(C_{iM}\neq C_{jM})\mathbb{E}\Big\{\big[f_{iM}(W_{iM},\theta^*_M)-\gamma^*_M\\
&-F_M(\theta^*_M)H_M(\theta^*_M)^{-1}m_{iM}(W_{iM},\theta^*_M)\big]\cdot\\
&\big[f_{jM}(W_{jM},\theta^*_M)-\gamma^*_M-F_M(\theta^*_M)H_M(\theta^*_M)^{-1}m_{jM}(W_{jM},\theta^*_M)\big]'\Big\},
\end{aligned}
\end{equation}
\begin{equation}
\begin{aligned}
\Delta^f_{ES,M}=&\frac{1}{|D_M|}\sum_{i\in D_M}\sum_{j\in D_M, j\neq i}\mathbbm{1}(C_{iM}\neq C_{jM})\Big\{\mathbb{E}\big[f_{iM}(W_{iM},\theta^*_M)-\gamma^*_M\\
&-F_M(\theta^*_M)H_M(\theta^*_M)^{-1}m_{iM}(W_{iM},\theta^*_M)\big]\cdot\\
&\mathbb{E}\Big[f_{jM}(W_{jM},\theta^*_M)-\gamma^*_M-F_M(\theta^*_M)H_M(\theta^*_M)^{-1}m_{jM}(W_{jM},\theta^*_M)\big]'\Big\},
\end{aligned}
\end{equation}
\begin{equation}
\begin{aligned}
\hat{V}_{f,SN}=&\frac{1}{|D_N|}\sum_{i\in D_M}\sum_{j\in D_M}R_{iM}R_{jM}\cdot \omega\bigg(\frac{\nu(i,j)}{b_M}\bigg)\Big\{\big[f_{iM}(W_{iM},\hat{\theta}_N)-\hat{\gamma}_N\\
&-\hat{F}_N(\hat{\theta}_N)\hat{H}_N(\hat{\theta}_N)^{-1}m_{iM}(W_{iM},\hat{\theta}_N)\big]\cdot\\
&\big[f_{jM}(W_{jM},\hat{\theta}_N)-\hat{\gamma}_N-\hat{F}_N(\hat{\theta}_N)\hat{H}_N(\hat{\theta}_N)^{-1}m_{jM}(W_{jM},\hat{\theta}_N)\big]'\Big\},
\end{aligned}
\end{equation}
and 
\begin{equation}
\begin{aligned}
V_{f,E}=&\frac{1}{|D_M|}\sum_{i\in D_M}\sum_{j\in D_M}\omega\bigg(\frac{\nu(i,j)}{b_M}\bigg)\Big\{\mathbb{E}\big[f_{iM}(W_{iM},\theta^*_M)-\gamma^*_M\\
&-F_M(\theta^*_M)H_M(\theta^*_M)^{-1}m_{iM}(W_{iM},\theta^*_M)\big]\cdot\\
&\mathbb{E}\big[f_{iM}(W_{iM},\theta^*_M)-\gamma^*_M-F_M(\theta^*_M)H_M(\theta^*_M)^{-1}m_{iM}(W_{iM},\theta^*_M)\big]'\Big\},
\end{aligned}
\end{equation}
where
\begin{equation}
F_M(\theta)=\frac{1}{|D_M|}\sum_{i\in D_M}\mathbb{E}\big[\nabla_\theta f_{iM}(W_{iM},\theta)\big]
\end{equation}
and
\begin{equation}
\hat{F}_N(\theta)=\frac{1}{|D_N|}\sum_{i\in D_M}R_{iM}\nabla_\theta f_{iM}(W_{iM},\theta).
\end{equation}

\begin{definition}
The random function $g_{iM}(W_{iM},\theta)$ is said to be Lipschitz in the parameter $\theta$ on $\Theta$ if there is $h(u)\downarrow 0$ as $u\downarrow 0$ and $b(\cdot): \mathcal{W}\to R$ such that $\sup _{M, i\in D_M}\mathbb{E}\big[|b_{iM}(W_{iM})|\big]<\infty$, and for all $\tilde{\theta},\theta\in\Theta$, $\big|g_{iM}(W_{iM},\tilde{\theta})-g_{iM}(W_{iM},\theta)\big|\leq b_{iM}(W_{iM}) h(\|\tilde{\theta}-\theta\|)$, $i\in D_M, M\geq 1$.
\end{definition}
 
We impose the following regularity conditions for the theorems in the paper. 

\begin{assumption}\label{assumpA1}
Suppose that $\frac{1}{|D_N|}\sum _{i\in D_M}R_{iM}\cdot m_{iM}(W_{iM},\hat{\theta}_N)=o_p(|D_N|^{-1/2})$ and 
(\romannumeral 1) let $Q_M(\theta) =\frac{1}{|D_M|}\sum _{i\in D_M}\mathbb{E}\big[q_{iM}(W_{iM},\theta)\big]$. $\{Q_M(\theta)\}$ has identifiably unique minimizers $\{\theta_M^*\}$ on $\Theta$ as in Definition 3.2 in \cite{gallant1988unified}; (\romannumeral 2) $\Theta$ is compact; 
(\romannumeral 3) $\theta^*_M\in int(\Theta)$ uniformly in $M$; 
(\romannumeral 4) $\sup _{M, i\in D_M}\mathbb{E}\Big[\sup _{\theta\in\Theta}|q_{iM}(W_{iM},\theta)/c_{iM}|^2\Big]<\infty$, where $\{c_{iM}, i\in D_M, M\geq 1\}$ is a sequence of nonrandom positive constants; 
(\romannumeral 5) $q_{iM}(W_{iM},\theta)/c_{iM}$ is Lipschitz in $\theta$ on $\Theta$; 
(\romannumeral 6) $q_{iM}(w,\theta)$ is twice continuously differentiable on $int(\Theta)$ for all $w$ in the support of $W_{iM}$, $\forall\ i,M$;  
(\romannumeral 7) $\sup _{M, i\in D_M}\mathbb{E}\Big[\sup _{\theta\in\Theta}\left\|\nabla_\theta m_{iM}(W_{iM},\theta)/c_{iM}\right\|^2\Big]<\infty$; 
(\romannumeral 8) $\sup _{M, i\in D_M}\mathbb{E}\Big[\sup _{\theta\in\Theta}\left\|m_{iM}(W_{iM},\theta)/c_{iM}\right\|^{2+\delta}\Big]<\infty$ for some $\delta>0$; 
\\(\romannumeral 9) $\inf_M|D_M|^{-1}J_M^{-2}\lambda_{min}(S_M)>0$, where $J_M=\max_{i\in D_M}\{c_{iM},d_{iM}\}$ and $\lambda_{min}(\cdot)$ stands for the smallest eigenvalue; 
%(\romannumeral 10) the NED scaling factors of the random field $m=\{m_{iM}(W_{iM},\theta), i\in D_M, M\geq 1\}$ satisfy $\sup_{M,i\in D_M}c_{iM}^{-1}d_{iM}\leq C<\infty$; 
(\romannumeral 10) $\nabla_\theta m_{iM}(W_{iM},\theta)/c_{iM}$ is Lipschitz in $\theta$ on $\Theta$; 
(\romannumeral 11) $H_M(\theta^*_M)/J_M$ is nonsingular;
(\romannumeral 12) for any fixed $s>0$, there exists a positive constant $C$ such that for any $M$ and every nonempty set $K\subseteq D_M$, $\mathbb{E}\Big[\big(\sum_{i\in K}m^s_{iM}\big)^2\Big]\geq C \sum_{i\in K}\mathbb{E}\big[(m^s_{iM})^2\big]$, where $m^s_{iM}=\frac{R_{iM}}{\sqrt{\rho_{uM}\rho_{cM}}}\mathbb{E}\big(m_{iM}(W_{iM},\theta^*_M)/J_M|\mathcal{F}_{iM}(s)\big)$.
\end{assumption}

\begin{assumption}\label{assumpA2}
(\romannumeral 1) The NED scaling factors of the random field $\{m_{iM}(W_{iM},\theta)m_{jM}(W_{jM},\theta), i,j\in D_M, M\geq 1\}$ satisfy $\sup_{M,i\in D_M}c_{iM}^{-2}d'_{iM}\leq C<\infty$; 
(\romannumeral 2) $\sup _{M, i\in D_M}\mathbb{E}\Big[\sup _{\theta\in\Theta}\left\|m_{iM}(W_{iM},\theta)/c_{iM}\right\|^{2r}\Big]<\infty$ for $r$ in Assumption \ref{assump6}.
%(\romannumeral 3) $\sup _{M, i\in D_M}\mathbb{E}\Big[\sup _{\theta\in\Theta}\left\|\nabla_\theta m_{iM}(W_{iM},\theta)/c_{iM}\right\|^2\Big]<\infty$.
\end{assumption}

\begin{assumption}\label{assumpA3}
(\romannumeral 1) $f_{iM}(w,\theta)$ is continuously differentiable on $int(\Theta)$ for all $w$ in the support of $W_{iM}$, $\forall\ i,M$; 
(\romannumeral 2) $\sup_{M, i\in D_M}\mathbb{E}\Big[\sup_{\theta\in\Theta}\left\|f_{iM}(W_{iM},\theta)/c_{iM}\right\|^{2r}\Big]<\infty$ for $r$ in Assumption \ref{assump6}; 
(\romannumeral 3) the NED scaling factors of the random field $\{f_{iM}f_{jM}, i,j\in D_M, M\geq 1\}$ satisfy $\sup_{M,i\in D_M}c_{iM}^{-2}d'_{iM}\leq C<\infty$, where $f_{iM}=f_{iM}(W_{iM},\theta)-\gamma^*_M-F_M(\theta^*_M)H_M(\theta^*_M)^{-1}m_{iM}(W_{iM},\theta)$;
(\romannumeral 4) $\inf_M|D_M|^{-1}J_M^{-2}\lambda_{min}(V_{f,M})>0$, where $J_M=\max_{i\in D_M} \{c_{iM},d_{iM}\}$; 
\\(\romannumeral 5) $\sup_{M, i\in D_M}\mathbb{E}\Big[\sup_{\theta\in\Theta}\left\|\nabla_\theta f_{iM}(W_{iM},\theta)\right\|^2\Big]<\infty$; 
%(\romannumeral 6) $f_{iM}(W_{iM},\theta)/c_{iM}$ is Lipschitz in $\theta$ on $\Theta$ with $\sup _{M, i\in D_M}\mathbb{E}\big[|b_{iM}(W_{iM})|^2\big]<\infty$; 
(\romannumeral 6) $f_{iM}(W_{iM},\theta)/c_{iM}$ is Lipschitz in $\theta$ on $\Theta$;
(\romannumeral 7) $\nabla_\theta f_{iM}(W_{iM},\theta)/c_{iM}$ is Lipschitz in $\theta$ on $\Theta$;
(\romannumeral 8) for any fixed $s>0$, there exists a positive constant $C$ such that for any $M$ and every nonempty set $K\subseteq D_M$, $\mathbb{E}\Big[\big(\sum_{i\in K}f^s_{iM}\big)^2\Big]\geq C \sum_{i\in K}\mathbb{E}\big[(f^s_{iM})^2\big]$, where $f^s_{iM}=\frac{R_{iM}}{\sqrt{\rho_{uM}\rho_{cM}}}\mathbb{E}\big[\big(f_{iM}(W_{iM},\theta^*_M)-\gamma^*_M-F_M(\theta^*_M)H_M(\theta^*_M)^{-1}m_{iM}(W_{iM},\theta^*_M)\big)/J_M|\mathcal{F}_{iM}(s)\big]$.
\end{assumption}

\section{Proof}

\begin{lemma}\label{lemma1}
Under Assumptions 1-3, $\frac{|D_N|}{|D_M|\rho_{uM}\rho_{cM}}\overset{p}\to 1$.
\end{lemma}

\noindent
\textbf{Proof:}

\begin{equation}
\frac{|D_N|}{|D_M|\rho_{uM}\rho_{cM}}=\frac{\sum_{i\in D_M}R_{iM}}{|D_M|\rho_{uM}\rho_{cM}}
\end{equation}
Since 
\begin{equation}
\mathbb{E}\bigg(\frac{\sum_{i\in D_M}R_{iM}}{|D_M|\rho_{uM}\rho_{cM}}\bigg)=1
\end{equation}
and
\begin{align*}
\mathbb{V}\bigg(\frac{\sum_{i\in D_M}R_{iM}}{|D_M|\rho_{uM}\rho_{cM}}\bigg)& \leq \frac{\sum_{g=1}^{G_M}\mathbb{E}\big[(\sum_{i\in D_{gM}}R_{iM})^2\big]}{|D_M|^2\rho_{uM}^2\rho_{cM}^2}\\
&=\frac{\sum_{g=1}^{G_M}\mathbb{E}\big[\sum_{i\in D_{gM}}R_{iM}+\sum_{i\in D_{gM}, j\in D_{gM}, j\neq i}R_{iM}R_{jM}\big]}{|D_M|^2\rho_{uM}^2\rho_{cM}^2}\\
&\leq \frac{C\sum_{g=1}^{G_M}\sum_{i\in D_{gM}}\rho_{uM}\rho_{cM}}{|D_M|^2\rho_{uM}^2\rho_{cM}^2}\\
&= \frac{C}{|D_M|\rho_{uM}\rho_{cM}}\to 0,
\tag{\stepcounter{equation}\theequation}
\end{align*} 
Lemma \ref{lemma1} follows from the mean square convergence. 

\begin{lemma}\label{lemma2}
(Weak Law of Large Numbers)
Suppose that $W=\{W_{iM}, i\in D_M, M\geq 1\}$ is $L_1$-NED on $U=\{U_{iM},i\in T_M, M\geq 1\}$ with the scaling factors $d_{iM}$ and $\sup_{M, i\in D_M}\mathbb{E}\norm{W_{iM}/c_{iM}}^2<\infty$, where $\{c_{iM}, i\in D_M, M\geq 1\}$ is a sequence of nonrandom positive constants. Under Assumptions \ref{assump1}-\ref{assump4} and Assumption \ref{assump5}(\romannumeral 2), 
\begin{equation*}
\frac{1}{J_M|D_N|}\sum_{i\in D_M}R_{iM}W_{iM}-\frac{1}{J_M|D_M|}\sum_{i\in D_M}EW_{iM}\overset{L_1}\to 0, 
\end{equation*}
where $J_M=\max_{i\in D_M}\max\{c_{iM},d_{iM}\}$.
\end{lemma}

\textbf{Proof:}
%The proof is based on the modification of the proof of Theorem 1 in \cite{jenish2012spatial}.
Define $Y_{iM}=W_{iM}/J_M$. 
\begin{align*}
&\frac{1}{J_M|D_N|}\sum_{i\in D_M}R_{iM}W_{iM}-\frac{1}{J_M|D_M|}\sum_{i\in D_M}EW_{iM}\\
=&\frac{|D_M|\rho_{uM}\rho_{cM}}{|D_N|}\frac{1}{|D_M|}\sum_{i\in D_M}\frac{R_{iM}}{\rho_{uM}\rho_{cM}}Y_{iM}-\frac{1}{|D_M|}\sum_{i\in D_M}EY_{iM}\\
=&\big(1+o_p(1)\big)\frac{1}{|D_M|}\sum_{i\in D_M}\frac{R_{iM}}{\rho_{uM}\rho_{cM}}Y_{iM}-\frac{1}{|D_M|}\sum_{i\in D_M}EY_{iM}
\tag{\stepcounter{equation}\theequation}
\end{align*}
Hence, it suffices to show that 
\begin{equation}
\frac{1}{|D_M|}\sum_{i\in D_M}\bigg(\frac{R_{iM}}{\rho_{uM}\rho_{cM}}Y_{iM}-EY_{iM}\bigg)\overset{L_1}\to 0.
\end{equation}
We can write 
\begin{align}
&\lim_{M\to\infty}\Bigg\||D_M|^{-1}\sum_{i\in D_M}\bigg(\frac{R_{iM}}{\rho_{uM}\rho_{cM}}Y_{iM}-EY_{iM}\bigg)\Bigg\|_1\\
=&\lim_{s\to\infty}\lim_{M\to\infty}\Bigg\||D_M|^{-1}\sum_{i\in D_M}\bigg(\frac{R_{iM}}{\rho_{uM}\rho_{cM}}Y_{iM}-EY_{iM}\bigg)\Bigg\|_1\\
\label{eq1} \leq & \lim_{s\to\infty}\limsup_{M\to\infty}\Bigg\||D_M|^{-1}\sum_{i\in D_M}\bigg(\frac{R_{iM}}{\rho_{uM}\rho_{cM}}Y_{iM}-\frac{R_{iM}}{\rho_{uM}\rho_{cM}}E\big(Y_{iM}|\mathcal{F}_{iM}(s)\big)\bigg)\Bigg\|_1\\ 
\label{eq2} &+\lim_{s\to\infty}\lim_{M\to\infty}\Bigg\||D_M|^{-1}\sum_{i\in D_M}\bigg(\frac{R_{iM}}{\rho_{uM}\rho_{cM}}E\big(Y_{iM}|\mathcal{F}_{iM}(s)\big)-EY_{iM}\bigg)\Bigg\|_1,
%\tag{\stepcounter{equation}\theequation}
\end{align} 
where $\mathcal{F}_{iM}(s)=\sigma(U_{jM}; j\in T_M: \nu(i,j) \leq s)$.

Note that 
\begin{align*}
&\mathbb{E}\Bigg\||D_M|^{-1}\sum_{i\in D_M}\bigg(\frac{R_{iM}}{\rho_{uM}\rho_{cM}}Y_{iM}-\frac{R_{iM}}{\rho_{uM}\rho_{cM}}E\big(Y_{iM}|\mathcal{F}_{iM}(s)\big)\bigg)\Bigg\| \\
\leq & |D_M|^{-1}\sum_{i\in D_M}E\bigg\|\frac{R_{iM}}{\rho_{uM}\rho_{cM}}\Big[Y_{iM}-E\big(Y_{iM}|\mathcal{F}_{iM}(s)\big)\Big]\bigg\|\\
=&|D_M|^{-1}\sum_{i\in D_M}E\Big\|Y_{iM}-E\big(Y_{iM}|\mathcal{F}_{iM}(s)\big)\Big\|\\
\leq & \sup_{M,i\in D_M} E\Big\|Y_{iM}-E\big(Y_{iM}|\mathcal{F}_{iM}(s)\big)\Big\|\\
\leq & \sup_{M,i\in D_M} J_M^{-1}d_{iM}\psi(s)\leq \psi(s).
\tag{\stepcounter{equation}\theequation}
\end{align*}
Therefore, (\ref{eq1}) is zero as $\lim_{s\to\infty}\psi(s)=0$.

Next, we show that for fixed $s>0$, (\ref{eq2}) is zero. 
Denote $V_{iM}^s = \frac{R_{iM}}{\rho_{uM}\rho_{cM}}\mathbb{E}\big(Y_{iM}|\mathcal{F}_{iM}(s)\big)$. Obverse that $\mathbb{E}(V_{iM}^s)=\mathbb{E}(Y_{iM})$.
For each fixed $s$, $\mathbb{E}\big(Y_{iM}|\mathcal{F}_{iM}(s)\big)$ is a measurable function of $\{U_{jM}; j\in T_M: \nu(i,j)\leq s\}$.
Since the cluster sizes are bounded $\{R_{iM}\}$ is $m$-dependent.
As a result, the maximal correlation coefficient on $\{V_{iM}^s\}$ also satisfies $\displaystyle\lim_{r\to\infty}\bar{\rho}(r)<1$ for each fixed $s$.

By Theorem 2.1 in \cite{bradley2017central}, 
\begin{align*}
\mathbb{V}\bigg(\sum_{i\in D_M}V^s_{iM}\bigg)=&\mathbb{E}\Bigg[\norm{\sum_{i\in D_M}(V^s_{iM}-\mathbb{E}(Y_{iM}))}^2\Bigg]\\
\leq &C\sum_{i\in D_M}\mathbb{E}\Big[\norm{V^s_{iM}-\mathbb{E}(Y_{iM})}^2\Big]\leq C\sum_{i\in D_M}\mathbb{E}\big(\norm{V^s_{iM}}^2\big)\\
=&C\sum_{i\in D_M}\mathbb{E}\bigg\{\frac{R_{iM}}{\rho_{uM}^2\rho_{cM}^2}\norm{\mathbb{E}(Y_{iM}|F_{iM}(s))}^2\bigg\}\\
\leq& C\sum_{i\in D_M}\frac{1}{\rho_{uM}\rho_{cM}}\mathbb{E}(\norm{Y_{iM}}^2).
\tag{\stepcounter{equation}\theequation}
\end{align*}
Therefore, 
\begin{equation}
\mathbb{V}\bigg(\frac{1}{|D_M|}\sum_{i\in D_M}V^s_{iM}\bigg)\leq \frac{C}{|D_M|\rho_{uM}\rho_{cM}}\frac{1}{|D_M|}\sum_{i\in D_M}\mathbb{E}(\norm{Y_{iM}}^2)\to 0,
\end{equation}
as $|D_M|\rho_{uM}\rho_{cM}\to \infty$. The result follows by the Liapunov's inequality.

\begin{lemma}\label{lemma3}
Under Assumptions \ref{assump1}, \ref{assump2}, and \ref{assump4}, suppose (\romannumeral 1) $a_{iM}(W_{iM},\theta)$ is Lipschitz in the $\theta$ on $\Theta$; 
(\romannumeral 2) $\sup_{M,i\in D_M}\mathbb{E}\big[\sup _{\theta\in\Theta}\|a_{iM}(W_{iM},\theta)\|^p\big]<\infty$ for some $p>1$. Then (1) Let $A_N(\theta) = \frac{1}{|D_N|}\sum_{i\in D_M}R_{iM}a_{iM}(W_{iM},\theta)$. $\left\|A_N(\tilde{\theta})-A_N(\theta)\right\|\leq B_N h(\|\tilde{\theta}-\theta\|)$, where $B_N =\frac{1}{|D_N|}\sum_{i\in D_M}R_{iM}\cdot b_{iM}(W_{iM})=O_p(1)$;
and (2) $A_M(\theta) =\frac{1}{|D_M|}\sum_{i\in D_M}\mathbb{E}\big[a_{iM}(W_{iM},\theta)\big]$ is uniformly equicontinuous.
\end{lemma}

\noindent
\textbf{Proof:}
We first show result (1).
\begin{align*}
&\left\|A_N(\tilde{\theta})-A_N(\theta)\right\|\\
=&\left\|\frac{1}{|D_N|}\sum_{i\in D_M}R_{iM}\big[a_{iM}(W_{iM},\tilde{\theta})-a_{iM}(W_{iM},\theta)\big]\right\|\\
\leq&\frac{1}{|D_N|}\sum_{i\in D_M}R_{iM}\left\|a_{iM}(W_{iM},\tilde{\theta})-a_{iM}(W_{iM},\theta)\right\|\\
\leq&\frac{1}{|D_N|}\sum_{i\in D_M}R_{iM}\cdot b_{iM}(W_{iM})h(\|\tilde{\theta}-\theta\|)\\
=&B_N h(\|\tilde{\theta}-\theta\|), 
\tag{\stepcounter{equation}\theequation}
\end{align*}
where
\begin{equation}
B_N =\frac{1}{|D_N|}\sum_{i\in D_M}R_{iM}\cdot b_{iM}(W_{iM})=\frac{|D_M|\rho_{uM}\rho_{cM}}{|D_N|}\frac{1}{|D_M|}\sum_{i\in D_M}\frac{R_{iM}}{\rho_{uM}\rho_{cM}}b_{iM}(W_{iM}).
\end{equation}
For all $\epsilon>0$, let $b_\epsilon=C/\epsilon$ for some $C<\infty$,
\begin{align*}
&P\bigg(\bigg|\frac{1}{|D_M|}\sum_{i\in D_M}\frac{R_{iM}}{\rho_{uM}\rho_{cM}}b_{iM}(W_{iM})\bigg|\geq b_\epsilon\bigg)\\
\leq&\mathbb{E}\bigg(\bigg|\frac{1}{|D_M|}\sum_{i\in D_M}\frac{R_{iM}}{\rho_{uM}\rho_{cM}}b_{iM}(W_{iM})\bigg|\bigg)/b_\epsilon\\
\leq&\frac{1}{|D_M|}\sum_{i\in D_M}\mathbb{E}\Big(\frac{R_{iM}}{\rho_{uM}\rho_{cM}}\Big)\mathbb{E}\Big[\big|b_{iM}(W_{iM})\big|\Big]/b_\epsilon\\
\leq &\sup _{M, i\in D_M}\mathbb{E}\Big[\big|b_{iM}(W_{iM})\big|\Big]/b_\epsilon<C/b_\epsilon= \epsilon.
\tag{\stepcounter{equation}\theequation}
\end{align*}
Because of Lemma \ref{lemma1} and the continuous mapping theorem, $\frac{|D_M|\rho_{uM}\rho_{cM}}{|D_N|}\overset{p}\to 1$.
Hence, $B_N=O_p(1)$.

\iffalse
The steps below follow the proof of Corollary 2.2 in \cite{newey1991uniform}.
Consider $\epsilon,\eta>0$ there is $K$ such that for all $N\geq N_0(\epsilon,\eta)$, $P(B_N>\epsilon K)<\eta$. Define $\Delta_N(\epsilon,\eta)=B_N/K$. Choose $\delta$ small enough that $h(u)<1/K$ for all $0\leq u<\delta$ and let $\mathcal{N}(\theta,\epsilon,\eta)=\{\tilde{\theta}\in \Theta: \|\tilde{\theta}-\theta\|<\delta\}$. Then $\sup_{\mathcal{N}}\norm{A_N(\tilde{\theta})-A_N(\theta)}\leq B_N \sup_{0\leq u<\delta}h(u)\leq \Delta_N(\epsilon,\eta)$ for each $\theta\in \Theta$. Therefore, $A_N(\theta)$ is stochastically equicontinuous. 
\fi

Next, we show $\{A_M(\theta)\}$ is uniformly equicontinuous. The proof is based on slight modification of the proof of Theorem 2 in \cite{jenish2009central}.
\begin{align*}
&\sup_{\theta\in\Theta}\sup_{\tilde{\theta}\in B(\theta,\delta)}\norm{A_M(\tilde{\theta})-A_M(\theta)}\\
\leq&\frac{1}{|D_M|}\sum_{i\in D_M}\sup_{\theta\in\Theta}\sup_{\tilde{\theta}\in B(\theta,\delta)}\norm{\mathbb{E}\big[a_{iM}(W_{iM},\tilde{\theta})-a_{iM}(W_{iM},\theta)\big]}\\
\leq&\frac{1}{|D_M|}\sum_{i\in D_M}\mathbb{E}\sup_{\theta\in\Theta}\sup_{\tilde{\theta}\in B(\theta,\delta)}\norm{a_{iM}(W_{iM},\tilde{\theta})-a_{iM}(W_{iM},\theta)}\\
=&\frac{1}{|D_M|}\sum_{i\in D_M} \mathbb{E}Y_{iM}(\delta),
\tag{\stepcounter{equation}\theequation}
\end{align*}
where $Y_{iM}(\delta) = \sup_{\theta\in\Theta}\sup_{\tilde{\theta}\in B(\theta,\delta)}\|a_{iM}(W_{iM},\tilde{\theta})-a_{iM}(W_{iM},\theta)\|$.
Define $f_{iM}=\sup_{\theta\in\Theta}\|a_{iM}(W_{iM},\theta)\|$.
Given condition (\romannumeral 2), there exists $k=k(\epsilon)<\infty$ for some $\epsilon>0$ such that 
\begin{equation}
\limsup_{M\to\infty} \frac{1}{|D_M|}\sum_{i\in D_M}\mathbb{E}(f_{iM}\mathbbm{1}(f_{iM}>k))<\frac{\epsilon}{6}.
\end{equation}
Under condition (\romannumeral 1), $a_{iM}(W_{iM},\theta)$ is $L_0$ stochastically equicontinuous on $\Theta$ by Proposition 1 in \cite{jenish2009central}.
Hence, we can find some $\delta=\delta(\epsilon)$ such that 
\begin{equation}
\limsup_{M\to\infty}\frac{1}{|D_M|}\sum_{i\in D_M} P(Y_{iM}(\delta)>\epsilon/3)\leq \frac{\epsilon}{6k}.
\end{equation}
\begin{align*}
&\limsup_{M\to\infty}\frac{1}{|D_M|}\sum_{i\in D_M} \mathbb{E}Y_{iM}(\delta)\\
\leq &\epsilon/3+\limsup_{M\to\infty}\frac{1}{|D_M|}\sum_{i\in D_M} \mathbb{E}Y_{iM}(\delta)\mathbbm{1}(Y_{iM}(\delta)>\epsilon/3,f_{iM}>k)\\
&+\limsup_{M\to\infty}\frac{1}{|D_M|}\sum_{i\in D_M} \mathbb{E}Y_{iM}(\delta)\mathbbm{1}(Y_{iM}(\delta)>\epsilon/3,f_{iM}\leq k)\\
\leq & \epsilon/3+2\limsup_{M\to\infty}\frac{1}{|D_M|}\sum_{i\in D_M}\mathbb{E}f_{iM}\mathbbm{1}(f_{iM}>k)+2k\frac{1}{|D_M|}\sum_{i\in D_M}\limsup_{M\to\infty}P(Y_{iM}(\delta)>\epsilon/3) =\epsilon
\tag{\stepcounter{equation}\theequation}
\end{align*}
As a result, $\limsup_{M\to\infty}\sup_{\theta\in\Theta}\sup_{\tilde{\theta}\in B(\theta,\delta)}\norm{A_M(\tilde{\theta})-A_M(\theta)}\to 0$ as $\delta\to 0$.

\begin{lemma}\label{lemma4}
(Central Limit Theorem) Let $W=\{W_{iM}, i\in D_M, M\geq 1\}$ be a real valued zero-mean random field that is $L_2$-NED on $U=\{U_{iM},i\in T_M, M\geq 1\}$ with the scaling factors $d_{iM}$ and the NED coefficients $\psi(s)$. Define $L_M=\sum _{i\in D_M}\frac{R_{iM}}{\sqrt{\rho_{uM}\rho_{cM}}}W_{iM}$ and $\sigma^2_M=\mathbb{V}(L_M)$. Suppose $W$ satisfies: (\romannumeral 1) $\sup _{M, i\in D_M}\mathbb{E}|W_{iM}/c_{iM}|^{2+\delta}<\infty$ for some $\delta>0$, where $\{c_{iM}, i\in D_M, M\geq 1\}$ is a sequence of nonrandom positive constants; (\romannumeral 2) for any fixed $s>0$, there exists a positive constant $C$ such that for any $M$ and every nonempty set $K\subseteq D_M$, $\mathbb{E}\Big[\big(\sum_{i\in K}\xi^s_{iM}\big)^2\Big]\geq C \sum_{i\in K}\mathbb{E}\big[(\xi^s_{iM})^2\big]$, where $\xi^s_{iM}=\frac{R_{iM}}{\sqrt{\rho_{uM}\rho_{cM}}}\mathbb{E}\big(W_{iM}/J_M|\mathcal{F}_{iM}(s)\big)$ and $J_M=\max _{i\in D_M}\{c_{iM}, d_{iM}\}$; (\romannumeral 3) $\inf _M|D_M|^{-1}J_M^{-2}\sigma^2_M>0$; (\romannumeral 4) the NED coefficient $\psi(s)$ is of size $-d$. %(\romannumeral 5) NED scaling factors satisfy $\sup _{M,i\in D_M}c_{iM}^{-1}d_{iM}\leq C<\infty$. 
Under Assumptions \ref{assump1}-\ref{assump5}, 
\begin{equation*}
\sigma^{-1}_ML_M\overset{d}\to \mathcal{N}(0,1).
\end{equation*}
\end{lemma}

\noindent
\textbf{Proof:}
%The proof is based on modification of the proof of Theorem 2 in \cite{jenish2012spatial}.
Define $Y_{iM}=W_{iM}/J_M$.
\begin{align}
\sigma^{-1}_ML_M=&J_M\sigma^{-1}_M\sum_{i\in D_M}\frac{R_{iM}}{\sqrt{\rho_{uM}\rho_{cM}}}W_{iM}/J_M
=\sigma_{Y,M}^{-1}\sum_{i\in D_M}\frac{R_{iM}}{\sqrt{\rho_{uM}\rho_{cM}}}Y_{iM}\\
=&\sigma_{Y,M}^{-1}\sum_{i\in D_M}\frac{R_{iM}}{\sqrt{\rho_{uM}\rho_{cM}}}\mathbb{E}\big(Y_{iM}|\mathcal{F}_{iM}(s)\big)\\
&+\sigma_{Y,M}^{-1}\sum_{i\in D_M}\frac{R_{iM}}{\sqrt{\rho_{uM}\rho_{cM}}}\Big[Y_{iM}-\mathbb{E}\big(Y_{iM}|\mathcal{F}_{iM}(s)\big)\Big]
\end{align}
From now on, let $L_M$ denote $\sum _{i\in D_M}\frac{R_{iM}}{\sqrt{\rho_{uM}\rho_{cM}}}Y_{iM}$ and $\sigma^2_M$ denote the variance of $\sum _{i\in D_M}\frac{R_{iM}}{\sqrt{\rho_{uM}\rho_{cM}}}Y_{iM}$. 

Define $\xi^s_{iM}=\frac{R_{iM}}{\sqrt{\rho_{uM}\rho_{cM}}}\mathbb{E}\big(Y_{iM}|\mathcal{F}_{iM}(s)\big)$ and $\eta^s_{iM}=\frac{R_{iM}}{\sqrt{\rho_{uM}\rho_{cM}}}\Big[Y_{iM}-\mathbb{E}\big(Y_{iM}|\mathcal{F}_{iM}(s)\big)\Big]$. Let $\norm{Y}_2=\sup_{M,i\in D_M}\norm{Y_{iM}}_2$. 
$\forall\ i\in D_M$,
\begin{align*}
 \norm{\xi^s_{iM}}_2=&\bigg(\mathbb{E}\frac{R_{iM}}{\rho_{uM}\rho_{cM}}\bigg)^{1/2}\norm{\mathbb{E}\big(Y_{iM}|\mathcal{F}_{iM}(s)\big)}_2 \leq \norm{Y_{iM}}_2\leq \norm{Y}_2<\infty
 \tag{\stepcounter{equation}\theequation}
\end{align*}
and 
\begin{align*}
\norm{\eta^s_{iM}}_2=&\bigg(\mathbb{E}\frac{R_{iM}}{\rho_{uM}\rho_{cM}}\bigg)^{1/2}\norm{Y_{iM}-\mathbb{E}\big(Y_{iM}|\mathcal{F}_{iM}(s)\big)}_2\\
\leq & \norm{Y_{iM}}_2+\norm{\mathbb{E}\big(Y_{iM}|\mathcal{F}_{iM}(s)\big)}_2\\
\leq& 2\norm{Y_{iM}}_2<\infty.
 \tag{\stepcounter{equation}\theequation}
\end{align*}
Furthermore, 
\begin{align*}
\sup_{M,i\in D_M}\norm{\eta^s_{iM}}_2=&\sup_{M,i\in D_M}\bigg(\mathbb{E}\frac{R_{iM}}{\rho_{uM}\rho_{cM}}\bigg)^{1/2}\norm{Y_{iM}-\mathbb{E}\big(Y_{iM}|\mathcal{F}_{iM}(s)\big)}_2\\
\leq &\sup_{M,i\in D_M}J_M^{-1}d_{iM}\psi(s)\leq \psi(s).
\tag{\stepcounter{equation}\theequation}
\end{align*}
 
For each fixed $s>0$, $\mathbb{E}\big(Y_{iM}|\mathcal{F}_{iM}(s)\big)$ is a measurable function of $U_{iM}$. Also, $R_{iM}$ is $m$-dependent. Hence, for the random field $\{\xi^s_{iM}\}$, $\bar{\alpha}(r)\to 0$ as $r\to \infty$ and $\displaystyle\lim_{r\to\infty}\bar{\rho}(r)<1$. 

Now, decompose $\frac{R_{iM}}{\sqrt{\rho_{uM}\rho_{cM}}}Y_{iM}$ as 
\begin{equation}
\frac{R_{iM}}{\sqrt{\rho_{uM}\rho_{cM}}}Y_{iM}=\xi_{iM}^s+\eta_{iM}^s.
\end{equation}
Applying Theorem 2.1 in \cite{bradley2017central}, 
\begin{equation}
\mathbb{E}\bigg|\sum_{i\in D_M}\xi^s_{iM}\bigg|^2\leq C\sum_{i\in D_M}\mathbb{E}(\xi^s_{iM})^2\leq C|D_M|\norm{Y}_2^2.
\end{equation}
By the Cauchy-Schwartz inequality,
\begin{equation}\label{eq11}
\big|Cov(\xi_{iM}^s,\eta_{jM}^s)\big|\leq \norm{\xi_{iM}^s}_2\norm{\eta_{jM}^s}_2\leq C\psi(s)
\end{equation}
and
\begin{equation}\label{eq12}
\big|Cov(\eta_{iM}^s,\eta_{jM}^s)\big|\leq \norm{\eta_{iM}^s}_2\norm{\eta_{jM}^s}_2 \leq C\psi(s).
\end{equation}

Let $h=r/3$ with $r$ in the definition of the maximal correlation coefficient in Assumption \ref{assump5}. Then, for sufficiently large $r$,
\begin{align*}
\sigma^2_M=&\mathbb{V}(L_M)=\mathbb{V}\bigg(\sum_{i \in D_M}\xi^h_{iM}+\sum_{i \in D_M}\eta^h_{iM}\bigg)\\
=&\mathbb{V}\bigg(\sum_{i \in D_M}\xi^h_{iM}\bigg)+\mathbb{V}\bigg(\sum_{i \in D_M}\eta^h_{iM}\bigg)+2Cov\bigg(\sum_{i \in D_M}\xi^h_{iM},\sum_{i \in D_M}\eta^h_{iM}\bigg)\\
\leq& \mathbb{E}\bigg[\Big(\sum_{i \in D_M}\xi^h_{iM}\Big)^2\bigg]+\sum_{i\in D_M}\mathbb{V}(\eta^h_{iM})+\sum_{i,j\in D_M,i\neq j}\big|Cov(\eta^h_{iM}, \eta^h_{jM})\big|\\
&+2\sum_{i,j\in D_M, i\neq j}\big|Cov(\xi^h_{iM}, \eta^h_{jM})\big|+2\sum_{i\in D_M}\big|Cov(\xi^h_{iM}, \eta^h_{iM})\big|\\
\leq & C_1|D_M|+C_2\sum_{i\in D_M}\bigg[\sum_{j\in D_M: j\neq i, \nu(i,j)\leq h}\psi(h)+\sum_{j\in D_M: \nu(i,j)> h}\psi(h)\bigg]\\
\leq & C_1|D_M|+C_2\sum_{i\in D_M}\bigg[C_3h^d\psi(h)+\sum_{j\in D_M: \nu(i,j)> h}\psi(h)\bigg]\\
\leq &C|D_M|,
\tag{\stepcounter{equation}\theequation}
\end{align*}
given the last term $\sum_{j\in D_M: \nu(i,j)> h}\psi(h)$ approaches zero as $h\to\infty$. 

Under condition (\romannumeral 3), 
\begin{equation}
\inf_M|D_M|^{-1}\sigma^2_M>0. 
\end{equation}
Hence, there exists $0<B<\infty$ such that for all $M$,
\begin{equation}\label{eqb32}
B|D_M|\leq \sigma^2_M.
\end{equation}
As a result, we have 
\begin{equation}
B|D_M|\leq \sigma^2_M\leq C|D_M|.
\end{equation}

Similarly, let $\lambda^m_{iM}=\frac{R_{iM}}{\sqrt{\rho_{uM}\rho_{cM}}}\mathbb{E}\Big[Y_{iM}-\mathbb{E}\big(Y_{iM}|\mathcal{F}_{iM}(s)\big)|\mathcal{F}_{iM}(m)\Big]$ and $\gamma^m_{iM}=\eta^s_{iM}-\lambda^m_{iM}$.
We have $\norm{\lambda^m_{iM}}_2\leq \norm{\eta^s_{iM}}_2<\infty$ and $\norm{\gamma^m_{iM}}_2\leq 2\norm{\eta^s_{iM}}_2<\infty$. In addition, 
\begin{align*}
\sup_{M,i\in D_M}\norm{\gamma^m_{iM}}_2=&\sup_{M,i\in D_M}\bigg(\mathbb{E}\frac{R_{iM}}{\rho_{uM}\rho_{cM}}\bigg)^{1/2}\bigg\|Y_{iM}-\mathbb{E}\big(Y_{iM}|\mathcal{F}_{iM}(s)\big)\\
&-\mathbb{E}\big(Y_{iM}|\mathcal{F}_{iM}(m)\big)+\mathbb{E}\Big[\mathbb{E}\big(Y_{iM}|\mathcal{F}_{iM}(s)\big)|\mathcal{F}_{iM}(m)\Big]\bigg\|_2\\
=&
\begin{cases}
\sup_{M,i\in D_M}\norm{Y_{iM}-\mathbb{E}\big(Y_{iM}|\mathcal{F}_{iM}(m)\big)}_2\leq C\psi(m), & m\geq s\\
\sup_{M,i\in D_M}\norm{Y_{iM}-\mathbb{E}\big(Y_{iM}|\mathcal{F}_{iM}(s)\big)}_2\leq C\psi(s)\leq C\psi(m), & m< s.
\end{cases}
\tag{\stepcounter{equation}\theequation}
\end{align*}

\begin{align*}
\tilde{\sigma}^2_{M,s}=&\mathbb{V}\bigg(\sum_{i\in D_M}\eta^s_{iM}\bigg)=\mathbb{V}\bigg(\sum_{i\in D_M}\lambda^h_{iM}+\sum_{i\in D_M}\gamma^h_{iM}\bigg)\\
=& \mathbb{V}\bigg(\sum_{i\in D_M}\lambda^h_{iM}\bigg)+\mathbb{V}\bigg(\sum_{i\in D_M}\gamma^h_{iM}\bigg)+2Cov\bigg(\sum_{i\in D_M}\lambda^h_{iM}, \sum_{i\in D_M}\gamma^h_{iM}\bigg)\\
\leq &\mathbb{E}\bigg[\Big(\sum_{i\in D_M}\lambda^h_{iM}\Big)^2\bigg]+\sum_{i\in D_M}\mathbb{V}(\gamma^h_{iM})+\sum_{i,j\in D_M, i\neq j}\big|Cov(\gamma_{iM}^h, \gamma_{jM}^h)\big|\\
&+2\sum_{i,j\in D_M, i\neq j}\big|Cov(\lambda_{iM}^h, \gamma_{jM}^h)\big|+2\sum_{i\in D_M}\big|Cov(\lambda_{iM}^h, \gamma_{iM}^h)\big|\\
\leq &C_1\sum_{i\in D_M}\norm{\lambda^h_{iM}}^2_2+\sum_{i\in D_M}\norm{\gamma^h_{iM}}^2_2+2\sum_{i\in D_M}\norm{\lambda^h_{iM}}_2\norm{\gamma^h_{iM}}_2\\
&+2\sum_{i,j\in D_M, i\neq j}\norm{\lambda^m_{iM}}_2\norm{\gamma^h_{jM}}_2+\sum_{i,j\in D_M,i\neq j}\norm{\gamma^h_{iM}}_2\norm{\gamma_{jM}^h}_2\\
\leq & C_2|D_M|\norm{\eta^s_{iM}}_2+C_3\sum_{i\in D_M}\bigg[\sum_{j\in D_M: j\neq i, \nu(i,j)\leq h}\psi(h)\norm{\eta^s_{iM}}_2+\sum_{j\in D_M: \nu(i,j)>h}\psi(h)\norm{\eta^s_{iM}}_2\bigg]\\
\leq &C|D_M|\psi(s)
\tag{\stepcounter{equation}\theequation}
\end{align*}
Hence, 
\begin{equation}\label{eqb36}
\lim_{s\to\infty}\limsup_{M\to\infty}\frac{\tilde{\sigma}^2_{M,s}}{\sigma^2_M}\leq C\lim_{s\to\infty}\psi(s)=0.
\end{equation}

We now show that for any fixed $s>0$, ${\xi^s_{iM}}$ satisfies the central limit theorem for an $\alpha$-mixing process. The proof is based on the modification of the proof of Theorem 1.1 in \cite{bradley2017central}. We first verify the Lindeberg condition. It suffices to verify the Liapunov's condition. Let $\sigma^2_{M,s}=\mathbb{V}\Big(\sum_{i \in D_M}\xi^s_{iM}\Big)$. 
By repeated use of the Minkowski's inequality, we have
\begin{equation}
|\sigma_M-\tilde{\sigma}_{M,s}|\leq \sigma_{M,s}.
\end{equation} 
Because of (B.35), there exists $s_*$ such that for all $s>s_*$,
\begin{equation}
|D_M|^{-1}\tilde{\sigma}^2_{M,s}\leq C\psi(s)\leq B/2.
\end{equation}
Therefore, 
\begin{equation}
\inf_M |D_M|^{-1}\sigma^2_{M,s}\geq \inf_M |D_M|^{-1}\sigma_M^2-\sup_M |D_M|^{-1}\tilde{\sigma}^2_{M,s}\geq B-B/2=B/2>0.
\end{equation}
Consider $M$ large enough so that $\sigma^2_{M,s}/|D_M|>0$.
\begin{align*}
&\sum_{i\in D_M}\mathbb{E}\big[|\xi^s_{iM}/\sigma_{M,s}|^{2+\delta}\big]=\sum_{i\in D_M}\mathbb{E}\Bigg[\bigg|\frac{R_{iM}}{(\sigma_{M,s}/\sqrt{|D_M|})\sqrt{|D_M|\rho_{uM}\rho_{cM}}}\mathbb{E}\big(Y_{iM}|\mathcal{F}_{iM}(s)\big)\bigg|^{2+\delta}\Bigg]\\
\leq&\frac{\sum_{i\in D_M}\rho_{uM}\rho_{cM}\mathbb{E}\big(|Y_{iM}|^{2+\delta}\big)}{(\sigma_{M,s}/\sqrt{|D_M|})^{2+\delta}(|D_M|\rho_{uM}\rho_{cM})^{1+\delta/2}}\\
=&\frac{1}{(\sigma_{M,s}/\sqrt{|D_M|})^{2+\delta}(|D_M|\rho_{uM}\rho_{cM})^{\delta/2}}\frac{1}{|D_M|}\sum_{i\in D_M}\mathbb{E}\big(|Y_{iM}|^{2+\delta}\big)\to 0
\tag{\stepcounter{equation}\theequation}
\end{align*}  
as $|D_M|\rho_{uM}\rho_{cM}\to \infty$ as $M\to \infty$.

To allow for irregularly spaced lattice, we modify the proof of Theorem 1.1 in \cite{bradley2017central} in the following way. Change (4.10) in Step 5 to $L_1^{(n_q)}>q^2m_q^d$. 
Correspondingly, change (4.12) to 
\begin{equation}
\Bigg|\sum^{q^2m_q^d}_{j=q+1}\Big(s_{p(n_q,j)}^{(n_q)}\Big)^2-\sum^{q^2m_q^d}_{j=q+1}\lambda_j\Bigg|\leq \frac{1}{q}.
\end{equation}
In (4.13), replace $q^2m_q$ with $q^2m_q^d$. In Step 10, using Lemma A.1 (\romannumeral 2) in \cite{jenish2009central}, the cardinality of $\Gamma_3^{(q)}$ is $Cqm_q^d$.

The rest of the proof is the same as part 5 of the proof of Theorem 2 in \cite{jenish2012spatial}.

\bigskip
\noindent
\textbf{Proof of Theorem \ref{thm1}:}

%All we need to do is to verify the uniform convergence of the objective functions on $\Theta$.
We first show that $\hat{\theta}_N-\theta^*_M\overset{p}\to \textbf{0}$.

Denote $Q_N(\theta) = \frac{1}{|D_N|}\sum_{i\in {D_M}} R_{iM}q_{iM}(W_{iM}, \theta)$.
First note that 
\begin{align*}
Q_N(\theta)&=\frac{|D_M|\rho_{uM}\rho_{cM}}{|D_N|}\frac{1}{|D_M|}\sum_{i\in D_M}\frac{R_{iM}}{\rho_{uM}\rho_{cM}}q_{iM}(W_{iM},\theta)\\
&=\big(1+o_p(1)\big)\frac{1}{|D_M|}\sum_{i\in D_M}\frac{R_{iM}}{\rho_{uM}\rho_{cM}}q_{iM}(W_{iM},\theta).
\tag{\stepcounter{equation}\theequation}
\end{align*}
Define $J_M=\max_{i\in D_M} \{c_{iM}, d_{iM}\}$.
Hence, it is sufficient to show that for each $\theta\in\Theta$
\begin{equation}\label{eq3}
\left\|\frac{1}{J_M|D_M|}\sum_{i\in D_M}\frac{R_{iM}}{\rho_{uM}\rho_{cM}}q_{iM}(W_{iM},\theta)-\frac{1}{J_M}Q_M(\theta)\right\|\overset{p}\to 0,
\end{equation}
which holds by applying Lemma \ref{lemma2}.
Next, 
\begin{equation}
\sup_{\theta\in\Theta}\frac{1}{J_M}|Q_N(\theta)-Q_M(\theta)|=o_p(1)
\end{equation}
follows from Corollary 2.2 in \cite{newey1991uniform}, Lemma \ref{lemma3}, and (\ref{eq3}).
As a result, consistency follows, e.g., from \cite{gallant1988unified}.

For asymptotic normality, I start by verifying that 
\begin{equation}
\sum_{i\in D_M}\mathbb{E}\big[m_{iM}(W_{iM},\theta^*_M)/J_M\big]=\bm{0},
\end{equation}
which holds by Lemma 3.6 in \cite{newey1994large} and Jensen's inequality.

By the element-by-element mean value expansion around $\theta^*_M$,
\begin{equation}\label{eq6}
\begin{aligned}
&o_p(|D_N|^{-1/2})=\frac{1}{|D_N|}\sum_{i\in D_M}R_{iM}\cdot m_{iM}(W_{iM},\hat{\theta}_N)\\
=&\frac{1}{|D_N|}\sum_{i\in D_M}R_{iM}\cdot m_{iM}(W_{iM},\theta^*_M)+J_M\frac{1}{J_M|D_N|}\sum_{i\in D_M}R_{iM} \nabla_\theta m_{iM}(W_{iM},\check{\theta})(\hat{\theta}_N-\theta^*_M),
\end{aligned}
\end{equation}
where $\check{\theta}$ lies on the line segment connecting $\theta^*_M$ and $\hat{\theta}_N$.

I first show 
\begin{equation}\label{eq4}
\frac{1}{J_M}\hat{H}_N(\check{\theta}) = \frac{1}{J_M|D_N|}\sum_{i\in D_M}R_{iM} \nabla_\theta m_{iM}(W_{iM},\check{\theta})=\frac{1}{J_M}H_M(\theta^*_M)\big(I_k+o_p(1)\big).
\end{equation}
We can write
\begin{equation}
\begin{aligned}
\frac{1}{J_M}\hat{H}_N(\theta)=&\frac{|D_M|\rho_{uM}\rho_{cM}}{|D_N|}\frac{1}{J_M|D_M|}\sum_{i\in D_M}\frac{R_{iM}}{\rho_{uM}\rho_{cM}}\nabla_\theta m_{iM}(W_{iM},\theta)\\
=&\big(1+o_p(1)\big)\frac{1}{J_M|D_M|}\sum_{i\in D_M}\frac{R_{iM}}{\rho_{uM}\rho_{cM}}\nabla_\theta m_{iM}(W_{iM},\theta).
\end{aligned}
\end{equation}
Applying Lemma \ref{lemma2}, 
\begin{equation}
\left\|\frac{1}{J_M|D_M|}\sum_{i\in D_M}\frac{R_{iM}}{\rho_{uM}\rho_{cM}}\nabla_\theta m_{iM}(W_{iM},\theta)-\frac{1}{J_M}H_M(\theta)\right\|\overset{p}\to \bm{0}.
\end{equation}
By Lemma \ref{lemma3} above and Corollary 2.2 in \cite{newey1991uniform},
\begin{align*}
&\left\|\frac{1}{J_M}\hat{H}_N(\check{\theta})-\frac{1}{J_M}H_M(\theta^*_M)\right\|\\
= & \left\|\frac{1}{J_M}\big(\hat{H}_N(\check{\theta})-H_M(\check{\theta})+H_M(\check{\theta})-H_M(\theta^*_M)\big)\right\|  \\
\leq& \sup_{\theta\in\Theta}\frac{1}{J_M}\left\|\hat{H}_N(\theta)-H_M(\theta)\right\|+\frac{1}{J_M}\left\|H_M(\check{\theta})-H_M(\theta^*_M)\right\|\overset{p}\to \bm{0}.
\tag{\stepcounter{equation}\theequation}
\end{align*}

(\ref{eq4}) implies
\begin{equation}\label{eq5}
J_M\hat{H}_N(\check{\theta})^{-1}=J_MH_M(\theta^*_M)^{-1}(I_k+o_p(1)).
\end{equation}
Using (\ref{eq5}), (\ref{eq6}) can be written as 
\begin{equation}\label{eq8}
\begin{aligned}
S_M^{-1/2}H_M(\theta^*_M)\sqrt{|D_N|}(\hat{\theta}_N-\theta^*_M)=&-S_M^{-1/2}\frac{1}{\sqrt{|D_N|}}\sum_{i\in D_M}R_{iM}\cdot m_{iM}(W_{iM},\theta^*_M)\\
&-S_M^{-1/2}o_p(1)\frac{1}{\sqrt{|D_N|}}\sum_{i\in D_M}R_{iM}\cdot m_{iM}(W_{iM},\theta^*_M)+o_p(1).
\end{aligned}
\end{equation}

We can write
\begin{equation}\label{eq7}
\begin{aligned}
&\frac{1}{\sqrt{|D_N|}}\sum_{i\in D_M}R_{iM}\cdot m_{iM}(W_{iM},\theta)\\
=&\sqrt{\frac{|D_M|\rho_{uM}\rho_{cM}}{|D_N|}}\frac{1}{\sqrt{|D_M|}}\sum_{i\in D_M}\frac{R_{iM}}{\sqrt{\rho_{uM}\rho_{cM}}}m_{iM}(W_{iM},\theta)\\
=&\big(1+o_p(1)\big)\frac{1}{\sqrt{|D_M|}}\sum_{i\in D_M}\frac{R_{iM}}{\sqrt{\rho_{uM}\rho_{cM}}}m_{iM}(W_{iM},\theta)
\end{aligned}
\end{equation}
Plug (\ref{eq7}) into (\ref{eq8}), we have 
\begin{align*}
&S_M^{-1/2}H_M(\theta^*_M)\sqrt{|D_N|}(\hat{\theta}_N-\theta^*_M)\\
=&-S_M^{-1/2}\frac{1}{\sqrt{|D_M|}}\sum_{i\in D_M}\frac{R_{iM}}{\sqrt{\rho_{uM}\rho_{cM}}}m_{iM}(W_{iM},\theta^*_M)\\
&-S_M^{-1/2}\frac{1}{\sqrt{|D_M|}}\sum_{i\in D_M}\frac{R_{iM}}{\sqrt{\rho_{uM}\rho_{cM}}}m_{iM}(W_{iM},\theta^*_M)\cdot o_p(1)+o_p(1).
\tag{\stepcounter{equation}\theequation}
\end{align*}

By Lemma \ref{lemma4} above and Corollary 1 in \cite{jenish2012spatial},
\begin{equation}\label{eq9}
\begin{aligned}
&S_M^{-1/2}\frac{1}{\sqrt{|D_M|}}\sum_{i\in D_M}\frac{R_{iM}}{\sqrt{\rho_{uM}\rho_{cM}}}m_{iM}(W_{iM},\theta^*_M)\\
=&S_M^{-1/2}\frac{1}{\sqrt{|D_M|}}\sum_{i\in D_M}\bigg(\frac{R_{iM}}{\sqrt{\rho_{uM}\rho_{cM}}}m_{iM}(W_{iM},\theta^*_M)-\sqrt{\rho_{uM}\rho_{cM}}\mathbb{E}\big[m_{iM}(W_{iM},\theta^*_M)\big]\bigg)\\
\overset{d}\to &\mathcal{N}(\textbf{0},I_k)
\end{aligned}
\end{equation}

Because of (\ref{eq9}), 
\begin{align*}
S_M^{-1/2}H_M(\theta^*_M)\sqrt{|D_N|}(\hat{\theta}_N-\theta^*_M)=&-S_M^{-1/2}\frac{1}{\sqrt{|D_M|}}\sum_{i\in D_M}\frac{R_{iM}}{\sqrt{\rho_{uM}\rho_{cM}}}m_{iM}(W_{iM},\theta^*_M)
\\
&+o_p(1)O_p(1)+o_p(1)\overset{d}\to \mathcal{N}(\textbf{0},I_k).
\tag{\stepcounter{equation}\theequation}
\end{align*}

\begin{lemma}\label{lemma5}
Suppose (\romannumeral 1) $X=\{X_{iM}, i\in D_M, M\geq 1\}$ is $L_2$-NED on $U=\{U_{iM}, i\in T_M, M\geq 1\}$ with the scaling factors $d_{iM}$ and the NED coefficients $\psi(s)$ of size $-2d(r-1)/(r-2)$ for some $r>2$; (\romannumeral 2) $\sup_{M, i\in D_M}\mathbb{E}\big(|X_{iM}/c_{iM}|^{2r}\big)<\infty$, where $\{c_{iM}, i\in D_M, M\geq 1\}$ is a sequence of nonrandom positive constants.\\
Define $W_{iM}=\sum_{j\in D_M: j\neq i, \nu(i,j)\leq b_M}\omega\Big(\frac{\nu(i,j)}{b_M}\Big)\Big[\frac{R_{iM}R_{jM}}{\rho_{uM}\rho_{cM}} X_{iM}X_{jM}-\mathbb{E}\Big(\frac{R_{iM}R_{jM}}{\rho_{uM}\rho_{cM}}X_{iM}X_{jM}\Big)\Big]$.
Under Assumptions \ref{assump1}-\ref{assump4}, Assumption \ref{assump5}(\romannumeral 2), and Assumption \ref{assump7},
\begin{equation*}
\frac{1}{J_M^2|D_M|}\sum_{i\in D_M}W_{iM}\overset{p}\to 0,
\end{equation*}
where $J_M=\max_{i\in D_M}\max \{c_{iM}, \sqrt{d'_{iM}}\}$ and $d'_{iM}$ is the NED scaling factor of the random field $\{X_{iM}X_{jM}, i,j\in D_M, M\geq 1 \}$.
\end{lemma}

\noindent
\textbf{Proof:}
Denote 
\begin{equation*}
Y_{iM}=\sum_{j\in D_M: j\neq i, \nu(i,j)\leq b_M}\omega\Big(\frac{\nu(i,j)}{b_M}\Big)\frac{R_{iM}R_{jM}}{\rho_{uM}\rho_{cM}} X_{iM}X_{jM}/J_M^2,
\end{equation*}
\begin{equation*}
\lambda^s_{iM}=\sum_{j\in D_M: j\neq i, \nu(i,j)\leq b_M}\omega\Big(\frac{\nu(i,j)}{b_M}\Big)\frac{R_{iM}R_{jM}}{\rho_{uM}\rho_{cM}}\mathbb{E}\big[X_{iM}X_{jM}/J_M^2|\mathcal{F}_{iM}(s)\big],
\end{equation*}
and $\eta^s_{iM}=Y_{iM}-\lambda^s_{iM}$. $\forall\ i\in D_M$,
\begin{align*}
\norm{\lambda^s_{iM}}_2\leq& \sum_{j\in D_M: j\neq i, \nu(i,j)\leq b_M}\bigg|\omega\Big(\frac{\nu(i,j)}{b_M}\Big)\bigg|\bigg[\mathbb{E}\frac{R_{iM}R_{jM}}{\rho^2_{uM}\rho^2_{cM}}\bigg]^{1/2}\Big[\mathbb{E}\Big(\mathbb{E}\big[X_{iM}X_{jM}/J_M^2|\mathcal{F}_{iM}(s)\big]\Big)^2\Big]^{1/2}\\
\leq &C_1\sum^{b_M}_{r=1}\sum_{\nu(i,j)\in [r,r+1)}\frac{1}{\sqrt{\rho_{cM}}}\leq C_2\sum^{b_M}_{r=1}r^{d-1}\frac{1}{\sqrt{\rho_{cM}}}\leq Cb_M\cdot b_M^{d-1}\frac{1}{\sqrt{\rho_{cM}}}=C\frac{b_M^d}{\sqrt{\rho_{cM}}}
\tag{\stepcounter{equation}\theequation}
\end{align*}
and 
\begin{equation}
\norm{\eta^s_{iM}}_2\leq C\frac{b_M^d}{\sqrt{\rho_{cM}}}.
\end{equation}
Applying Corollary 4.3 in \cite{gallant1988unified},
\begin{align*}
\sup_{M,i\in D_M}\norm{\eta^s_{iM}}_2\leq C\frac{b_M^d}{\sqrt{\rho_{cM}}}\psi(s),
\tag{\stepcounter{equation}\theequation}
\end{align*}
where $\psi(s)$ is of order $-d$.

Using a similar argument of (B.30),
\begin{align*}
\mathbb{V}\bigg(\sum_{i\in D_M}Y_{iM}\bigg)=&\mathbb{V}\bigg(\sum_{i \in D_M}\lambda^h_{iM}+\sum_{i \in D_M}\eta^h_{iM}\bigg)\\
=&\mathbb{V}\bigg(\sum_{i \in D_M}\lambda^h_{iM}\bigg)+\mathbb{V}\bigg(\sum_{i \in D_M}\eta^h_{iM}\bigg)+2Cov\bigg(\sum_{i \in D_M}\lambda^h_{iM},\sum_{i \in D_M}\eta^h_{iM}\bigg)\\
\leq& \sum_{i \in D_M}\mathbb{E}\big[\big(\lambda^h_{iM}\big)^2\big]+\sum_{i\in D_M}\mathbb{V}(\eta^h_{iM})+\sum_{i,j\in D_M,i\neq j}\big|Cov(\eta^h_{iM}, \eta^h_{jM})\big|\\
&+2\sum_{i,j\in D_M, i\neq j}\big|Cov(\lambda^h_{iM}, \eta^h_{jM})\big|+2\sum_{i\in D_M}\big|Cov(\lambda^h_{iM}, \eta^h_{iM})\big|\\
\leq & C_1|D_M|\frac{b_m^{2d}}{\rho_{cM}}+C_2\frac{b_m^{2d}}{\rho_{cM}}\sum_{i\in D_M}\bigg[\sum_{j\in D_M: j\neq i, \nu(i,j)\leq h}\psi(h)+\sum_{j\in D_M: \nu(i,j)> h}\psi(h)\bigg]\\
\leq &C|D_M|\frac{b_m^{2d}}{\rho_{cM}}.
\tag{\stepcounter{equation}\theequation}
\end{align*}

Using Chebyshev's inequality, for any $\eta>0$,
\begin{align*}
&P\Bigg(\bigg|\frac{1}{J_M^2|D_M|}\sum_{i\in D_M}W_{iM}\bigg| \geq \eta\Bigg)\leq \frac{1}{\eta^2|D_M|^2}\mathbb{V}\Big(\sum_{i\in D_M}Y_{iM}\Big)\\
\leq & \frac{C}{\eta^2} \frac{b_M^{2d}}{|D_M|\rho_{cM}}=o(1).
\tag{\stepcounter{equation}\theequation}
\end{align*}
Hence, the proof.

\bigskip
\noindent
\textbf{Proof of Theorem \ref{thm3}:}

Since (\ref{eq5}) holds by replacing $\check{\theta}$ with $\hat{\theta}_N$, 
\begin{equation}
J_M\hat{H}_N(\hat{\theta}_N)^{-1}=J_MH_M(\theta^*_M)^{-1}(I_k+o_p(1)),
\end{equation}
where $J_M=\max_{i \in D_M} \{c_{iM},d_{iM}\}$ and $d_{iM}$ is the maximum of the NED scaling factors of the random fields used in the proof.

Notice that 
\begin{align*}
S_M=&\frac{1}{|D_M|}\sum_{i\in D_M}\sum_{j\in D_M}\mathbb{E}\Bigg\{\bigg(\frac{R_{iM}}{\sqrt{\rho_{uM}\rho_{cM}}}m_{iM}(W_{iM},\theta^*_M)-\sqrt{\rho_{uM}\rho_{cM}}\mathbb{E}\big[m_{iM}(W_{iM},\theta^*_M)\big]\bigg)\\
&\cdot\bigg(\frac{R_{jM}}{\sqrt{\rho_{uM}\rho_{cM}}}m_{jM}(W_{jM},\theta^*_M)-\sqrt{\rho_{uM}\rho_{cM}}\mathbb{E}\big[m_{jM}(W_{jM},\theta^*_M)\big]\bigg)\Bigg\}\\
=&\frac{1}{|D_M|}\sum_{i\in D_M}\sum_{j\in D_M}\mathbb{E}(\tilde{m}_{iM}\tilde{m}_{jM}),
\tag{\stepcounter{equation}\theequation}
\end{align*}
where 
\begin{equation}
\tilde{m}_{iM}=\frac{R_{iM}}{\sqrt{\rho_{uM}\rho_{cM}}}m_{iM}(W_{iM},\theta^*_M)-\sqrt{\rho_{uM}\rho_{cM}}\mathbb{E}\big[m_{iM}(W_{iM},\theta^*_M)\big].
\end{equation}
%The proof below is a generalization of \cite{quah1990improved} and Theorem 6.8 in \cite{gallant1988unified} to random fields under finite populations.
The proof below is a generalization of Theorem 6.8 in \cite{gallant1988unified} to random fields under finite populations.
\begin{equation}\label{eq14}
\begin{aligned}
&\norm{\hat{S}_N(\hat{\theta}_N)-S_M-\rho_{uM}\rho_{cM}S_E}/J_M^2\\
\leq &\norm{\hat{S}_N(\hat{\theta}_N)-\hat{S}_N(\theta^*_M)}/J_M^2+\norm{\hat{S}_N(\theta^*_M)-\tilde{S}_M-\rho_{uM}\rho_{cM}S_E}/J_M^2\\
&+\norm{\tilde{S}_M-S_M}/J_M^2,
\end{aligned}
\end{equation}
where 
\begin{align*}
\tilde{S}_M=&\frac{1}{|D_M|}\sum_{i\in D_M}\mathbb{E}(\tilde{m}_{iM}\tilde{m}_{iM}')+\frac{1}{|D_M|}\sum_{i\in D_M}\sum_{j\in D_M: j\neq i, \nu(i,j)\leq b_M}\omega\bigg(\frac{\nu(i,j)}{b_M}\bigg)\mathbb{E}(\tilde{m}_{iM}\tilde{m}_{jM}').
\tag{\stepcounter{equation}\theequation}
\end{align*}

Since any sequence of symmetric matrices $\{A_N\}$ converges to a symmetric matrix $\{A_0\}$ if and only if $c'A_Nc\to c'A_0c$ for any vectors c, we can reach our conclusion by taking an arbitrary linear combination of $m_{iM}(W_{iM},\theta)$. From now on, we focus on the case of scalar $m_{iM}(W_{iM},\theta)$.

For the last term in the right-hand side of (\ref{eq14}), 
\begin{equation}\label{eq31}
\begin{aligned}
|\tilde{S}_M-S_M|/J_M^2\leq& \frac{1}{|D_M|}\Bigg|\sum_{i\in D_M}\sum_{j\in D_M: j\neq i, \nu(i,j)\leq b_M}\bigg[\omega\bigg(\frac{\nu(i,j)}{b_M}\bigg)-1\bigg]\cdot\mathbb{E}(\tilde{m}_{iM}\tilde{m}_{jM})\Bigg|/J_M^2\\
&+\frac{1}{|D_M|}\Bigg|\sum_{i\in D_M}\sum_{j\in D_M: \nu(i,j)> b_M}\mathbb{E}(\tilde{m}_{iM}\tilde{m}_{jM})\Bigg|/J_M^2.
\end{aligned}
\end{equation}
The first term in the right-hand side of (\ref{eq31}) goes to zero because of Assumption \ref{assump7}.
For the second term in the right-hand side of (\ref{eq31}),
\begin{equation}\label{eq18}
\begin{aligned}
&\frac{1}{|D_M|}\Bigg|\sum_{i\in D_M}\sum_{j\in D_M: \nu(i,j)> b_M}\mathbb{E}(\tilde{m}_{iM}\tilde{m}_{jM})\Bigg|/J_M^2\\
=&\frac{1}{|D_M|}\Bigg|\sum_{i\in D_M}\sum_{j\in D_M}\mathbb{E}(\tilde{m}_{iM}\tilde{m}_{jM})\Bigg|/J_M^2-\frac{1}{|D_M|}\Bigg|\sum_{i\in D_M}\sum_{j\in D_M, \nu(i,j)\leq b_M}\mathbb{E}(\tilde{m}_{iM}\tilde{m}_{jM})\Bigg|/J_M^2.
\end{aligned}
\end{equation}
Applying a similar argument in (B.30), $\frac{1}{|D_M|}\sum_{i\in D_M}\sum_{j\in D_M}\mathbb{E}(\tilde{m}_{iM}\tilde{m}_{jM}/J_M^2)$ is bounded. As $b_M\to \infty$ along with $M\to\infty$, the two sums on the right-hand side of (\ref{eq18}) approach to the same limit. 
As a result, 
\begin{equation}
|\tilde{S}_M-S_M|/J_M^2=o(1).
\end{equation}

In terms of the first term in the right hand side of (\ref{eq14}), take a mean value expansion of $\hat{S}_N(\hat{\theta}_N)$ around $\theta^*_M$. Let $\check{\theta}$ denote the mean value from this expansion. 
\begin{align*}
&|\hat{S}_N(\hat{\theta}_N)-\hat{S}_N(\theta^*_M)|/J_M^2\\
=&\Bigg|(\hat{\theta}_N-\theta^*_M)\frac{1}{|D_N|}\sum_{i\in D_M}\sum_{j\in D_M}R_{iM}R_{jM}\cdot  \omega\bigg(\frac{\nu(i,j)}{b_M}\bigg)\\
&\cdot\big[\nabla_\theta m_{iM}(W_{iM},\check{\theta})m_{jM}(W_{jM},\check{\theta})+ m_{iM}(W_{iM},\check{\theta})\nabla_\theta m_{jM}(W_{jM},\check{\theta})\big]/J_M^2\Bigg|\\
\leq & C_1 \big|\sqrt{|D_N|}(\hat{\theta}_N-\theta^*_M)\big|\frac{1}{|D_N|}\sum_{i\in D_N}\frac{1}{\sqrt{|D_N|}}\sum_{r=1}^{b_M} \sum_{j\in D_N: \nu(i,j)\in [r,r+1)}\sup_{\theta\in\Theta}\big|\nabla_\theta m_{iM}(W_{iM},\theta)\cdot m_{jM}(W_{jM},\theta)/J_M^2\big|\\
\leq& C \big|\sqrt{|D_N|}(\hat{\theta}_N-\theta^*_M)\big|\frac{1}{\sqrt{|D_N|}}\sum_{r=1}^{b_M} r^{d-1}\frac{1}{|D_N|}\sum_{i\in D_N}\sup_{\theta\in\Theta}\big|\nabla_\theta m_{iM}(W_{iM},\theta)\cdot m_{jM}(W_{jM},\theta)/J_M^2\big|
\tag{\stepcounter{equation}\theequation}
\end{align*}
Since 
\begin{align*}
&\mathbb{E}\Big[\frac{1}{|D_N|}\sum_{i\in D_N}\sup_{\theta\in\Theta}\big|\nabla_\theta m_{iM}(W_{iM},\theta)\cdot m_{jM}(W_{jM},\theta)/J_M^2\big|\Big]\\
\leq & \sup_{M,i\in D_M}\mathbb{E}\Big[\sup_{\theta\in\Theta}\big|\nabla_\theta m_{iM}(W_{iM},\theta)\cdot m_{jM}(W_{jM},\theta)/J_M^2\big|\Big]\\
\leq & \sup_{M,i\in D_M}\mathbb{E}\Big[\sup_{\theta\in\Theta}\big|\nabla_\theta m_{iM}(W_{iM},\theta)/c_{iM}\big|^2\Big]^{1/2}\cdot \sup_{M,i\in D_M}\mathbb{E}\Big[\sup_{\theta\in\Theta}\big| m_{iM}(W_{iM},\theta)/c_{iM}\big|^2\Big]^{1/2}<\infty,
\tag{\stepcounter{equation}\theequation}
\end{align*}
\begin{equation}
\frac{1}{|D_N|}\sum_{i\in D_N}\sup_{\theta\in\Theta}\big|\nabla_\theta m_{iM}(W_{iM},\theta)\cdot m_{jM}(W_{jM},\theta)/J_M^2\big|=O_p(1)
\end{equation}
by Markov's inequality. 
Given $b_M=o\big((|D_M|\rho_{uM}\rho_{cM})^{1/2d}\big)$, 
$\frac{1}{\sqrt{|D_N|}}\sum_{r=1}^{b_M} r^{d-1}=o_p(1)$.
Also, $\sqrt{|D_N|}(\hat{\theta}_N-\theta^*_M)=O_p(1)$ by Theorem \ref{thm1}. Hence, $|\hat{S}_N(\hat{\theta}_N)-\hat{S}_N(\theta^*_M)|/J_M^2=o_p(1)$.

Let us focus on the second term in the right-hand side of (\ref{eq14}) now.
\begin{align*}
\hat{S}_N(\theta^*_M)=&\frac{1}{|D_N|}\sum_{i\in D_M}\sum_{j\in D_M}R_{iM}R_{jM}\cdot \omega\bigg(\frac{\nu(i,j)}{b_M}\bigg)m_{iM}(W_{iM},\theta^*_M)m_{jM}(W_{jM},\theta^*_M)\\
=&\frac{|D_M|\rho_{uM}\rho_{cM}}{|D_N|}\frac{1}{|D_M|}\sum_{i\in D_M}\sum_{j\in D_M}\frac{R_{iM}R_{jM}}{\rho_{uM}\rho_{cM}} \omega\bigg(\frac{\nu(i,j)}{b_M}\bigg)m_{iM}(W_{iM},\theta^*_M)m_{jM}(W_{jM},\theta^*_M)\\
=&\big(1+o_p(1)\big)\frac{1}{|D_M|}\sum_{i\in D_M}\sum_{j\in D_M}\frac{R_{iM}R_{jM}}{\rho_{uM}\rho_{cM}} \omega\bigg(\frac{\nu(i,j)}{b_M}\bigg)m_{iM}(W_{iM},\theta^*_M)m_{jM}(W_{jM},\theta^*_M)
\tag{\stepcounter{equation}\theequation}
\end{align*}
by Lemma \ref{lemma1}.
It suffices to prove that 
\begin{align*}
&\Bigg|\frac{1}{|D_M|}\sum_{i\in D_M}\sum_{j\in D_M}\frac{R_{iM}R_{jM}}{\rho_{uM}\rho_{cM}} \omega\bigg(\frac{\nu(i,j)}{b_M}\bigg)m_{iM}(W_{iM},\theta^*_M)m_{jM}(W_{jM},\theta^*_M)-\tilde{S}_M-\rho_{uM}\rho_{cM}S_E\Bigg|/J_M^2\\
=&\Bigg|\frac{1}{J_M^2|D_M|}\sum_{i\in D_M}\sum_{j\in D_M}\omega\bigg(\frac{\nu(i,j)}{b_M}\bigg) \Bigg\{\frac{R_{iM}R_{jM}}{\rho_{uM}\rho_{cM}} m_{iM}(W_{iM},\theta^*_M)m_{jM}(W_{jM},\theta^*_M)\\
&-\mathbb{E}\bigg[\frac{R_{iM}R_{jM}}{\rho_{uM}\rho_{cM}} m_{iM}(W_{iM},\theta^*_M)m_{jM}(W_{jM},\theta^*_M)\bigg]\Bigg\}\Bigg|\\
\leq &\Bigg|\frac{1}{J_M^2|D_M|}\sum_{i\in D_M}\Big\{\frac{R_{iM}}{\rho_{uM}\rho_{cM}}m_{iM}(W_{iM},\theta^*_M)^2-\mathbb{E}\big[m_{iM}(W_{iM},\theta^*_M)^2\big]\Big\}\Bigg|\\
&+\Bigg|\frac{1}{J_M^2|D_M|}\sum_{i\in D_M}\sum_{j\in D_M: j\neq i, \nu(i,j)\leq b_M}\omega\bigg(\frac{\nu(i,j)}{b_M}\bigg) \Bigg\{\frac{R_{iM}R_{jM}}{\rho_{uM}\rho_{cM}} m_{iM}(W_{iM},\theta^*_M)m_{jM}(W_{jM},\theta^*_M)\\
&-\mathbb{E}\bigg[\frac{R_{iM}R_{jM}}{\rho_{uM}\rho_{cM}} m_{iM}(W_{iM},\theta^*_M)m_{jM}(W_{jM},\theta^*_M)\bigg]\Bigg\}\Bigg|
=o_p(1).
\tag{\stepcounter{equation}\theequation}
\end{align*}
Applying Lemma \ref{lemma2}, the first term in the right-hand side of the last inequality of (B.74) converges to zero in probability. Applying Lemma \ref{lemma5},  the second term in the right-hand side of the last inequality of (B.74) converges to zero in probability as well. 

Hence,
\begin{align*}
&\big|\hat{V}_{SN}-(V_M+\rho_{uM}\rho_{cM}V_E)\big|\\
=&\big|\big(H_M(\theta^*_M)^{-1}+o_p(1)\big)\big(S_M+\rho_{uM}\rho_{cM}S_E+o_p(1)\big)\big(H_M(\theta^*_M)^{-1}+o_p(1)\big)-(V_M+\rho_{uM}\rho_{cM}V_E)\big|\\
=&o_p(1).
\tag{\stepcounter{equation}\theequation}
\end{align*}

% pay attention to the transformation of the NED scaling factor 
\bigskip
\noindent
\textbf{Proof of Theorem \ref{thm4}:}

To simplify notation, from now on, we treat all functions as the original functions divided by $J_M$.

First, 
using similar arguments in the proof of Theorem \ref{thm1}, 
\begin{equation}
(\hat{\gamma}_N-\gamma^*_M)\overset{p}\to \textbf{0}.
\end{equation}

%From now on, define $\tilde{f}_{iM}(\theta)=f_{iM}(W_{iM},\theta)/J_M$, $\nabla_\theta\tilde{f}_{iM}(\theta)=\nabla_\theta f_{iM}(W_{iM},\theta)/J_M$, and $\tilde{V}_{f,M}=V_{f,M}/J_M$.
By the mean value expansion around $\theta^*_M$,
\begin{equation}\label{eq21}
\begin{aligned}
&V_{f,M}^{-1/2}\frac{1}{\sqrt{|D_N|}}\sum_{i\in D_M}R_{iM}f_{iM}(W_{iM},\hat{\theta}_N)\\
=&V_{f,M}^{-1/2}\frac{1}{\sqrt{|D_N|}}\sum_{i\in D_M}R_{iM}f_{iM}(W_{iM},\theta^*_M)\\
&+V_{f,M}^{-1/2}\frac{1}{|D_N|}\sum_{i\in D_M}R_{iM}\nabla_\theta f_{iM}(W_{iM},\check{\theta})\sqrt{|D_N|}(\hat{\theta}_N-\theta^*_M),
\end{aligned}
\end{equation}
where $\check{\theta}$ lies on the line segment connecting $\theta^*_M$ and $\hat{\theta}_N$.
Given Theorem \ref{thm1}, 
\begin{equation}
\sqrt{|D_N|}(\hat{\theta}_N-\theta^*_M)=O_p(1). 
\end{equation}
Further,
\begin{equation}
\hat{F}_N(\check{\theta})=F_M(\theta^*_M)+o_p(1).
\end{equation}
Therefore,
\begin{equation}\label{eq22}
\begin{aligned}
&V_{f,M}^{-1/2}\frac{1}{|D_N|}\sum_{i\in D_M}R_{iM}\nabla_\theta f_{iM}(W_{iM},\check{\theta})\sqrt{|D_N|}(\hat{\theta}_N-\theta^*_M)\\
=&V_{f,M}^{-1/2}F_M(\theta^*_M)\sqrt{|D_N|}(\hat{\theta}_N-\theta^*_M)+o_p(1).
\end{aligned}
\end{equation}

According to the mean value expansion in the proof of the asymptotic normality of \\$\sqrt{|D_N|}(\hat{\theta}_N-\theta^*_M)$,
\begin{equation}\label{eq23}
\sqrt{|D_N|}(\hat{\theta}_N-\theta^*_M)=-\frac{1}{\sqrt{|D_N|}}\sum_{i\in D_M}R_{iM}H_M(\theta^*_M)^{-1}m_{iM}(W_{iM},\theta^*_M)+o_p(1).
\end{equation}
Combining (\ref{eq21}), (\ref{eq22}), and (\ref{eq23}),
\begin{equation}\label{eq24}
\begin{aligned}
&V_{f,M}^{-1/2}\frac{1}{\sqrt{|D_N|}}\sum_{i\in D_M}R_{iM}f_{iM}(W_{iM},\hat{\theta}_N)\\
=&V_{f,M}^{-1/2}\frac{1}{\sqrt{|D_N|}}\sum_{i\in D_M}R_{iM}\big[f_{iM}(W_{iM},\theta^*_M)-F_M(\theta^*_M)H_M(\theta^*_M)^{-1}m_{iM}(W_{iM},\theta^*_M)\big]+o_p(1).
\end{aligned}
\end{equation}

Subtract $V_{f,M}^{-1/2}\sqrt{|D_N|}\gamma^*_M$ from both sides of (\ref{eq24}).
\begin{align*}
&V_{f,M}^{-1/2}\sqrt{|D_N|}\big(\hat{\gamma}-\gamma^*_M\big)\\
=&V_{f,M}^{-1/2}\frac{1}{\sqrt{|D_N|}}\sum_{i\in D_M}R_{iM}\big[f_{iM}(W_{iM},\theta^*_M)-\gamma^*_M-F_M(\theta^*_M)H_M(\theta^*_M)^{-1}m_{iM}(W_{iM},\theta^*_M)\big]+o_p(1)\\
=&V_{f,M}^{-1/2}\sqrt{\frac{|D_M|\rho_{uM}\rho_{cM}}{|D_N|}}\frac{1}{\sqrt{|D_M|}}\sum_{i\in D_M}\frac{R_{iM}}{\sqrt{\rho_{uM}\rho_{cM}}}\big[f_{iM}(W_{iM},\theta^*_M)-\gamma^*_M\\
&-F_M(\theta^*_M)H_M(\theta^*_M)^{-1}m_{iM}(W_{iM},\theta^*_M)\big]+o_p(1)\\
=&\big(1+o_p(1)\big)V_{f,M}^{-1/2}\frac{1}{\sqrt{|D_M|}}\sum_{i\in D_M}\frac{R_{iM}}{\sqrt{\rho_{uM}\rho_{cM}}}\big[f_{iM}(W_{iM},\theta^*_M)-\gamma^*_M\\
&-F_M(\theta^*_M)H_M(\theta^*_M)^{-1}m_{iM}(W_{iM},\theta^*_M)\big]+o_p(1)\tag{\stepcounter{equation}\theequation}
\end{align*}
Observe that $\forall\ \theta\in\Theta$
\begin{align*}
&\sup_{M,i\in D_M}\mathbb{E}\Big[\left\|\big[f_{iM}(W_{iM},\theta)-\gamma^*_M-F_M(\theta^*_M)H_M(\theta^*_M)^{-1}m_{iM}(W_{iM},\theta)\big]\right\|^{2+\delta}\Big]\\
\leq&\sup_{M,i\in D_M}\Bigg\{\Big[\mathbb{E}\Big(\sup_{\theta\in\Theta}\left\|f_{iM}(W_{iM},\theta)\right\|^{2+\delta}\Big)\Big]^{1/(2+\delta)}+\left\|\gamma^*_M\right\|\\
&+C\left\|F_M(\theta^*_M)\right\|\Big[\mathbb{E}\Big(\sup_{\theta\in\Theta}\left\|m_{iM}(W_{iM},\theta)\right\|^{2+\delta}\Big)\Big]^{1/(2+\delta)}\Bigg\}^{2+\delta}<\infty \tag{\stepcounter{equation}\theequation}
\end{align*}
for some $\delta>0$ by Minkowski's inequality and Jensen's inequality.
By Lemma \ref{lemma4} above and Corollary 1 in \cite{jenish2012spatial},
\begin{equation}
\begin{aligned}
&V_{f,M}^{-1/2}\frac{1}{\sqrt{|D_M|}}\sum_{i\in D_M}\Bigg\{\frac{R_{iM}}{\sqrt{\rho_{uM}\rho_{cM}}}\big[f_{iM}(W_{iM},\theta^*_M)-\gamma^*_M-F_M(\theta^*_M)H_M(\theta^*_M)^{-1}m_{iM}(W_{iM},\theta^*_M)\big]\\
&-\sqrt{\rho_{uM}\rho_{cM}}\Big[\mathbb{E}\big(f_{iM}(W_{iM},\theta^*_M)\big)-\gamma^*_M-F_M(\theta^*_M)H_M(\theta^*_M)^{-1}\mathbb{E}\big(m_{iM}(W_{iM},\theta^*_M)\big)\Big]\Bigg\}\overset{d}\to \mathcal{N}(\textbf{0},I_q).
\end{aligned}
\end{equation}

The proof of Theorem \ref{thm4}(2) follows closely from the proof of Theorem \ref{thm3} and hence is omitted.

\bibliography{finitepop_spatial}

\begin{thebibliography}{25}
\providecommand{\natexlab}[1]{#1}
\providecommand{\url}[1]{\texttt{#1}}
\providecommand{\urlprefix}{URL }
\expandafter\ifx\csname urlstyle\endcsname\relax
  \providecommand{\doi}[1]{doi:\discretionary{}{}{}#1}\else
  \providecommand{\doi}{doi:\discretionary{}{}{}\begingroup
  \urlstyle{rm}\Url}\fi
\providecommand{\eprint}[2][]{\url{#2}}

\bibitem[{Abadie et~al.(2017)Abadie, Athey, Imbens, and
  Wooldridge}]{abadie2017should}
Abadie, A., Athey, S., Imbens, G.W., and Wooldridge, J. (2017), When should you
  adjust standard errors for clustering? Tech. rep., National Bureau of
  Economic Research.

\bibitem[{Abadie et~al.(2020)Abadie, Athey, Imbens, and
  Wooldridge}]{abadie2020sampling}
Abadie, A., Athey, S., Imbens, G.W., and Wooldridge, J.M. (2020),
  Sampling-based versus design-based uncertainty in regression analysis.
  \emph{Econometrica} 88(1), 265--296.

\bibitem[{Aliprantis and Hartley(2015)}]{aliprantis2015blowing}
Aliprantis, D. and Hartley, D. (2015), Blowing it up and knocking it down: The
  local and city-wide effects of demolishing high concentration public housing
  on crime. \emph{Journal of Urban Economics} 88, 67--81.

\bibitem[{Bojinov et~al.(2021)Bojinov, Rambachan, and
  Shephard}]{bojinov2021panel}
Bojinov, I., Rambachan, A., and Shephard, N. (2021), Panel experiments and
  dynamic causal effects: A finite population perspective. \emph{Quantitative
  Economics} 12(4), 1171--1196.

\bibitem[{Bradley and Tone(2017)}]{bradley2017central}
Bradley, R.C. and Tone, C. (2017), A central limit theorem for non-stationary
  strongly mixing random fields. \emph{Journal of Theoretical Probability}
  30(2), 655--674.

\bibitem[{Gallant and White(1988)}]{gallant1988unified}
Gallant, A.R. and White, H. (1988), \emph{A unified theory of estimation and
  inference for nonlinear dynamic models}. Blackwell.

\bibitem[{Houde(2012)}]{houde2012spatial}
Houde, J.F. (2012), Spatial differentiation and vertical mergers in retail
  markets for gasoline. \emph{American Economic Review} 102(5), 2147--2182.

\bibitem[{Hudgens and Halloran(2008)}]{hudgens2008toward}
Hudgens, M.G. and Halloran, M.E. (2008), Toward causal inference with
  interference. \emph{Journal of the American Statistical Association}
  103(482), 832--842.

\bibitem[{Jenish and Prucha(2009)}]{jenish2009central}
Jenish, N. and Prucha, I.R. (2009), Central limit theorems and uniform laws of
  large numbers for arrays of random fields. \emph{Journal of Econometrics}
  150(1), 86--98.

\bibitem[{Jenish and Prucha(2012)}]{jenish2012spatial}
Jenish, N. and Prucha, I.R. (2012), On spatial processes and asymptotic
  inference under near-epoch dependence. \emph{Journal of Econometrics} 170(1),
  178--190.

\bibitem[{Leung(2022)}]{leung2022causal}
Leung, M.P. (2022), Causal inference under approximate neighborhood
  interference. \emph{Econometrica} 90(1), 267--293.

\bibitem[{Manski(2013)}]{manski2013identification}
Manski, C.F. (2013), Identification of treatment response with social
  interactions. \emph{The Econometrics Journal} 16(1), S1--S23.

\bibitem[{M{\'o}ricz et~al.(2008)M{\'o}ricz, Stadtm{\"u}ller, and
  Thalmaier}]{moricz2008strong}
M{\'o}ricz, F., Stadtm{\"u}ller, U., and Thalmaier, M. (2008), Strong laws for
  blockwise m-dependent random fields. \emph{Journal of Theoretical
  Probability} 21(3), 660--671.

\bibitem[{M{\"u}ller and Watson(2022{\natexlab{a}})}]{muller2022spatiala}
M{\"u}ller, U.K. and Watson, M.W. (2022{\natexlab{a}}), Spatial correlation
  robust inference. \emph{Econometrica} .

\bibitem[{M{\"u}ller and Watson(2022{\natexlab{b}})}]{muller2022spatialb}
M{\"u}ller, U.K. and Watson, M.W. (2022{\natexlab{b}}), Spatial correlation
  robust inference in linear regression and panel models. \emph{Journal of
  Business \& Economic Statistics} pp. 1--15.

\bibitem[{Newey(1991)}]{newey1991uniform}
Newey, W.K. (1991), Uniform convergence in probability and stochastic
  equicontinuity. \emph{Econometrica} 59, 1161--1167.

\bibitem[{Newey and McFadden(1994)}]{newey1994large}
Newey, W.K. and McFadden, D. (1994), Large sample estimation and hypothesis
  testing. \emph{Handbook of Econometrics} 4, 2111--2245.

\bibitem[{Neyman(1923)}]{neyman1923application}
Neyman, J. (1923), On the application of probability theory to agricultural
  experiments. essay on principles. section 9. \emph{Annals of Agricultural
  Sciences} 10, 1--51.

\bibitem[{Pinkse et~al.(2007)Pinkse, Shen, and Slade}]{pinkse2007central}
Pinkse, J., Shen, L., and Slade, M. (2007), A central limit theorem for
  endogenous locations and complex spatial interactions. \emph{Journal of
  Econometrics} 140(1), 215--225.

\bibitem[{Rambachan and Roth(2022)}]{rambachan2020design}
Rambachan, A. and Roth, J. (2022), Design-based uncertainty for
  quasi-experiments. \emph{arXiv preprint arXiv:2008.00602} .

\bibitem[{S{\"a}vje(2021)}]{savje2021causal}
S{\"a}vje, F. (2021), Causal inference with misspecified exposure mappings.
  Tech. rep., Yale University.

\bibitem[{S{\"a}vje et~al.(2021)S{\"a}vje, Aronow, and
  Hudgens}]{savje2021average}
S{\"a}vje, F., Aronow, P.M., and Hudgens, M.G. (2021), Average treatment
  effects in the presence of unknown interference. \emph{The Annals of
  Statistics} 49(2), 673--701.

\bibitem[{Xu(2021{\natexlab{a}})}]{xu2021asymptotic}
Xu, R. (2021{\natexlab{a}}), Asymptotic properties of m-estimators with finite
  populations under cluster sampling and cluster assignment. Tech. rep.,
  Rutgers University.

\bibitem[{Xu(2021{\natexlab{b}})}]{xu2021potential}
Xu, R. (2021{\natexlab{b}}), Potential outcomes and finite-population inference
  for m-estimators. \emph{The Econometrics Journal} 24(1), 162--176.

\bibitem[{Xu and Lee(2019)}]{xu2019theoretical}
Xu, X. and Lee, L.f. (2019), Theoretical foundations for spatial econometric
  research. \emph{Regional Science and Urban Economics} 76, 2--12.

\end{thebibliography}

\end{document}